\documentclass[pre,aps,amsmath,amssymb,amsfonts,floatfix,superscriptaddress,showpacs,twocolumn,nofootinbib]{revtex4-1}
\usepackage{graphicx}
\usepackage{color}
\usepackage{bbm}
\usepackage[utf8]{inputenc}
\usepackage{dcolumn}
\usepackage{hyperref}
\usepackage[caption=false]{subfig}
\usepackage[english]{babel}
\usepackage{indentfirst}
\usepackage{textcomp}
\hypersetup{colorlinks=true,breaklinks,linkcolor=blue,urlcolor=blue,citecolor=blue}

\usepackage{tikz}
\usetikzlibrary{calc}
\usetikzlibrary{decorations.pathmorphing}
\usetikzlibrary{decorations.pathreplacing}

\renewcommand{\i}{{\ensuremath{i}}}
\newcommand\Let{\mathrel{\mathop:\!\!=}}

\newcommand{\iom}{{\ensuremath{\i\omega}}}
\newcommand{\inu}{{\ensuremath{\i\nu}}}
\newcommand{\kv}{\ensuremath{\mathbf{k}}}

\newcommand{\qv}{\ensuremath{\mathbf{q}}}
\newcommand{\rv}{\ensuremath{\mathbf{r}}} 
\newcommand{\ev}{\ensuremath{\mathbf{e}}}
\newcommand{\dv}{\ensuremath{\boldsymbol{\delta}}} 
\newcommand{\Av}{\ensuremath{\mathbf{A}}}
\newcommand{\jv}{\ensuremath{\mathbf{j}}}
\newcommand{\vc}[1]{\ensuremath{\mathbf{#1}}}
\newcommand{\abs}[1]{\ensuremath{\lvert#1\rvert}}
\newcommand{\Ttau}{\ensuremath{T_{\tau}}}

\newcommand{\av}[1]{\ensuremath{\left\langle #1 \right\rangle}}

\newcommand{\up}{\ensuremath{\uparrow}}
\newcommand{\dn}{\ensuremath{\downarrow}}
\newcommand{\tti}{\ensuremath{\tilde{t}}}

\def \Im {\mathop {\rm Im}}
\def \Re {\mathop {\rm Re}}

\begin{document}

\title{Collective Charge Excitations of Strongly Correlated Electrons,\\
Vertex Corrections and Gauge Invariance}

\author{Hartmut Hafermann}
\affiliation{Institut de Physique Th\'eorique (IPhT), CEA, CNRS, 91191 Gif-sur-Yvette, France}

\author{Erik G. C. P. van Loon}
\affiliation{Radboud University Nijmegen, Institute for Molecules and Materials, NL-6525 AJ Nijmegen, The Netherlands}

\author{Mikhail I. Katsnelson}
\affiliation{Radboud University Nijmegen, Institute for Molecules and Materials, NL-6525 AJ Nijmegen, The Netherlands}

\author{Alexander I. Lichtenstein}
\affiliation{I. Institut f\"ur Theoretische Physik, Universit\"at Hamburg, Jungiusstra\ss e 9, D-20355 Hamburg, Germany}

\author{Olivier Parcollet}
\affiliation{Institut de Physique Th\'eorique (IPhT), CEA, CNRS, 91191 Gif-sur-Yvette, France}

\begin{abstract}
We consider the collective, long-wavelength charge excitations in correlated media in presence of short- and long-range forces.
As an example for the case of a short-range interaction, we examine the two-dimensional Hubbard model within dynamical mean-field theory (DMFT). It is shown explicitly that the DMFT susceptibility including vertex corrections respects the Ward identity and yields a manifestly gauge-invariant response in finite dimensions.
For computing the susceptibility, we use a different expression and establish its formal equivalence to the standard DMFT formula. It allows for a more stable analytical continuation. We find a zero-sound mode expected for short-range forces. The relation between the vertex corrections, gauge invariance, and the appearance of the collective modes is discussed.
Long-range forces are treated within extended dynamical mean-field theory.
In order to obtain a gauge-invariant response, it is necessary to additionally incorporate some \emph{non local} vertex corrections into the polarization. In doing so, we obtain plasmons in the three-dimensional Hubbard model.
The plasma frequency is determined by the (single-particle) density distribution as a consequence of gauge invariance. We compare this result with the plasma frequency extracted from the analytical continuation of the susceptibility. It is in good agreement with the prediction from the gauge-invariance condition.
\end{abstract}


\maketitle

\clearpage

\section{Introduction}

In the last two decades, dynamical mean-field theory (DMFT)~\cite{Georges96} and its various cluster~\cite{Maier05} and diagrammatic~\cite{Toschi07,Rubtsov08,Rubtsov09,Rubtsov12,Rohringer13} extensions have emerged as promising and useful tools to solve several aspects of strongly correlated fermion problems.
For example, DMFT has shed new light on the Mott transition problem~\cite{Georges96} and cluster extensions of DMFT have successfully described some aspects of high-temperature superconductors (see Ref.~\onlinecite{Gull13} for a recent example). 
Furthermore, DMFT is now routinely used in combination with density functional theory 
to provide an \emph{ab initio} electronic-structure method for strongly correlated systems~\cite{Kotliar06}.
For long-range interactions, the extended DMFT (EDMFT)~\cite{Si96,Kajueter96,Smith00,Chitra01} as well as more refined GW+DMFT approaches~\cite{Sun02,Biermann03,Ayral12,Ayral13} have been developed.

Two-particle quantities and response functions can also be computed within the DMFT theoretical framework~\cite{Georges96}.
For example, systematic computations of antiferromagnetic or superconducting susceptibilities
from high to low temperatures have been used to locate continuous phase transitions in cluster extensions of DMFT (see, e.g., Refs.~\onlinecite{Jarrell93,Kent05,Maier05-2}).
More recently, interest in two-particle vertex functions is increasing~\cite{Rohringer12, Kinza13, Huang13}. A new generation of approaches~\cite{Toschi07,Rubtsov08,Rubtsov12,Rohringer13} has emerged which uses certain two-particle functions in the self-consistency condition itself.
From a technical point of view, the task of computing these two-particle quantities is  
significantly more challenging than the computation of the single-particle quantities used in the DMFT 
self-consistency loop, because of the need to include vertex corrections.
However, due to the advent of continuous-time quantum Monte Carlo solvers~\cite{Rubtsov05,Werner06,Gull08},
they can now be computed reliably and up to high precision~\cite{Gull11}.

In this paper, we study in particular the long-wavelength collective charge excitations in the Hubbard and the extended Hubbard models in presence of short-range and long-range interactions, respectively. For short-range interactions, we use a regular DMFT scheme and obtain a zero sound mode in the metallic regime which persists up to the Mott transition.
For long-range interactions, we use a simplified version of the dual boson method and obtain a plasmon mode.
In the latter case, we discuss the failure of EDMFT to properly describe the low-energy excitations, even at a qualitative level.
In both the short- and the long-range cases, we obtain a low energy analysis similar to the standard textbook weak-coupling random phase approximation (RPA) analysis.

From a technical point of view, in order to compute the two-particle response, we employ a formula inspired from recent work on the dual boson approach~\cite{Rubtsov12}. While we prove it to be mathematically equivalent to the standard computation procedure of computing the DMFT susceptibility~\cite{Georges96},
it turns out to yield a much better numerical accuracy in practice.

Moreover, we show that a proper and complete treatment of the (non local) vertex corrections in the DMFT framework is essential for the correctness of the result at low energy. 
Simplified approximations (like a simple bubble approximation) lead to qualitatively wrong results. 
We trace the origin of this difficulty in constructing simple approximations to the role of gauge invariance and the associated Ward identities.
It has been known since the 1960's that gauge invariance is closely related to the collective modes~\cite{Nambu60} and 
the criteria for obtaining conserving approximations that respect gauge invariance have been formulated at the time~\cite{Baym61,Baym62}.
For weak coupling, these requirements are fulfilled within the RPA. 
However, the description of correlated systems requires a frequency-dependent self-energy. Designing gauge invariant approximations is much less straightforward in this case, because dynamical vertex corrections are required. We check explicitly that the DMFT susceptibility, which includes these corrections, fulfills the Ward identity in finite dimensions.
While our methodology can straightforwardly be generalized to treat the magnetic (spin) excitations and we expect vertex corrections to be important in general, we focus here on the charge excitations.

The paper is organized as follows: We first introduce the model in Sec.~\ref{sec:model}.
In Sec.~\ref{sec:shortrange}, we consider the case of short-range forces on the level of DMFT. We first recall DMFT and the standard calculation of susceptibilities and then introduce the new formula for the response function that follows from the dual boson approach. Results for the charge response are discussed in detail in Sec.~\ref{sec:results} and are compared to the RPA. We then discuss gauge invariance and show explicitly that it is fulfilled within DMFT.
In Sec.~\ref{sec:longrange}, we address the case of a long-range interaction. We show that EDMFT does not provide a valid description of plasmons. By including non-local vertex corrections into the polarization within the dual boson approximation, we demonstrate that the polarization obtains the proper momentum dependence required by gauge invariance. The energy of the appearing collective mode is compared with the plasma frequency and thereby identified as a plasmon mode.
A detailed derivation of the employed relations and a proof of the equivalence of the dual boson formula and the DMFT susceptibility are provided in the Appendix.

\section{Model}
\label{sec:model}

In the following, we consider the extended Hubbard model in finite dimensions. In particular, we focus on the two-dimensional square and three-dimensional cubic lattices. The model is described by the Hamiltonian
\begin{align}
\label{hmlt}
H = &- \tti \sum_{\rv\dv\sigma} \left(c^\dagger_{\rv\sigma}c_{\rv-\dv\sigma} + c^\dagger_{\rv-\dv\sigma}c_{\rv\sigma}\right)\notag\\
&+ U\sum_{\rv} n_{\rv\up} n_{\rv\dn}+ \frac{1}{2} \sum_{\rv\rv'} V(\rv-\rv') n_{\rv}n_{\rv'}.
\end{align}
Here, $\rv$ denote the \emph{discrete} positions of the lattice sites and the sum over $\dv$ implies a sum over the displacement vectors $\dv=a(1,0,0)$, $a(0,1,0)$ in two dimensions and additionally $a(0,0,1)$ in the three-dimensional case, respectively.
For simplicity, we restrict ourselves to nearest-neighbor hopping $\tilde{t}$ only. The tilde is used to distinguish from the symbol $t$ which is used for time.  The lattice spacing $a$ is set to unity in the following.
We further denote spin by $\sigma=\uparrow,\downarrow$ and $n=n_{\uparrow}+n_{\downarrow}$.
In the above, we have written the local Hubbard interaction with Coulomb repulsion $U$ explicitly. The last term contains the non local part of the interaction, which may be long-ranged. Its Fourier transform will be denoted $V(\qv)$. For the Hubbard model, $V(\qv)=0$. The energy unit is chosen such that $4\tti =1$ in both two and three dimensions and all results are obtained at temperature $T=0.02$.

\section{Short-range forces}
\label{sec:shortrange}

In order to address the collective excitations in presence of short-range forces and strong correlations, we consider the two-dimensional Hubbard model, which can be treated within dynamical mean-field theory (DMFT).

First, we briefly recall the DMFT procedure and the calculation of the susceptibilities in DMFT, as they can be found in the review of Ref.~\onlinecite{Georges96}.

\subsection{Recollection of DMFT}

In DMFT, the lattice problem \eqref{hmlt} with $V\equiv 0$ is mapped onto a local quantum impurity problem subject to a self-consistency condition. The lattice Green's function has the form
\begin{align}
\label{glat}
G_{\nu}(\kv) = \frac{1}{\inu+\mu-\varepsilon_{\kv}-\Sigma_{\nu}},
\end{align}
where $\varepsilon_{\kv}$ is the Fourier transform of the hopping, $\Sigma_{\nu}$ is the local but frequency dependent electronic self-energy and $\nu$ stands for the discrete Matsubara frequencies $\nu_{n}=(2n+1)\pi T$ with $T$ denoting temperature. Here and in the following it is convenient to write frequency labels as subscripts to obtain a more condensed notation.
We further consider the paramagnetic case and spin labels are omitted. In DMFT, the self-energy is a functional of the local Green's function only and has to be determined self-consistently. In practice, it is obtained from the solution of an Anderson impurity model, which, starting from an initial guess, is solved repeatedly until the following self-consistency condition is fulfilled:
\begin{align}
\label{dmftsc}
g_{\nu}= \frac{1}{N}\sum_{\kv}G_{\nu}(\kv).
\end{align}
It relates the local part of the lattice Green's function to the impurity Green's function denoted $g_{\nu}$ and implicitly determines the self-energy.

\subsection{DMFT susceptibility}
\label{dmftsusc}

\begin{figure}[t]
\includegraphics[scale=0.4,angle=0]{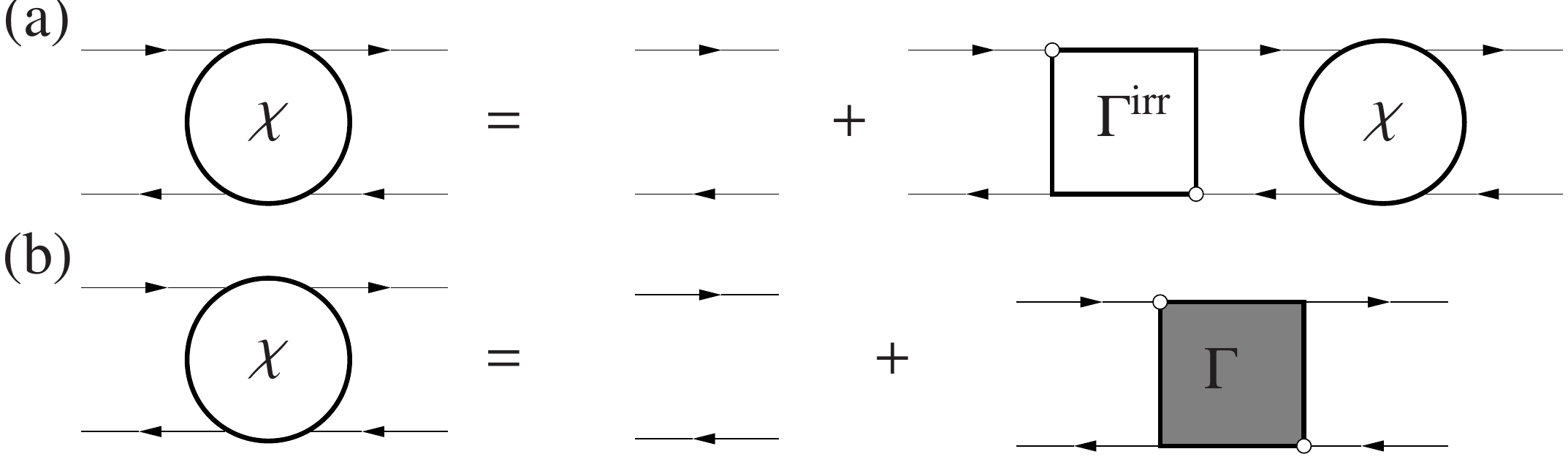} 
\caption{\label{fig:susc} Diagrammatic representation of (a) the Bethe-Salpeter equation for the generalized susceptibility $\chi$ and (b) the relation of the generalized susceptibility and the reducible vertex function $\Gamma$. Lines are fully dressed propagators.
}
\end{figure}

Response functions may be computed once a self-consistent solution to the DMFT equations has been found.
Here and in the remainder of the paper, we will focus on the charge response.
The charge susceptibility is given by the connected part of the density-density correlation function\footnote{We define the susceptibility with a minus sign relative to the convention used in Ref.~\onlinecite{Georges96}.}
\begin{align}
\label{chilatdef}
\chi_{\omega}(\qv) \Let -\av{n_{\omega}(\qv)n_{\omega}(\qv)}_{\text{con.}},
\end{align}
where $n=\sum_{\sigma} n_{\sigma}$ is the operator of the total density.
The charge susceptibility is expressed in terms of the generalized susceptibility $\chi_{\nu\nu'\omega}(\qv)$ as
\begin{align}
\chi_{\omega}(\qv) = 2T^{2}\sum_{\nu\nu'}\chi_{\nu\nu'\omega}(\qv),
\end{align}
where the factor of $2$ stems from the spin degeneracy.
The generalized susceptibility in turn is the solution to an integral equation which involves an irreducible vertex. Defining
\begin{align}
\label{chinuomega}
\chi^{0}_{\nu\omega}(\qv) = \frac{1}{N}\sum_{\kv}G_{\nu+\omega}(\kv+\qv)G_{\nu}(\kv),
\end{align}
this equation reads as
\begin{align}
\label{gensusc}
\chi_{\nu\nu'\omega}(\qv) =& \frac{1}{T} \chi^{0}_{\nu\omega}(\qv)\delta_{\nu\nu'} -\chi^{0}_{\nu\omega}(\qv)\,T\!\sum_{\nu''}\Gamma^{\text{irr}}_{\nu\nu''\omega} \chi_{\nu''\nu'\omega}(\qv).
\end{align}
It is depicted in graphically Fig.~\ref{fig:susc} (a). In DMFT, the \emph{irreducible} vertex $\Gamma^{\text{irr}}$ is given by the irreducible vertex of the impurity $\gamma^{\text{irr}}$, i.e. $\Gamma^{\text{irr}}\equiv\gamma^{\text{irr}}$, and hence local. In practice it is extracted from the impurity model on the final DMFT iteration by inverting the local Bethe-Salpeter equation
\begin{align}
[\gamma_{\omega}^{-1}]_{\nu\nu'} = [\gamma^{\text{irr}\,-1}_{\omega}]_{\nu\nu'}^{-1} + T\chi^{0}_{\nu\omega}\delta_{\nu\nu'}.
\end{align}
Here $\chi^{0}_{\nu\omega}=g_{\nu+\omega}g_{\nu}$ and the reducible impurity vertex $\gamma$ is defined through
\begin{align}
\gamma_{\nu\nu'\omega}^{\sigma\sigma'} \Let \frac{\av{c_{\nu\sigma}c^{*}_{\nu+\omega,\sigma}c_{\nu'+\omega,\sigma'}c^{*}_{\nu'\sigma'}}- \chi^{0\,\sigma\sigma'}_{\nu\nu'\omega}}{g_{\nu\sigma}g_{\nu+\omega,\sigma}g_{\nu'+\omega\sigma'}g_{\nu'\sigma'}}
\end{align}
and
\begin{align}
\chi^{0\,\sigma\sigma'}_{\nu\nu'\omega} \Let (g_{\nu\sigma}g_{\nu'\sigma'}\delta_{\omega} - g_{\nu+\omega,\sigma}g_{\nu\sigma}\delta_{\nu\nu'}\delta_{\sigma\sigma'})/T
\end{align}
(see also Appendix \ref{app:proof}).

\begin{figure}[b]
\includegraphics[scale=0.4,angle=0]{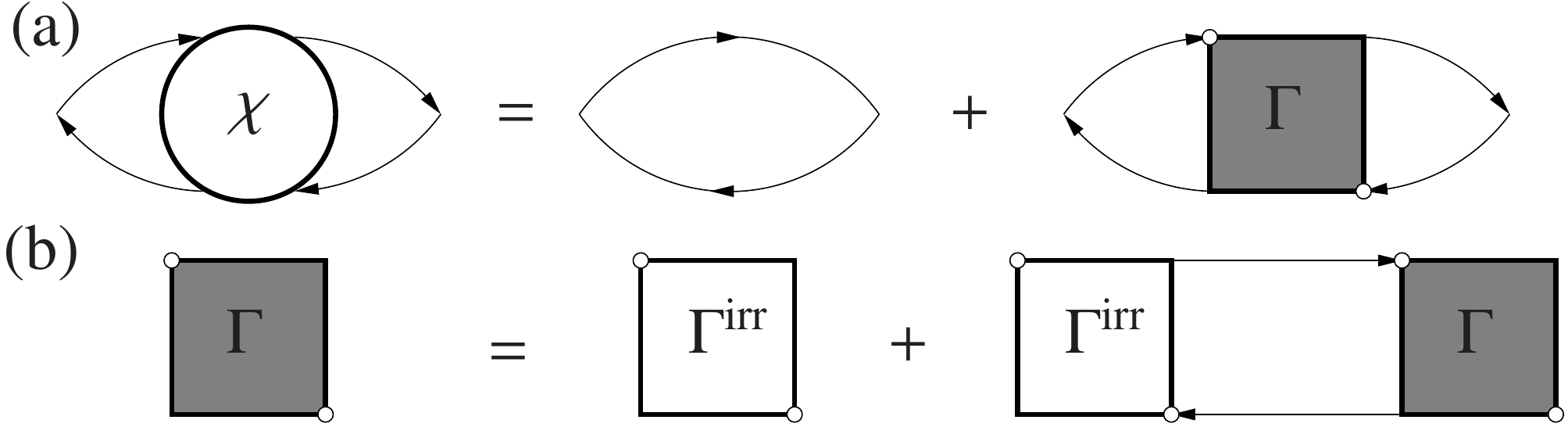} 
\caption{\label{fig:bse} Diagrammatical representation of (a) the susceptibility and (b) the Bethe-Salpeter equation for the vertex function. Lines are fully dressed propagators.
}
\end{figure}

In an equivalent formulation, the susceptibility is expressed in terms of the (reducible) vertex function of the lattice as follows:
\begin{align}
\label{chi}
\chi_{\omega}(\qv) = 2T\sum_{\nu}\chi^{0}_{\nu\omega}(\qv) - 2T^{2}\sum_{\nu\nu'} \chi^{0}_{\nu\omega}(\qv)\Gamma_{\nu\nu'\omega}(\qv) \chi^{0}_{\nu'\omega}(\qv).
\end{align}
This relation is graphically depicted in Fig.~\ref{fig:bse} (a).
The vertex function is obtained as the solution to the integral equation
\begin{align}
\label{bse}
\Gamma_{\nu\nu'\omega}(\qv) = \Gamma_{\nu\nu'\omega}^{\text{irr}} - T\sum_{\nu''} \Gamma_{\nu\nu''\omega}^{\text{irr}}\, \chi^{0}_{\nu''\omega}(\qv) \Gamma_{\nu''\nu'\omega}(\qv),
\end{align}
which is called the Bethe-Salpeter equation [BSE, see Fig.~\ref{fig:bse} (b)].
Diagrammatically, the BSE corresponds to the infinite sum of ladder-like diagrams to the vertex function with DMFT Green's functions as rails and the irreducible vertex appearing as rungs of the ladders. This can be seen by iterating it. Its physical content are the repeated particle-hole scattering processes which give rise to the collective excitations of the system. The BSE has the formal solution
\begin{align}
[\Gamma_{\omega}^{-1}(\qv)]_{\nu\nu'} = [\Gamma_{\omega}^{\text{irr}\,-1}]_{\nu\nu'} + T\chi_{\nu\omega}^{0}(\qv)\delta_{\nu\nu'}.
\end{align}
The lattice vertex depends on the transferred momentum $\qv$ only due to the locality of the irreducible vertex.

The equivalence between the two approaches is readily established using the representation of the generalized susceptibility in terms of the vertex function,
\begin{align}
\label{chigamma}
\chi_{\nu\nu'\omega}(\qv) =& \frac{1}{T} \chi^{0}_{\nu\omega}(\qv)\delta_{\nu\nu'}
 -\chi^{0}_{\nu\omega}(\qv)\Gamma_{\nu\nu'\omega}(\qv)\chi^{0}_{\nu'\omega}(\qv)
\end{align}
as shown in Fig.~\ref{fig:susc} (b). Inserting \eqref{chigamma} into the right-hand side of Eq.~\eqref{gensusc} and using the Bethe-Salpeter equation for the vertex function \eqref{bse} again recovers \eqref{chigamma}.

\subsection{Alternative expression for the susceptibility}
\label{sec:chialt}

In the following, we are primarily interested in the susceptibility on the real frequency axis. This requires an analytical continuation from Matsubara frequencies. The details of the analytical continuation are summarized in Appendix~\ref{app:pade}. Here, we discuss the calculation of the susceptibility on Matsubara frequencies.
Instead of a straightforward implementation of the equations discussed in the previous section, we employ the following approach, which leads to better results and less artifacts. It is a reformulation of the above equations and was first derived in the context of the dual boson approach~\cite{Rubtsov12}. In Appendix~\ref{app:proof}, we show that this formulation is exactly equivalent to the DMFT susceptibility [Eqs.~\eqref{chi}-\eqref{bse}].

In the alternative formulation, the susceptibility is separated into a local impurity and a lattice contribution:
\begin{align}
\label{chialt}
\chi_{\omega}(\qv)=\chi_{\omega} + 2T^{2}\sum_{\nu\nu'}\chi_{\omega}\lambda_{\nu\omega} \tilde{\chi}_{\nu\nu'\omega}(\qv) \lambda_{\nu'+\omega,-\omega}\chi_{\omega}.
\end{align}
Here, $\chi_{\omega}\Let-\av{n_{\omega}n_{\omega}}_{\text{con.}}$ denotes the impurity susceptibility which includes local vertex corrections. The lattice vertex $\Gamma$ includes non-local vertex corrections.
The expression further involves the three-leg vertex of the impurity model $\lambda$ (see Appendix \ref{app:proof}) 
\begin{align}
\lambda_{\nu\omega}^{\sigma} &\Let \frac{-\av{c_{\nu\sigma}c^{*}_{\nu+\omega,\sigma}n_{\omega}} - g_{\nu\sigma}\av{n}\delta_{\omega}/T}{g_{\nu\sigma}g_{\nu+\omega,\sigma}\chi_{\omega}}
\end{align}
and we have defined
\begin{align}
\label{dualgensusc}
\tilde{\chi}_{\nu\nu'\omega}(\qv)= \frac{1}{T}\tilde{\chi}^{0}_{\nu\omega}(\qv)\delta_{\nu\nu'} -\tilde{\chi}^{0}_{\nu\omega}(\qv)\Gamma_{\nu\nu'\omega}(\qv)\tilde{\chi}^{0}_{\nu'\omega}(\qv).
\end{align}
In the above, $\tilde{\chi}_{\nu\omega}(\qv)$ denotes the \emph{non-local part} of the bubble:
\begin{align}
\label{chidual}
\tilde{\chi}^{0}_{\nu\omega}(\qv) = \chi^{0}_{\nu\omega}(\qv) - \chi^{0}_{\nu\omega},
\end{align}
where in turn the impurity bubble is defined as $\chi^{0}_{\nu\omega}\Let g_{\nu}g_{\nu+\omega}$.
We note that the lattice contribution in Eq.~\eqref{chialt} contains a local part, which includes contributions from long-range collective excitations. 
When summed over $\qv$, the first term in Eq.~\eqref{dualgensusc} vanishes,\footnote{The local part of the dual bubble [Eq.~\ref{chidual}] vanishes, which can be seen by summing \ref{chinuomega} over $\qv$ and replacing $(1/N)\sum_{\kv}G_{\nu}(\kv)$ by $g_{\nu}$ according to~\ref{dmftsc}.
} while the second in general does not.
In DMFT, the local part of the lattice susceptibility and the impurity susceptibility differ.

The lattice vertex $\Gamma$ is the same as in the DMFT susceptibility [Eq.~\eqref{chi}] and can be obtained from the irreducible vertex through the Bethe-Salpeter equation \eqref{bse}. It is convenient to combine the latter with the impurity BSE for the irreducible vertex. Using the the non-local part of the bubble, the resulting BSE reads as
\begin{align}
\label{dualbse}
[\Gamma_{\omega}^{-1}(\qv)]_{\nu\nu'} = [\gamma_{\omega}^{-1}]_{\nu\nu'} + T\tilde{\chi}^{0}_{\nu\omega}(\qv)\delta_{\nu\nu'},
\end{align}
where $\gamma$ is the \emph{reducible} impurity vertex.
There is hence no need to explicitly compute the irreducible impurity vertex $\gamma^{\text{irr}}$. This is similar to the dual fermion approach~\cite{Brener08}.
Such a reformulation is important to avoid unphysical singularities in the low-frequency behavior of the irreducible two-particle vertex~\cite{Janis14}, which occur in the proximity of the metal insulator transition~\cite{Schaefer13}.

\subsection{Results}
\label{sec:results}

\begin{figure}[t]
\includegraphics[scale=0.75,angle=0]{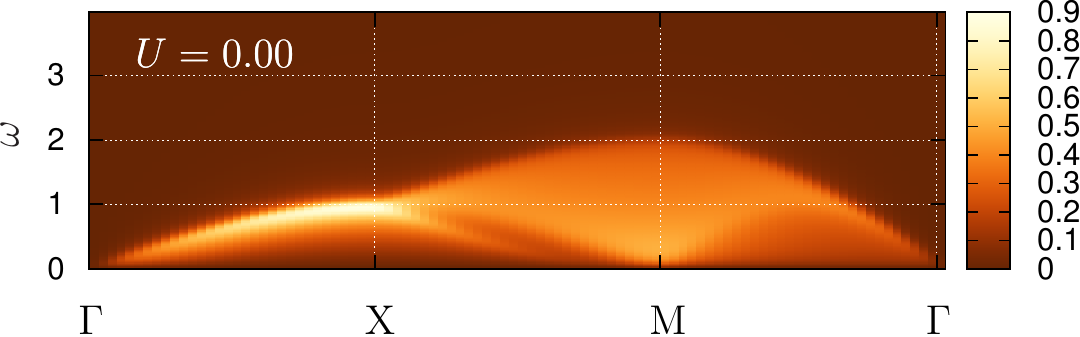}\\
\hspace{-4em}\includegraphics[scale=0.62,angle=0]{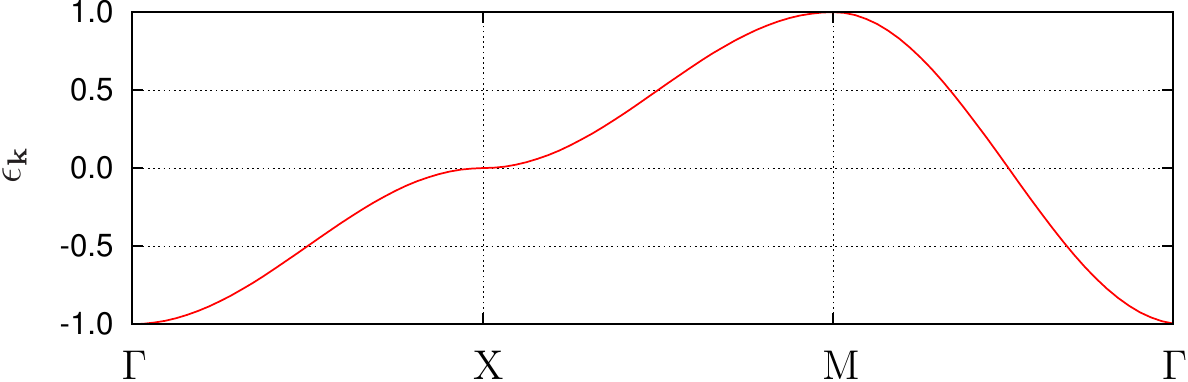}\\
\caption{\label{fig:xlat_nonint} (Color online) \emph{Top panel}: Lattice susceptibility $-\frac{1}{\pi}\Im \chi_{\omega}(\qv)$ for the half-filled noninteracting two-dimensional model (i.e. $U=0$) with full bandwidth $W=2$ and at finite temperature $T=0.02$, along a high-symmetry path in momentum space. $\Gamma$, $X$ and $M$ denote the wave vectors $(0,0)$, $(0,\pi)$, and $(\pi,\pi)$, respectively. \emph{Lower panel}: Noninteracting dispersion.}
\end{figure}

\begin{figure*}[t]
\includegraphics[scale=1.45,angle=0]{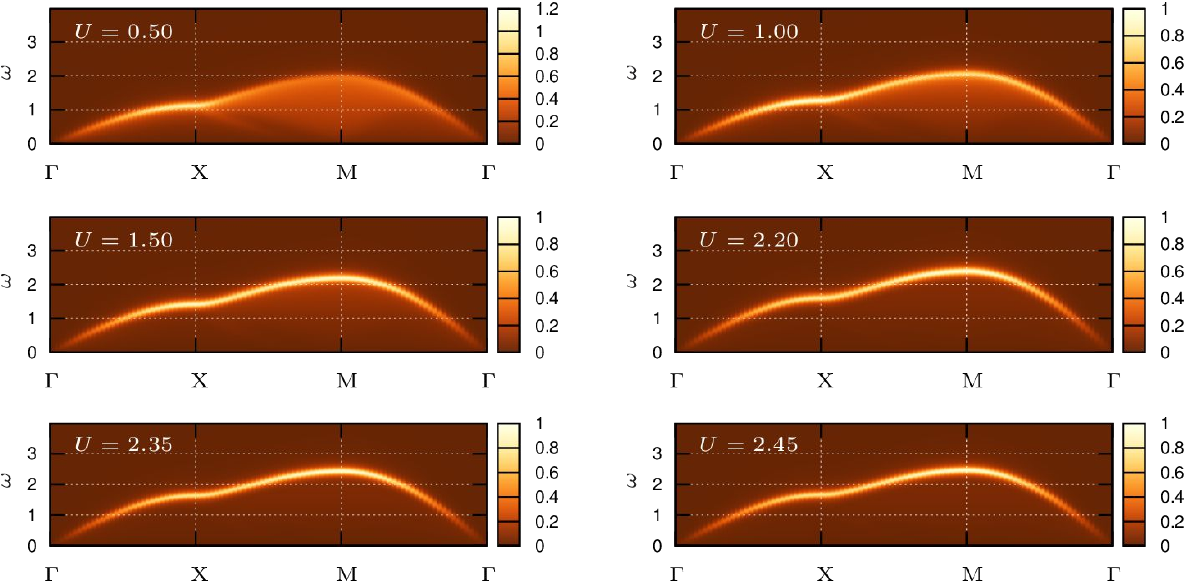} 
\caption{\label{fig:xlat_rpa} (Color online) RPA results for the imaginary part of the charge susceptibility $-\frac{1}{\pi}\Im \chi_{\omega}(\qv)$ of the half-filled Hubbard model for various values of $U$ and $T=0.02$.
}
\end{figure*}

\begin{figure}[b]
 \includegraphics[width=0.5\textwidth]{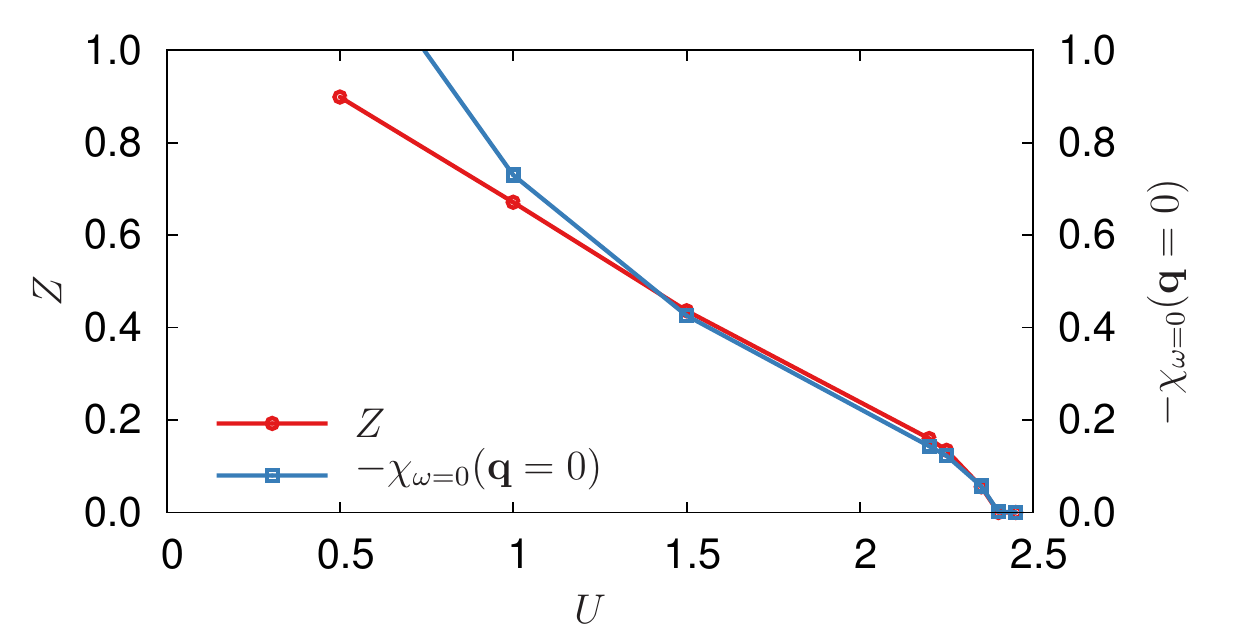}
 \caption{ \label{fig:z}(Color online) Quasi-particle weight $Z$ and $\lim_{q\to 0}\lim_{\omega\to 0}\chi_{\omega}(\qv)= -d n/d\mu$ as a function of $U$. Both vanish in the insulating phase. Close to the transition $Z$ is proportional to $dn/d\mu$.
 }
\end{figure}

\begin{figure}[b]
 \includegraphics[width=0.5\textwidth]{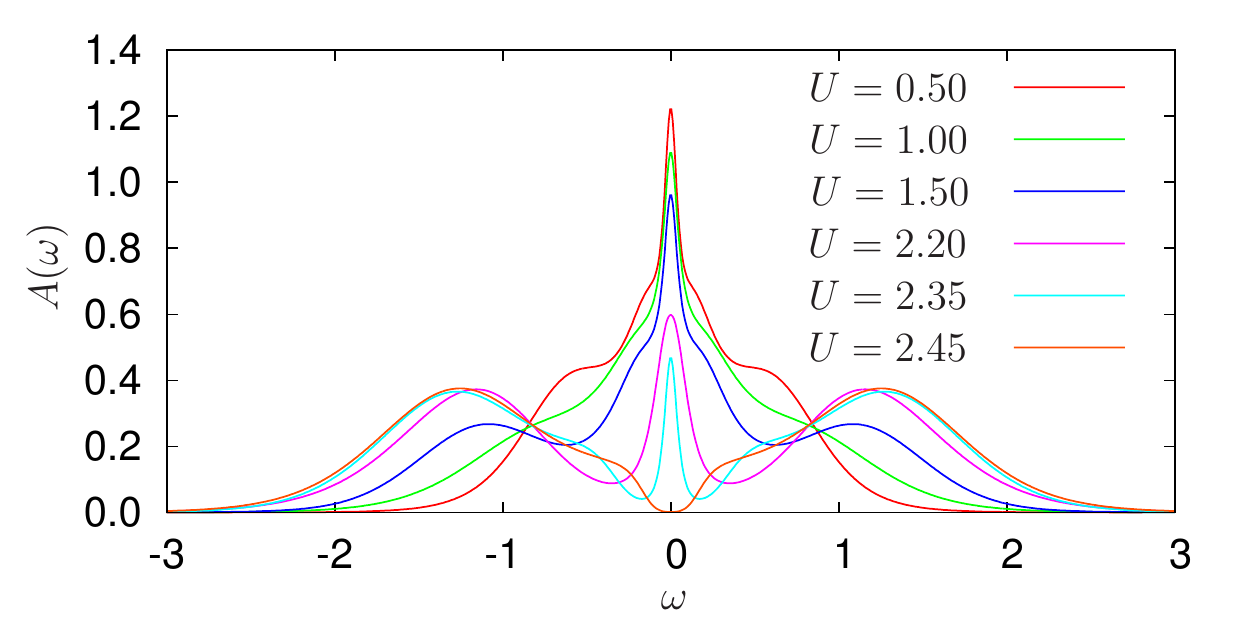}
 \caption{\label{fig:dos}(Color online). The finite temperature maximum entropy density of states for various values of $U$ at $T=0.02$, showing the evolution from a weakly to a moderately and strongly correlated system. The Mott transition occurs at $U\sim 2.36$.}
\end{figure}

Let us now turn to the results for the charge susceptibility in the two-dimensional Hubbard model. To set the stage for the discussion, we examine the noninteracting case first. In the upper panel of Fig.~\ref{fig:xlat_nonint} we plot the (negative) imaginary part of the noninteracting susceptibility
\begin{align}
\label{chi0}
\chi^{0}_{\omega}(\qv) = \frac{T}{N}\sum_{\kv\nu}G^{0}_{\nu+\omega}(\kv+\qv)G^{0}_{\nu}(\kv)
\end{align}
on real frequencies.
Its features are best understood in terms of the noninteracting dispersion $\epsilon_{\kv}$, which we plot in the panel below.
The maximum energy up to which one can see significant spectral weight is found at the M-point [$\qv=(\pi,\pi)$]. This wave vector connects maximum and minimum in the dispersion and its energy is correspondingly given by the bandwidth $W=2$. One can also see a structure of high intensity at very low energy and in the vicinity of the M-point, which is due to the nesting of the Fermi surface.
The maximal intensity at the M-point is found here.
The strongest overall response occurs at the X-point. Its dominant contribution in the convolution stems from those k-points for which the wave vector X connects two extremal points ($\Gamma$-X and X-M, respectively) and therefore corresponds to the energy $\omega=1$, which is equal to the half bandwidth. One can further see that the energy of the particle-hole excitations approaches zero in the long-wavelength limit.

\begin{figure*}[t]
\includegraphics[scale=1.45,angle=0]{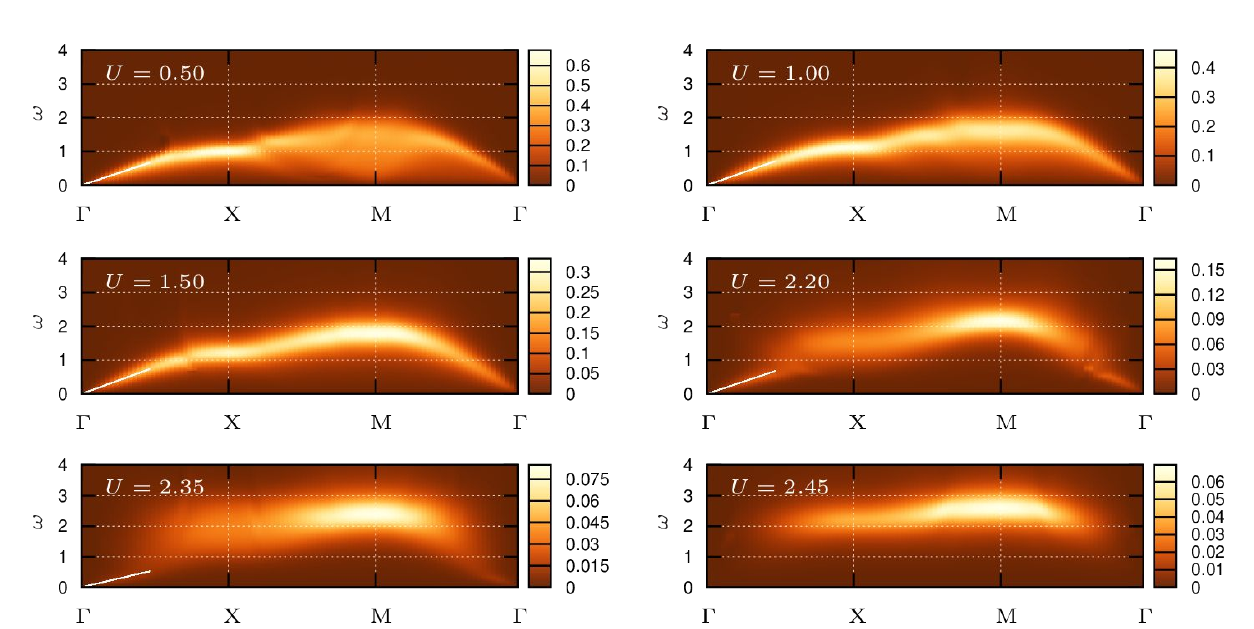} 
\caption{\label{fig:xlat_dmft} (Color online) Imaginary part $-\frac{1}{\pi}\Im \chi_{\omega}(\qv)$ of the DMFT charge susceptibility including vertex corrections obtained via analytical continuation using Pad\'e approximants.
The low energy dispersion obtained from a fit of the Matsubara data is indicated by a white line (cf. text).
}
\end{figure*}

In Figure~\ref{fig:xlat_rpa} we show results for the interacting case obtained from a standard random phase approximation (RPA) calculation. In RPA, the charge correlation function reads
\begin{align}
\label{RPA}
\chi^{\text{RPA}}_{\omega}(\qv)=\frac{\chi^{0}_{\omega}(\qv)}{1-U\chi^{0}_{\omega}(\qv)},
\end{align}
with $\chi^{0}$ defined in \eqref{chi0} [the minus sign in the denominator comes from the minus sign in the definition of $\chi_{\omega}(\qv)$, Eq.~\ref{chilatdef}.]
By construction, the RPA is of course only reliable for small values of $U$. Here we plot RPA results for larger values of the interaction for a comparison with the correlated case. For $U=0.5$, we observe a picture that is similar to the noninteracting case. The same structures are present also for larger values of $U$, albeit some of them, e.g. the structure which has an energy minimum at the M-point, become less visible.
As $U$ increases, the largest overall response shifts from X- to the to M-point above $U=1$. The collective excitation becomes better defined. Because the interaction is short-ranged, this collective mode --the zero-sound mode-- goes to zero energy in the long-wavelength limit for all values of $U$.
For values of $U$ larger than the bandwidth, the maximum energy at the M-point is determined by the energy scale $U$ instead of $W$.

Using DMFT, we can now investigate whether and how this physical picture is modified in a strongly correlated metal, close to a Mott transition.
For completeness, let us start by briefly showing some well-known aspects of the Mott transition in DMFT, as illustrated in Figs.~\ref{fig:z} and \ref{fig:dos}. 
Figure~\ref{fig:z} shows the corresponding quasiparticle weight $Z=(1 - d\Re \Sigma_{\omega}/d\omega)^{-1}$ as a function of $U$ computed using a polynomial extrapolation of the self-energy on Matsubara frequencies\footnote{We have used polynomials up to degree six. The results are converged for all values of $U$ for polynomials of degree five.}.
 On this lattice, the transition occurs at $U_{c}=2.36\pm 0.01$. $Z$ vanishes at the Mott transition, corresponding to a divergent effective mass~\cite{Georges96} $m^{*}/m\sim 1/Z$. The static homogeneous charge susceptibility $\lim_{\qv\to 0}\lim_{\omega\to 0}\chi(\omega,\qv) = -dn/d\mu$ is shown in the same figure.
It is proportional to the compressibility and therefore vanishes in the insulator.
In Fig.~\ref{fig:dos}, we plot the local density of states for different values of $U$. One can see a well-defined quasi-particle peak and the Hubbard bands at $\omega\sim U$. 
For values above the transition, the density of states exhibits a gap.

\begin{figure}[b]
\includegraphics[scale=0.7,angle=0]{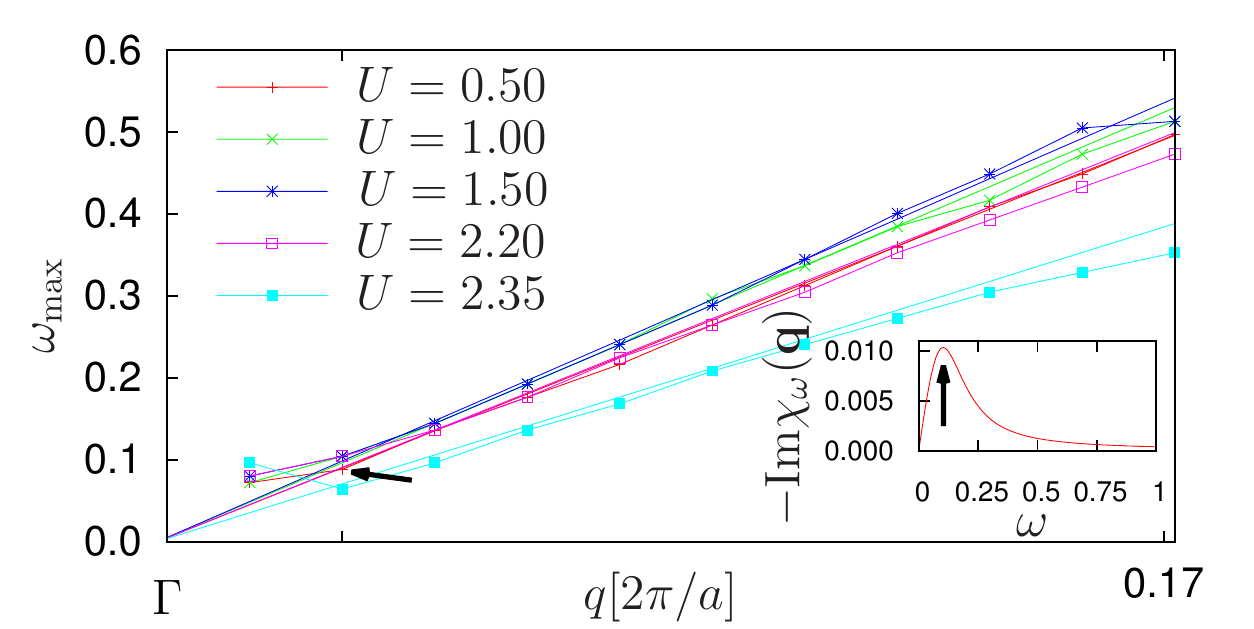} 
\caption{\label{fig:maxima} (Color online) Maxima of the charge susceptibility of Fig.~\ref{fig:xlat_dmft} (lines with symbols) and dispersion obtained from a fit of the Matsubara data (lines, cf. text) for small $\qv$-vectors up to about one third the way to the X-point indicating the slope of the zero-sound mode. Inset: cut at small fixed momentum $q=0.03125\, (2\pi/a)$ through the charge susceptibility for $U=2.2$. The arrow indicates the maximum which corresponds to the point marked by an arrow in the main panel.
}
\end{figure}

The DMFT susceptibility including vertex corrections is shown in Fig.~\ref{fig:xlat_dmft} for the same parameters as the RPA results in Fig.~\ref{fig:xlat_rpa}.
We obtain it by analytical continuation from Matsubara frequencies. In the weakly correlated regime, for $U$ up to about $1$, the results are similar to the RPA, albeit we observe a somewhat broader spectrum. In particular, one can see the minimum at the M-point, which is present up to at least $U\sim 1$. In the moderately correlated regime, $U\sim 1.5$, this feature is no longer resolved in our data, but the spectra retain a similar shape as in RPA, showing a well defined mode for all wave vectors. Its maximum at M is still approximately equal to the bandwidth. As the transition is approached, however, the spectrum changes substantially. It is considerably broadened and damped at the X-point, while at the M-point it gains relative intensity. For large interaction, the maximum at the M-point occurs at the scale of $U$.

The collective mode is visible all the way to the transition. Its frequency vanishes in the long-wavelength limit. This is expected for reasons we shall explain below.
We cannot strictly exclude even qualitative changes in particular of the high energy features in the spectra because of the ill-conditioned nature of the analytical continuation problem (see Appendix~\ref{app:pade} for a discussion of the analytical continuation procedure).
We can, however, further substantiate the results for the low energy collective mode directly from the Matsubara data: The polarization $\Pi_{\omega}(\qv)$ defined by
\begin{align}
\label{pidef}
\chi_{\omega}(\qv) = \frac{-\Pi_{\omega}(\qv)}{1+U\Pi_{\omega}(\qv)}
\end{align}
is a function of $q/\omega$ for small $q$, as explained in Sec.~\ref{sec:gauge}. We fit the Matsubara data with an expression of the form $\Pi_{\omega}(\qv) = -b (q/\iom_{m})^{2}/[1+c^{2}(q/\iom_{m})^{2}]$, for a small Matsubara frequency $m=3$, where $b$ and $c$ are the free parameters. We can readily analytically continue this function by letting $\iom\to\omega+i0^{+}$. The dispersion is defined by the zeros of the denominator in \eqref{pidef} and can be expressed in terms of the fit parameters as $\omega(q)=q\sqrt{c^{2}+b U}$. The thus obtained dispersion is indicated by a white line in Fig.~\ref{fig:xlat_dmft}.
In Fig.~\ref{fig:maxima}, we plot the maximum of the charge susceptibility for small wave vectors together with the linear dispersion obtained from the fits. The data are in good agreement showing  that the Pad\'e approximation is reliable within this energy range. There is no appreciable change in slope of the mode with increasing interaction. Although it is less visible in Fig.~\ref{fig:xlat_dmft} due to decreasing contrast close to the transition, it remains well defined as can be seen from a fixed momentum cut of the susceptibility shown in the inset of Fig.~\ref{fig:maxima}. In the insulator, this mode disappears as expected. All excitations acquire a minimum energy $\sim U$.

\begin{figure}[t]
\includegraphics[scale=0.75,angle=0]{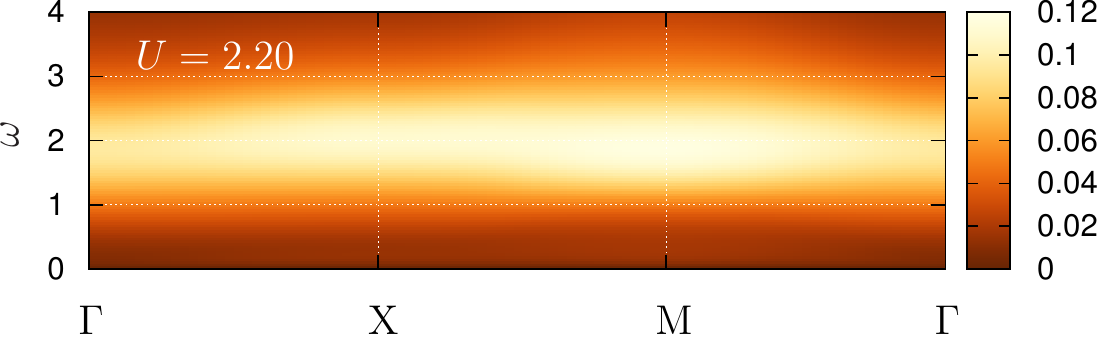} 
\caption{\label{fig:xlat_bubble} (Color online) $-\frac{1}{\pi}\Im \chi^{0}_{\omega}(\qv)$ in the bubble approximation, i.e. \emph{without vertex corrections} for local interaction $U=2.2$ and otherwise the same parameters as in Fig.~\ref{fig:xlat_dmft}. In the bubble approximation, the zero-sound mode disappears.
}
\end{figure}

We are now going to show that the {\it non-local} (reducible) vertex corrections play an essential role to obtain 
a correct description of the low energy physics, {\sl even at a very qualitative level}.
Because incorporating the full vertex corrections is technically demanding, they are often neglected
and the susceptibility is often approximated by a simple bubble approximation, i.e. by a product of \emph{interacting} DMFT Green's functions $\chi^{0}_{\omega}(q)=(T/N)\sum_{\kv\nu}G_{\nu+\omega}(\kv+\qv)G_{\nu}(\kv)$. 
In Fig.~\ref{fig:xlat_bubble}, we plot the susceptibility for $U=2.2$, obtained within this bubble approximation. It is essentially featureless for all wave vectors. In the long-wavelength limit, it exhibits spectral weight at \emph{finite} energy in contradiction to the foregoing and the standard textbook RPA approximation.
For small but finite $U$, the bubble approximation still exhibits a mode which goes, at least approximately, to zero in the long-wavelength limit.
In the correlated regime (larger $U$), however, the vertex corrections are essential for a qualitatively correct description of the collective excitations.
In the next section, we will relate the existence of the zero-sound to the gauge invariance and the associated Ward identities. The failure of the bubble approximation can be traced back to a violation of gauge invariance.

We note that in contrast to the susceptibility, the optical conductivity in DMFT is unaffected by the vertex corrections in the long-wavelength limit. Vertex corrections drop out of the conductivity as a consequence of the locality of the irreducible vertex and the inversion symmetry of the lattice (see Refs.~\onlinecite{Khurana90,Georges96}).

\begin{figure}[t]
\includegraphics[scale=0.75,angle=0]{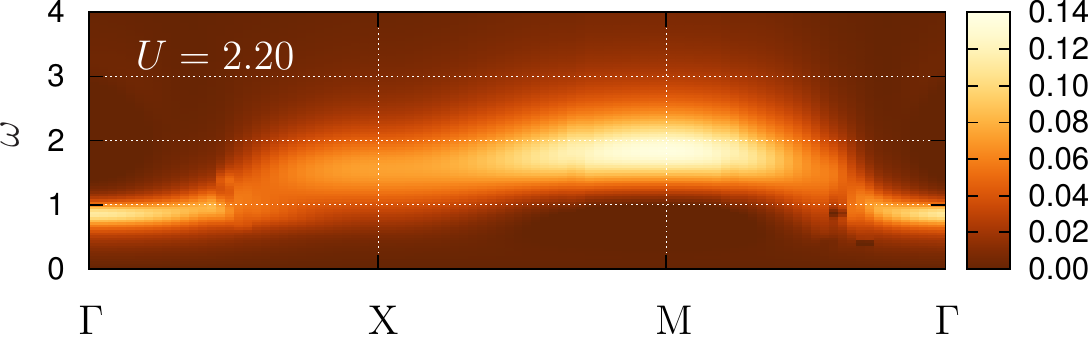} 
\caption{\label{fig:xlat_pi_bubble} (Color online) Lattice susceptibility $-\frac{1}{\pi}\Im \chi_{\omega}(\qv)$ obtained by neglecting non-local, long-range vertex corrections, more precisely by neglecting the second term in Eq.~\eqref{dualgensusc}. Parameters are the same as in Fig.~\ref{fig:xlat_bubble}. The plasmon-like mode at finite energy and long wavelengths is an artifact of the approximation. Local vertex corrections are not sufficient to restore the zero-sound mode.
}
\end{figure}

A natural question to ask at this stage is whether an approximation with only {\it local} vertex corrections could be sufficient to capture the zero-sound mode.
To gain more insight into this question, we consider a more advanced approximation than the bubble. Namely, we compute the susceptibility within an approximation that includes short-range vertex corrections, but neglects correlations from the \emph{reducible} vertex of the lattice $\Gamma$. To be precise, we compute it using Eqs.~\eqref{chialt}--\eqref{chidual}, however, neglecting the second term containing the lattice vertex $\Gamma$ in Eq.~\eqref{dualgensusc}. We note that this approximation is not exactly the same as a bubble approximation with local vertex corrections added. The important point here is that such an approximation neglects ladder diagrams containing many repeated particle-hole scattering processes on different lattice sites contained in $\Gamma$.
The result is shown in Fig.~\ref{fig:xlat_pi_bubble}. As expected, the long-wavelength properties are clearly not reproduced correctly in this approximation. It wrongly predicts a finite energy collective mode.
We note that in the long-wavelength limit, the ladder diagrams contribute at all orders. Low-order diagrams that describe charge correlation contain independent particle-hole propagation (described by the bubble) over large distances, which is unlikely. Therefore, it is physically clear that higher-order diagrams are important.
Features at finite wave vectors sufficiently far from the $\Gamma$-point are, however, remarkably well captured. Here, the short-range vertex corrections are necessary, but also appear to be sufficient to reproduce qualitative features.

\subsection{Gauge invariance}
\label{sec:gauge}

In this section, we discuss the consequences of gauge invariance. The failure of the bubble approximation to describe the low-energy collective modes can explicitly be traced back to its violation.
We will first check that the standard DMFT susceptibility~\cite{Georges96} as described in Sec.~\ref{dmftsusc} [Eqs.~\eqref{chi} and \eqref{bse}] and in the review Ref.~\cite{Georges96}
yields a response that obeys local charge conservation and leads to gauge-invariant results. We recall that for the response to be conserving, two conditions have to be fulfilled~\cite{Baym61,Baym62}: (i) The self-energy and the irreducible vertex function have to be given by functional derivatives of the Luttinger-Ward functional $\Phi[G_{ij}]$, i.e., $\Sigma_{ij}=\delta\Phi/\delta G_{ji}$ and $\Gamma_{ijkl}^{\text{irr}}=\delta^{2}\Phi/\delta G_{ji}\delta G_{lk}$.
(ii) The generalized susceptibility $L$ has to be constructed such that it gives the change in $G$ through a perturbation $A$ to linear order: $L \Let -\delta G/\delta A = G(\delta G^{-1}/\delta A) G$. With $G^{-1}=G_{0}^{-1}-A-\Sigma[G]$ and for approximations for which $\Sigma$ depends on the perturbation through $G$ only, this leads to the integral equation
\begin{align}
\label{chigen}
L = -GG -GG\frac{\delta \Sigma}{\delta A} &= -GG -GG\frac{\delta \Sigma}{\delta G} \frac{\delta G}{\delta A} \notag\\
&= -GG +GG\frac{\delta \Sigma}{\delta G} L.
\end{align}
Combined with condition (i), one identifies $\Gamma^{\text{irr}}\Let \delta\Sigma/\delta G=\delta^{2}\Phi/\delta G^{2}$.
Condition (ii) is evidently fulfilled in DMFT, since the above equation is equivalent to the integral equation \eqref{gensusc} (written in terms of $\chi\Let -L$ instead of $L$).
To address the first condition, we recall that in DMFT, the Luttinger-Ward functional depends on the site-diagonal Green's functions only~\cite{Georges96}, allowing for a decomposition in terms of local functionals, $\Phi[G_{i'j'}] = \sum_{i'} \phi[G_{i'i'}]$. This implies that the self-energy is local:
\begin{align}
\Sigma_{ij} = \frac{\delta\Phi[G_{i'j'}]}{\delta G_{ji}} = \frac{\delta\phi[G_{i'i'}]}{\delta G_{ii}}\delta_{ji}.
\end{align}
The same holds for the irreducible vertex function,
\begin{align}
\Gamma^{\text{irr}}_{ijkl} = \frac{\delta^{2}\Phi[G_{i'j'}]}{\delta G_{ji}\delta G_{lk}} =  \frac{\delta^{2}\phi[G_{l'l'}]}{\delta G_{ll}^{2}} \delta_{li}\delta_{lj}\delta_{lk}.
\end{align}
The functional $\phi[G_{i'i'}]$ as well as $\Sigma$ and $\Gamma^{\text{irr}}$ are generated from the impurity model subject to the self-consistency condition \eqref{dmftsc}. If the self-energy and vertex function are obtained numerically exactly from the solution of the impurity model, condition (i) is hence fulfilled. Note that the dimensionality of the lattice does not enter this argument. It therefore holds on finite-dimensional lattices, where DMFT is an approximation. It also holds for cluster extensions of dynamical mean-field theory.

We can make the argument more explicit by recalling that charge conservation is commonly expressed in terms of a  Ward identity. It can be viewed as the Green's function analog of the continuity equation and relates the single-particle Green's function to a vertex function. The above two conditions are sufficient for the Ward identity to be fulfilled, as can be shown by considering a perturbation which corresponds to a gauge transformation~\cite{Baym62}.

On a discrete lattice, gauge invariance and charge conservation can be preserved exactly, even for finite lattice spacing~\cite{Wilson74}. 
Projecting the continuum system onto a discrete Wannier basis under the assumption of weak and slowly varying fields leads to the gauge theory described here (see Appendix~\ref{app:current} and, e.g., Refs.~\onlinecite{Graf95,bluemer02}).
In Appendix~\ref{app:sec:ward}, we show that with the proper definition of the current and a suitable generalization of the notion of the derivative to the lattice, the Ward identity can be written
\begin{align}
\label{ward}
q^{F}_{\mu}\Gamma_{\mu}(k,q) = G^{-1}(k) - G^{-1}(k+q),
\end{align}
where we have introduced four-vector notation for clarity (only in this section). Summation over the time $(\mu=0)$ and spatial components $\mu=1,2,3$ is implied using the metric $(-1,1,1,1)$.
The corresponding continuity equation is $\partial n/\partial t + \nabla^{F}\cdot\vc{j} = 0$, where $\nabla^{F}$ denotes a \emph{forward} derivative. It corresponds to a finite difference expression owing to the discrete structure of the lattice. In the above, the main difference to the continuum case is the appearance of the momentum $q^{F}_{\mu}\equiv(\i\omega,\qv^{F})$ associated with a forward derivative. On a finite lattice of $N$ sites with periodic boundary conditions it has the spatial components $q^{F}_{\alpha}=-(\i/a)[\exp(iq_{\alpha}a)-1]$, where the index $q_{\alpha}$ takes on the discrete values $q_{\alpha}^{(n)}=2\pi n/N$, with $n$ integer.
In the above equation, $\Gamma_{\mu}$ is the renormalized current vertex, which describes the interaction of the interacting electrons with the electromagnetic field. The Ward identity hence relates a vertex function to the single-particle properties described by the Green's function $G$.

\begin{figure}[t]
\includegraphics[scale=0.4,angle=0]{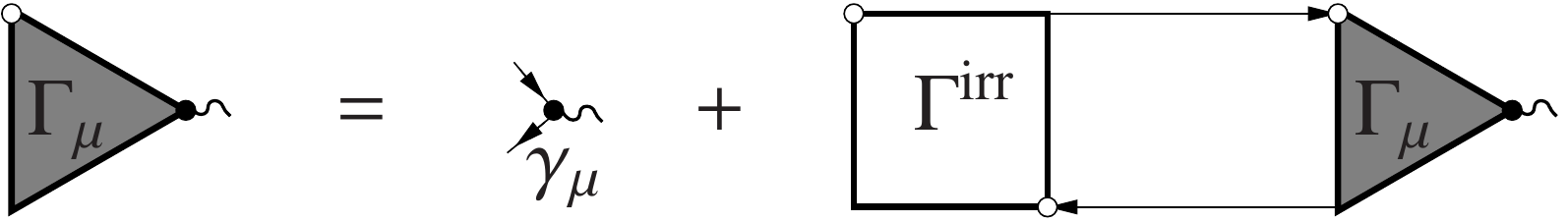} 
\caption{\label{fig:bsegmu} Diagrammatic representation of the Bethe-Salpeter equation for the renormalized current vertex $\Gamma_{\mu}$. Lines are fully dressed propagators.
}
\end{figure}

Let us now explicitly check the Ward identity in DMFT.
The current vertex obeys a ladder equation, which follows from the Bethe-Salpeter equation for the susceptibility [Eq.~\eqref{gensusc}]. In DMFT, the ladder equation for the current vertex reads as (see Fig.~\ref{fig:bsegmu} and Appendix \ref{app:Gmu}):
\begin{align}
\label{gammamuladder}
\Gamma_{\mu;\nu}(\kv,\qv) = &\gamma_{\mu}(\kv,\qv) - \frac{T}{N}\sum_{\nu'\kv'}\Gamma_{\nu\nu'\omega}^{\text{irr}}\notag
\\&\times G_{\nu'\sigma'}(\kv')G_{\nu'+\omega}(\kv'+\qv)\Gamma_{\mu;\nu'}(\kv',\qv).
\end{align}
Here, the \emph{bare} current vertex is given by
\begin{align}
\gamma_{\mu} = \left\{\begin{array}{ccl}
\i\tti a\left(e^{-\i(k_{\alpha}+q_{\alpha})}-e^{-\i k_{\alpha}}\right), & \mu=&\alpha=x,y,z\\
1, & \mu=&0
\end{array} \right.,
\end{align}
which itself obeys the Ward identity $q_{\mu}^{F}\gamma_{\mu}(k,q) = G_{0}^{-1}(k)-G_{0}^{-1}(k+q)$ with the noninteracting Green's function $G_{0}^{-1}(k)=\inu+\mu-\varepsilon_{\kv}$.
In order to demonstrate that the DMFT susceptibility is conserving, we have to show that the current vertex obeys the Ward identity \eqref{ward}. To this end, we form the quantity $q_{\mu}^{F}\Gamma_{\mu}(k,q)$ using \eqref{gammamuladder}:
\begin{align}
q_{\mu}^{F}\Gamma_{\mu;\nu}(\kv,\qv) =& q_{\mu}^{F}\gamma_{\mu}(\kv,\qv)- \frac{T}{N}\sum_{\nu'\kv'}\Gamma_{\nu\nu'\omega}^{\text{irr}}\notag\\
&\times G_{\nu'}(\kv')G_{\nu'+\omega}(\kv'+\qv)[q_{\mu}^{F}\Gamma_{\mu;\nu'}(\kv',\qv)].
\end{align}
Inserting the Ward identity \eqref{ward} for the interacting and noninteracting current vertices on both sides and using the definition of the Green's function \eqref{glat}, all momentum dependence cancels exactly. 
Since the irreducible vertex is local, the DMFT self-consistency condition \eqref{dmftsc} further allows us to express the local part of the lattice Green's function in terms of the impurity Green's function. We thus obtain the purely local equation
\begin{align}
\label{wardlocal}
\Sigma_{\nu+\omega} - \Sigma_{\nu} &= -T \sum_{\nu'}\Gamma_{\nu\nu'\omega}^{\text{irr}}\left[g_{\nu'+\omega}-g_{\nu'}\right],
\end{align}
which involves impurity quantities only.\footnote{The minus sign on the right-hand side of Eq.~\eqref{wardlocal} stems from the fact that our convention for the irreducible vertex is such that $\Gamma^{\text{irr}}=-U$ to lowest order in $U$.} We therefore conclude that the Ward identity is fulfilled if this equation is satisfied. It can be viewed as a local version of the Ward identity,\footnote{By writing Eq.~\eqref{wardlocal} in differential form, $\Gamma^{\text{irr}}$ is identified as the functional derivative $\delta\Sigma/\delta g$: The variation of a self-energy diagram is given by the functional derivative of $\Sigma$ with respect to $g$ times the variation in $g$.} which is fulfilled for the impurity model.
Gauge invariance and local charge conservation are hence guaranteed for the susceptibilities. This is completely in line with the previous argument: 
Charge conservation of the DMFT susceptibility follows if the self-energy and irreducible vertex function are determined from the exactly solvable impurity model.\footnote{For an approximate impurity solver, the susceptibility is conserving as long as Eq.~\eqref{wardlocal} is fulfilled.} While being an approximation in finite dimensions, the DMFT susceptibility preserves local charge conservation exactly, even on a finite dimensional lattice.

\begin{figure}[t]
\includegraphics[scale=0.7,angle=0]{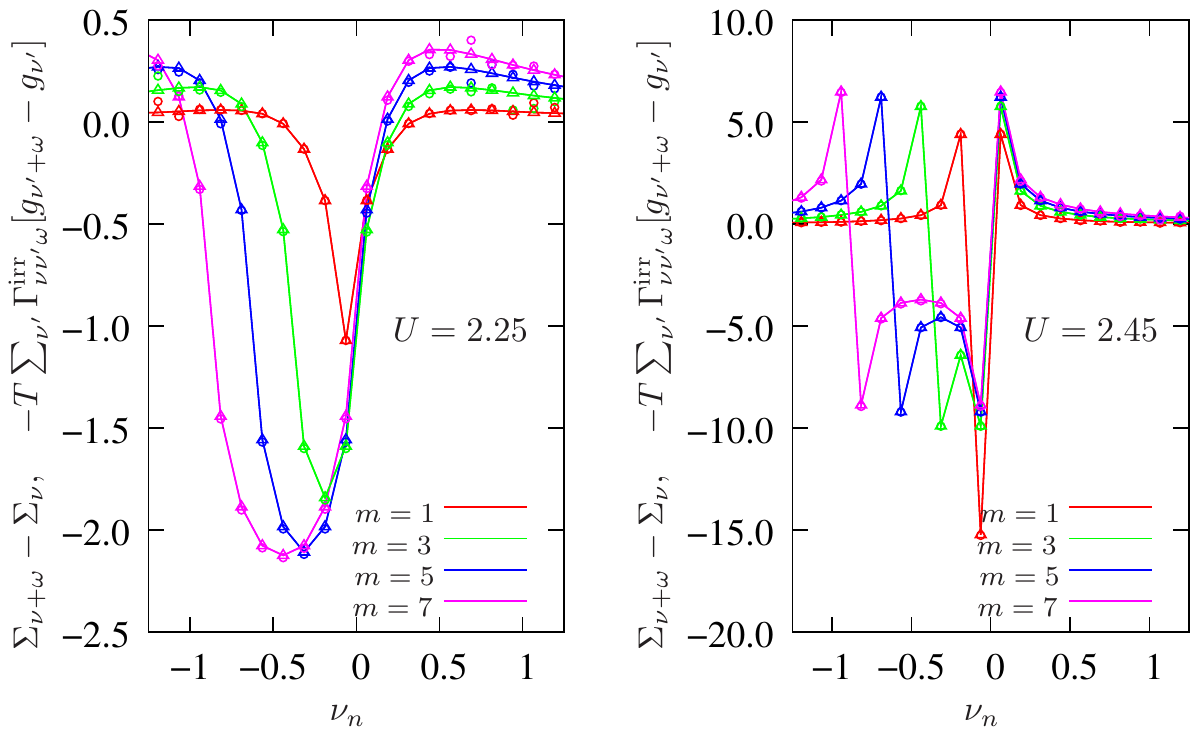} 
\caption{\label{fig:ward} (Color online) Numerical illustration of the fulfillment of the local Ward identity. Both sides of the equation \eqref{wardlocal} are plotted for different bosonic frequencies $\omega_{m}=2m\pi/\beta$ for the two-dimensional Hubbard model in DMFT for two different values of $U$ somewhat below (left) and above (right) the Mott transition. Results for the left-hand side of the equation, $\Sigma_{\nu+\omega}-\Sigma_{\nu}$, are marked by lines with triangles and for the right-hand side $-T\sum_{\nu'}\Gamma_{\nu\nu'\omega}^{\text{irr}}[g_{\nu'+\omega}-g_{\nu'}]$ by circles, showing good agreement. Note that for $\omega_{m}=0$, the equation is identically fulfilled as both sides vanish.
}
\end{figure}

In our calculations we obtain the vertex, self-energy, and Green's function by solving the impurity model numerically exactly.
That the Ward identity is indeed fulfilled numerically is illustrated in Fig.~\ref{fig:ward}, where we plot both sides of Eq.~\eqref{wardlocal} for different bosonic frequencies. In order to evaluate the frequency sum on the right-hand side, we have replaced the irreducible vertex by $-U$ above the frequency cutoff up to which it is calculated explicitly. The identity is evidently well fulfilled. For high frequencies, deviations occur which partly originate from the numerical noise which increases with frequency, as well as from the finite frequency cutoff of the vertex function.
The latter is computed by inverting a local BSE, which is affected by the finite frequency cutoff. In general, for a correct description of the collective excitations, the low-energy behavior of the vertex function is decisive, which is well captured in our calculations.
Note that the numerical error seen in this figure does not propagate into the calculation of the lattice susceptibility, because the irreducible vertex does not have to be computed (see Sec.~\ref{sec:chialt}).

We would like to point out the relation of the above to the self-consistent Hartree-Fock approximation and RPA~\cite{Mahan00}. In static mean-field theory, the above equations still hold with $\Sigma=U\av{n}$ and $\Gamma^{\text{irr}}=-U$. As a consequence, Eq.~\eqref{wardlocal} is identically fulfilled, showing that the Hartree-Fock approximation is conserving. The Bethe-Salpeter equation for the response function is then equivalent to the random phase approximation (RPA) for the susceptibility [Eq.~\eqref{RPA}].

As mentioned previously, it is known that the zero-sound mode is a consequence of gauge invariance. To see this, consider the electromagnetic response kernel $K_{\mu\nu}$ defined through
\begin{align}
J_{\mu}(q) = K_{\mu\nu}(q)A_{\nu}(q),
\end{align}
where $J_{\mu}$ is the expectation value of the current.
Gauge invariance and local charge conservation imply the two conditions (see Appendix \ref{app:k}):
\begin{align}
K_{\mu\nu}(q) q^{F}_{\nu} &= 0,\\
q^{F}_{\mu} K_{\mu\nu}(q) &= 0.
\end{align}
For the longitudinal response ($\qv$ pointing along direction $3$), one obtains $K_{03}q-\iom K_{00}=0$ and $K_{33}q-\iom K_{03}=0$. (In the following, we are interested in the long-wavelength behavior for which we may replace $q^{F}$ by $q$.)
The other components do not contribute by symmetry. Combining these two equations yields the continuity equation
\begin{align}
\label{kcontinuity}
K_{00}(q) = \frac{q^{2}}{(\i\omega)^{2}}K_{33}(q).
\end{align}
In the long-wavelength limit, only the diamagnetic contribution to the response kernel contributes to $K_{33}$, which is independent of $q$ and $\omega$ (see Appendix~\ref{app:k} and Ref.~\onlinecite{Nozieres64}). Hence, the susceptibility $\chi_{\omega}(\qv)=-K_{00}(q)/e^{2}$ is a function of the ratio $q/\omega$. The same holds for the polarization since it is related to $\chi$ through a simple geometric series:
\begin{align}
\chi_{\omega}(\qv) &= \frac{-\Pi_{\omega}(\qv)}{1+U\Pi_{\omega}(\qv)}.
\end{align}
The dispersion of the collective mode is determined by the poles of $\chi$, or as the solution to the equation $1+U\Pi_{\omega}(\qv)=0$. Since the interaction $U$ is constant (in general, a short-range interaction remains finite in the limit $q\to 0$), the solution to the above equation must be a kind of sound, i.e., $q/\omega=\text{const}$.
The analysis is the same as in the textbook RPA case, except that the polarization $\Pi_\omega(\qv)$ is a function produced
by the DMFT calculation instead of the Lindhardt function in the RPA case. The key fact is that for small momentum, $\Pi_\omega(\qv)$ is a function of $q/\omega$ and  
hence not a continuous function of $(\omega, q)$ at $(0,0)$. As shown above, this is a consequence of gauge invariance.
Moreover, every approximation that violates the Ward identity is likely to miss this singularity of the function, and will not be able 
to reproduce the correct low-energy behavior.

Since we use a quantum Monte Carlo impurity solver, it is useful being able to observe the restriction imposed by gauge invariance also on the level of the imaginary-time data. To this end, we rewrite \eqref{kcontinuity} as
\begin{align}
\label{kcontinuity2}
(\iom)^{2}K_{00}(q) = (q)^{2}K_{33}(q).
\end{align}
Using again that in the limit $q\to 0$ $K_{33}$ remains finite, we find the following condition imposed on the susceptibility:
\begin{align}
\label{gicond}
(\iom)^{2}\chi_{\omega}(\qv) \mathop{=}_{q\to 0} 0.
\end{align}
The charge susceptibility has to vanish for any finite frequency in this limit and hence also for subsequently taking the limit $\omega\to 0$. On the other hand, taking the limit $\omega\to 0$ first leads to the the static response $\lim_{q\to 0}\lim_{\omega\to 0}\chi_{\omega}(\qv)=-dn/d\mu$ which is finite in the metallic phase. 
Hence, the limits $\lim_{\omega\to 0}$ and $\lim_{q\to 0}$ do not commute, which implies a discontinuous jump in the susceptibility.
The Lindhardt bubble, corresponding to the noninteracting result, has the required property~\cite{Platzman73,Giuliani05}.
The non-commutativity of the susceptibility comes from the product of Green's functions $\chi^{0}_{kq}=G_{k}G_{k+q}$, which enters the vertex through the Bethe-Salpeter equation for the vertex function [Eq.~\eqref{bse}]. It is singular in this limit, because the poles of the Green's functions merge when $q\to 0$ (see, e.g., Ref.~\onlinecite{Nozieres64}, Chap. 6, Sec. 4).

\begin{figure}[t]
\includegraphics[width=0.5\textwidth]{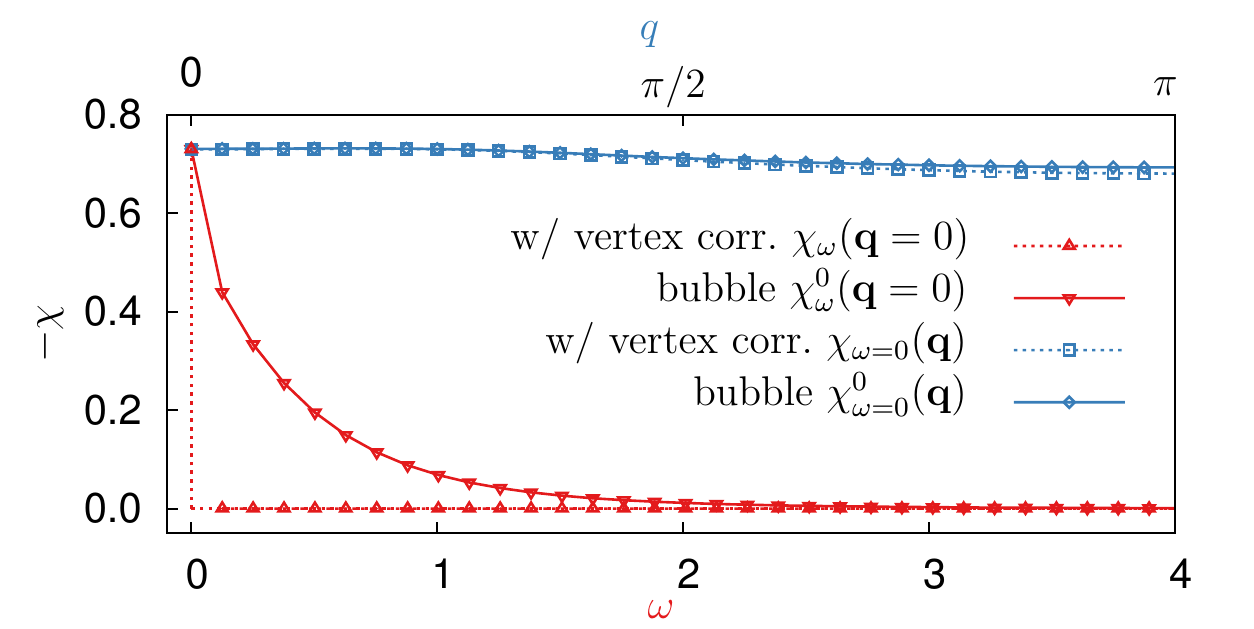}
\caption{(Color online). The DMFT susceptibility $\chi_{\omega}(\qv)$ for $\qv=0$ (red curve) and $\omega=0$ (blue curve), for $U=1$, $T=0.02$. As a consequence of gauge invariance, the limits $\omega\rightarrow 0$ and $\qv\rightarrow 0$ of $\chi_{\omega}(\qv)$ do not commute. For a frequency-dependent self-energy, this condition is respected by including vertex corrections. It is violated for the bubble approximation for which the limits commute.}
\label{fig:limits}
\end{figure}

Figure~\ref{fig:limits} shows that the susceptibility including vertex corrections indeed vanishes for finite frequencies and displays a discontinuity.\footnote{The condition~\eqref{gicond} can be used to benchmark the accuracy of the simulation. Deviations from zero occur due to Monte Carlo noise in the vertex function and an insufficient frequency cutoff.}
The bubble approximation computed from interacting Green's functions, on the other hand, clearly violates gauge invariance. The result is continuous, which explains the failure of the bubble approximation observed previously.

One can expect that the discontinuity (which can be used as a rigorous test for the implementation) will only be restored by summing an infinite number of diagrams beyond the bubble. Physically, it is clear that in order to describe the long-wavelength behavior of the two-particle excitations and the response functions, repeated particle-hole scattering generated through the Bethe-Salpeter equation is essential. At the same time, the Bethe-Salpeter equation accounts for the collective mode and ensures gauge invariance. In this sense, the collective excitations are key to the gauge-invariant character of the theory. A different way of seeing this is the fact that the effective quasiparticle interaction determined by the vertex function generates the back flow of electrons around a quasiparticle moving through the medium~\cite{Nozieres64}. This back flow is necessary to fulfill the continuity equation and hence to assure local charge conservation. In the insulator, $dn/d\mu=0$ because of the gap so that the discontinuity disappears and with it the zero-sound mode.

\section{Long-range interaction: the dual boson approach}
\label{sec:longrange}

Plasmons are long-wavelength excitations of the electron gas with a finite energy, which appear in presence of a Coulomb potential. Here, we use the following expression for a Coulomb-type potential in three dimensions,
\begin{align}
\label{Vq}
V(\qv) = \left\{\begin{array}{ccl}
e^{2}\frac{V}{q^{2}} & q\neq 0,\\
0 & q =0,
\end{array} \right.
\end{align}
where $e$ is the electron charge.
Setting the interaction to zero for $q=0$ corresponds to adding a homogeneous positively charged background which compensates the negative charge of the electron gas.

Non-local interactions can be treated on the basis of extended dynamical mean-field theory~\cite{Kajueter96,Smith00,Chitra00}. In EDMFT, the lattice model is mapped to a local impurity problem which contains a local, but retarded interaction $W_{\omega}$. The self-energy in EDMFT is determined through the self-consistency condition \eqref{dmftsc} as in DMFT. The dynamical interaction accounts for the dynamical screening of the local charge due to the non-local interaction $V(\qv)$. It is determined through an additional self-consistency condition
\begin{align}
\chi_{\omega} = \sum_{\kv}X_{\omega}(\kv),
\end{align}
which is written in terms of the two-particle propagator (the lattice susceptibility)
\begin{align}
\label{xedmft}
X_{\omega}(\qv) = \frac{1}{\chi_{\omega}^{-1}+W_{\omega}-V(\qv)}.
\end{align}
We denote it by $X_{\omega}(\qv)$ in order to distinguish it from the DMFT susceptibility $\chi_{\omega}(\qv)$. In the above, $\chi_{\omega}$ is the impurity charge susceptibility. This equation can be understood as follows. Consider the representation of the susceptibility in terms of the polarization $\Pi$:
\begin{align}
\label{xinpi}
X_{\omega}(\qv) &= -\Pi_{\omega}(\qv) + \Pi_{\omega}(\qv)V(\qv)\Pi_{\omega}(\qv) \mp\ldots\notag\\
&= \frac{-\Pi_{\omega}(\qv)}{1+V(\qv)\Pi_{\omega}(\qv)} = \frac{1}{-\Pi_{\omega}(\qv)^{-1}-V(\qv)},
\end{align}
which is a simple geometric series. $\Pi$ contains all diagrams irreducible with respect to the interaction $V_{\qv}$. In EDMFT, the polarization is obtained from the impurity model, which yields a non-perturbative, albeit local result.
The impurity susceptibility $\chi_{\omega}$ contains polarization diagrams. However, it cannot directly be used as the polarization because it contains diagrams reducible in $W$. Let us denote $\Pi^{W}_{\omega}=-\chi_{\omega}$. We can easily take out these reducible contributions by writing $\Pi^{W}_{\omega}$ as a geometric series: $\Pi^{W}_{\omega}=\Pi/(1+W_{\omega}\Pi_{\omega})$ or $\Pi^{-1}_{\omega} = (\Pi^{W}_{\omega})^{-1} -W_{\omega}=-\chi^{-1}_{\omega}-W_{\omega}$. Inserting this into \eqref{xinpi} recovers \eqref{xedmft}.

\begin{figure}[b]
 \includegraphics[width=0.45\textwidth]{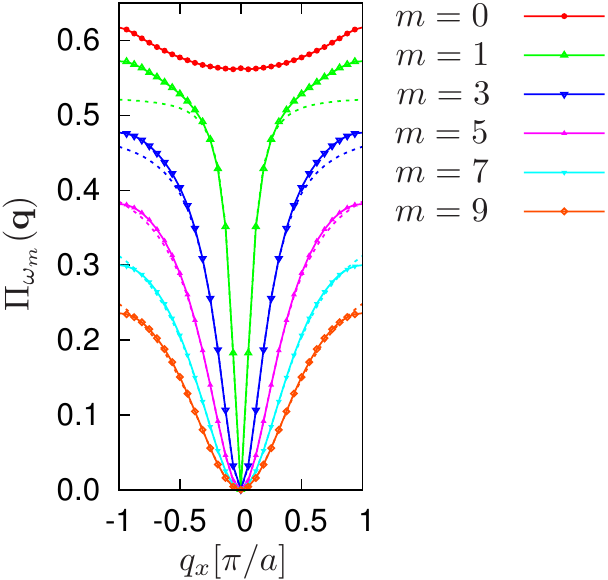}
 \caption{ \label{fig:polarization}(Color online) Polarization $\Pi_{\omega_{m}}(\qv)$ as a function of momentum $q_{x}$ for $q_{y}=0$ and different Matsubara frequencies $\omega_{m}$. For $\omega_{m}>0$ and small $q$, we clearly observe the $\sim q^{2}$ behavior of the polarization as required by gauge invariance. Dashed lines show fits to a $-b (q/\iom_{m})^{2}/[1+c^{2}(q/\iom_{m})^{2}]$ behavior for $\omega_{m}>0$ ($b$ and $c$ are fit parameters). When $q$ is small compared to the frequency, the $q^{2}$ behavior is clearly visible. For small $\omega_{m}$, it is less visible due to the finite momentum resolution.}
\end{figure}

In case of a Coulomb potential, the collective charge excitations are plasmons, whose dispersion relation $\omega(q)$ is solution of 
\begin{align}
 1+V(\qv)\Pi_{\omega}(\qv)=0.
\end{align}
In the standard RPA analysis (where $\Pi$ is just the Lindhardt function), for $q\rightarrow 0$, $\omega$ finite, 
one has (with $f$ some function, and $g$ the coupling constant) 
\begin{align}
\Pi_{\omega}(\qv) \sim g q^2 f(\omega) + \mathcal{O}(q^{4}),
\end{align}
which yields the plasmon dispersion relation $\omega(\qv) = \omega_p + a q ^2$ at small $q$, where $a$ is a constant and $\omega_p$ the plasma frequency.  $\omega_p$ is the solution of 
\begin{align}
   1+ g e^{2} V f(\omega_p) = 0.
\end{align}
From the previous discussion of gauge invariance, we know that the susceptibility vanishes in the long-wavelength limit for finite frequencies, $X_{\omega}(\qv\to 0)=0$. By virtue of \eqref{xinpi}, we expect the same behavior for the polarization, i.e., $\Pi(\qv\to 0,\omega)=0$. This behavior can be observed directly on the Matsubara data, as shown in Fig.~\ref{fig:polarization}: for finite Matsubara frequencies, the polarization vanishes in the long-wavelength limit. At small momenta, the data are well described by a function of the form $-b (q/\iom_{m})^{2}/[1+c^{2}(q/\iom_{m})^{2}]$. For $q$ small compared to $\iom_{m}$, the polarization is clearly proportional to $q^{2}$. One therefore expects that the standard RPA analysis still holds in the correlated regime.

In the EDMFT approximation, however, 
the polarization is computed from the local susceptibility and hence momentum independent. Therefore, the equation for the dispersion relation is
\begin{align}
1+\frac{e^{2}V\Pi_{\omega}}{q^{2}} = 0.
\end{align}
The polarization has to decrease as a function of large frequencies and to leading order we expect it to behave as $\Pi(\omega)\sim 1/\omega^{\alpha}$, with some $\alpha >0$. 
As a consequence, the frequency of the excitation will diverge in the long-wavelength limit: $\omega\sim 1/q^{2/\alpha}$. This behavior is shown in the left panel of Fig.~\ref{fig:epsinv_edmft}.
We note that \emph{a priori} a solution with a finite plasma frequency could exist if the polarization were to vanish for a finite $\omega$. We do not observe this in our calculations, however.
We therefore find that EDMFT does not provide a valid description of plasmons.

Let us now consider the dual boson approach~\cite{Rubtsov12}, which can be viewed as a diagrammatic expansion around extended dynamical mean-field theory. Additional details on the approach and its convergence properties can be found in Refs.~\onlinecite{Rubtsov12,vanLoon2014}. In this approach, the description of the collective modes amounts to replacing (\ref{xedmft}) by 
\begin{align}
\label{xdb}
X_{\omega}(\qv) = \frac{1}{\chi_{\omega}^{-1}(\qv) +W_{\omega} -V(\qv) },
\end{align}
where $\chi_{\omega}(\qv)$ is given by Eqs.~\eqref{chialt}--\eqref{chidual}, which includes non-local vertex corrections into the EDMFT susceptibility.
In EDMFT, there is an ambiguity of calculating the susceptibility. It is either given by the bosonic propagator \eqref{xedmft}, or can be obtained similarly as in DMFT. In the latter case, one computes the susceptibility from~\eqref{xdb} with $\chi_{\omega}(\qv)$ computed from the DMFT expressions of Sec.~\ref{dmftsusc}. In dual boson, the bosonic propagator is also given by~\eqref{xdb}, however with $\chi_{\omega}(\qv)$ given by the alternative expressions described in Sec.~\ref{sec:chialt}. As we have shown (see Appendix~\ref{app:proof}), the two expressions for $\chi_{\omega}(\qv)$ are equivalent, so that the dual boson approach resolves this ambiguity.

The resulting polarization $\Pi_{\omega}(\qv)=[-\chi_{\omega}^{-1}(\qv)-W_{\omega}]^{-1}$ depends on momentum, in contrast to EDMFT. It can be proven that this approach yields a gauge-invariant response in the long-wavelength limit~\cite{Rubtsov12}.
Let us now show that this approach indeed describes plasmons.

\subsection{Results}
\label{sec:lrresults}

\begin{figure}[t]
\includegraphics[width=0.5\textwidth]{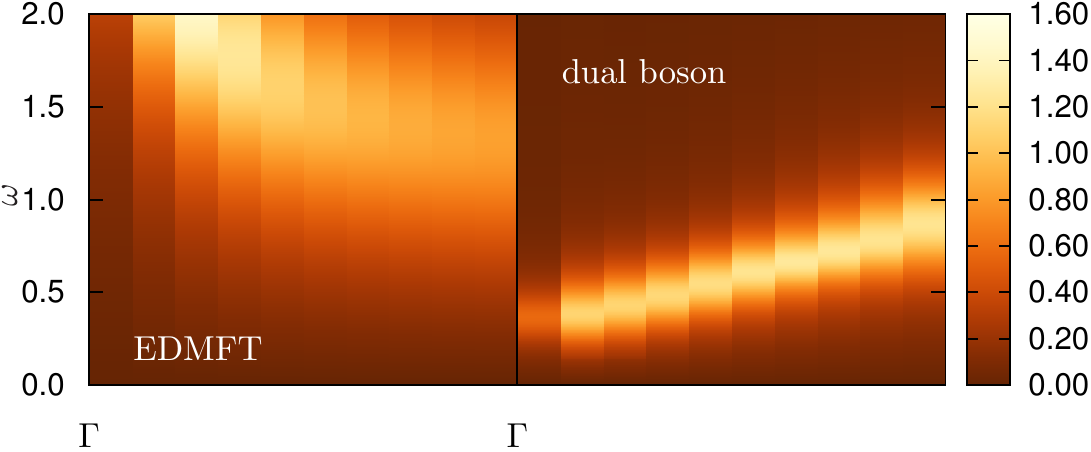}
\caption{\label{fig:epsinv_edmft} (Color online).  $-\Im \epsilon^{-1}(E,\qv)$ for small values wave vectors (up to $\abs{\qv}\sim 2\pi\cdot 0.34$) for $U=V=1$ and $T=0.02$. The energy of the collective mode diverges in EDMFT for $\abs{\qv}\to 0$ due to violation of gauge invariance. This unphysical behavior is corrected by including non-local vertex corrections, yielding a finite energy plasmon mode.
}
\end{figure}

In this section, we present results for the three-dimensional extended Hubbard model with the infinite-range potential \eqref{Vq} of strength $V$. We use a momentum-space discretization of $32\times 32\times 32$ $k$-points.\footnote{Momentum-dependent quantities including the vertex function are calculated on the irreducible part of the Brillouin zone only and the vertex function is stored for a single momentum at a time.} All results are for temperature $T=0.02$.

To begin with, we compare the physical content of the two approximations for the polarization operator which enter Eqs.~\eqref{xedmft} and \eqref{xdb}, respectively. To this end, we examine the \emph{inverse dielectric function}
\begin{align}
\epsilon^{-1}_{\omega}(\qv) = 1+V(\qv)X_{\omega}(\qv).
\end{align}
Here, $X_{\omega}(\qv)$ is analytically continued to the real axis using Pad\'e approximants. The dielectric function is experimentally accessible through electron energy-loss spectroscopy (EELS).
In Fig.~\ref{fig:epsinv_edmft}, we plot the inverse dielectric function for small momenta  and on real frequencies. As mentioned before, the energy of the collective mode diverges in the long-wavelength limit in EDMFT (left part of the figure). Including vertex corrections into the polarization, corrects this unphysical result and we observe a plasmon mode at finite energy in the long-wavelength limit (right panel). The dispersion of this mode is roughly consistent with a $q^{2}$ behavior.

\begin{figure}[b]
\includegraphics[width=0.5\textwidth]{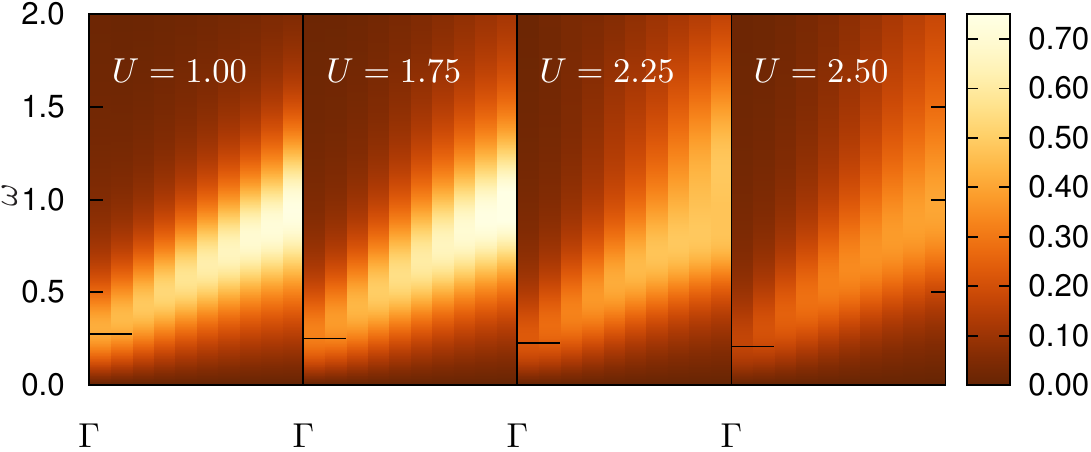}
\caption{(Color online)\label{fig:epsinv_3d}  $-\Im \epsilon^{-1}(E,\qv)$ for small values wave vectors up to $\abs{\qv}\sim 2\pi\cdot 0.34$ showing the plasmon mode in the three-dimensional Hubbard model with Coulomb interaction $V(\qv)\propto \frac{1}{\qv^2}$. Data is shown for $V=0.5$ and different values of $U$ at $T=0.02$. The plasma frequency corresponding to the maximum intensity of the peak for $q\to 0$ (indicated by the horizontal line) decreases with increasing $U$.
}

\end{figure}

We can study the dependence of this mode on the interaction strength. 
In Fig.~\ref{fig:epsinv_3d}, we plot the inverse dielectric function for different values of the local interaction $U$ but fixed strength of the long-range potential set to $V=0.5$. 
We see that with increasing on-site interaction, the mode shifts to lower energies, as indicated by the horizontal bars which mark the center of the peak at low $q$. The spectral weight also decreases with increasing interaction.

\subsection{Plasma frequency}
\label{sec:plasmafreq}

In order to show that the observed mode is indeed the collective plasmon excitation, we can compare the maximum of the spectral intensity to the plasma frequency for this model. The derivation of the plasma frequency is essentially based on gauge invariance. One can obtain the expression either starting from the continuity equation \eqref{kcontinuity} for the response kernel, or alternatively from the electrical conductivity and using the $f$-sum rule~\cite{anderson97}. Details of the derivation are given in Appendix~\ref{app:plasmafreq}. The result is
\begin{align}
\label{wp}
\omega_{p}^{2} = e^{2}a^{2}\tti V\mathcal{N},
\end{align}
where we have defined
\begin{align}
\label{n}
\mathcal{N}\Let \frac{2}{N}\sum_{\kv\sigma}\cos(k_{z}a)\av{n_{\kv\sigma}}.
\end{align}
Here we have assumed the field to be oriented along the $z$-axis. In order to rationalize this expression, we note that by linearizing the dispersion in the vicinity of the Fermi level and identifying the coefficient (the velocity) with $k/m$, we see that $\tti\sim 1/m$, where $m$ is the bare band mass. Further letting $a=1$, $V=4\pi$ and replacing $\mathcal{N}$ with $n$, it formally takes the same form as the plasma frequency in the continuum, $\omega_{p}^{2}=4\pi ne^{2}/m$. The appearance of $\mathcal{N}$ instead of the local density is a peculiarity of the lattice model and a consequence of the fact that the electromagnetic potential couples to the bonds rather than to the local charge density (see Appendix~\ref{app:current}).

In Fig.~\ref{fig:epsinv_cut}, we show low momentum cuts of the inverse dielectric function (the data for $V=0.5$ are the same as in the first panel of Fig.~\ref{fig:epsinv_3d}). They exhibit a well-defined peak, the position of which is well captured by the expression for the plasma frequency [Eq.~\eqref{wp}]. The agreement is remarkable given that the dielectric function has been obtained by analytical continuation. We emphasize that this coincidence is non trivial: 
the position of the peak (the energy of the collective mode) is determined by the two-particle properties of the system, while the plasma frequency is computed from single-particle properties (i.e., the density distribution) only. This relation is a consequence of gauge invariance, which is seen to be fulfilled in our calculation.
The connection of single- and two-particle properties is reminiscent of the Ward identity. 
We further note that the result for the plasma frequency is not restricted to our particular approximation, but applies to any approximation on a discrete lattice which respects gauge invariance (including RPA).

\begin{figure}[b]
\includegraphics[width=0.5\textwidth]{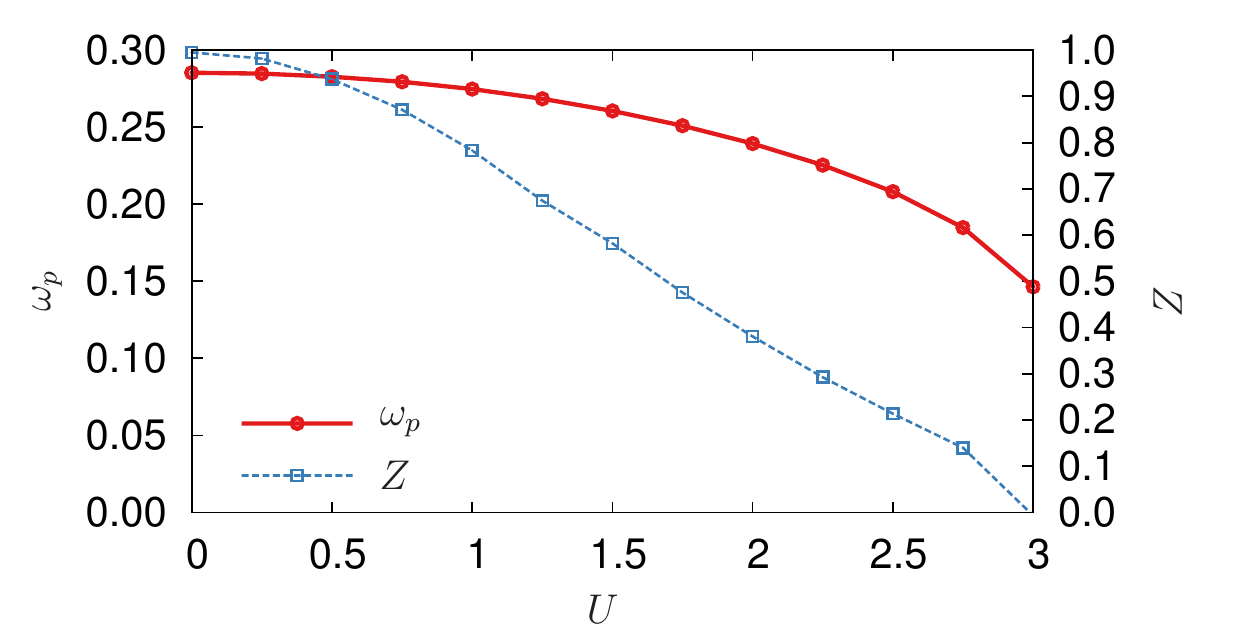}
\caption{\label{fig:wpvsu} (Color online) Dependence of the plasma frequency on the local interaction $U$ for fixed $V=0.5$ and $T=0.02$. The quasiparticle weight $Z$ is shown for comparison. The plasma frequency remains finite at the Mott transition.}
\end{figure}

The plasma frequency depends on the local interaction through $\mathcal{N}$. In Fig.~\ref{fig:wpvsu}, we plot the dependence of $\omega_{p}$ on $U$. The plasma frequency decreases with increasing interaction as observed in Fig.~\ref{fig:epsinv_cut}. In a simplified picture, the plasma frequency decreases because it is inversely proportional to the square root of the effective mass and the  effective mass increases with interaction. According to \eqref{n}, the frequency decreases because the density distribution becomes less momentum dependent as the interaction increases. In the insulator $\av{n_{\kv\sigma}}$ remains momentum dependent and hence the frequency remains finite, but the spectral weight drops to zero.

Clearly, the plasma frequency does not scale with the quasiparticle $Z$ (which also holds for a short-range interaction). This has the important implication that plasmons in strongly correlated systems are beyond Fermi-liquid theory where the quasiparticle contributions are considered dominant~\cite{Nozieres64,Abrikosov65,Platzman73}.
From the general theory of interacting Fermi systems~\cite{Nozieres64,Abrikosov65} it is known that there are two contributions to the occupation number $\av{n_{\kv\sigma}}$: the quasiparticle contribution, which originates from the pole of the electron Green's function and is proportional to $Z$, and a non-quasiparticle one, which stems from the branch cut of the Green's function and relates to its ``incoherent'' part. One can clearly see from Fig.~\ref{fig:wpvsu} that in the vicinity of the metal-insulator transition, the plasma frequency is mostly associated with the incoherent (non-quasiparticle) properties.\footnote{Within DMFT, the MIT is first-order~\cite{Georges96} so that strictly speaking the point $Z=0$ cannot be reached, but $Z$ near the transition is small so our conclusion remains valid.} The plasmons, however, remain well defined as one can see from
Figs.~\ref{fig:epsinv_3d} and \ref{fig:epsinv_cut}.

\begin{figure}[t]
\includegraphics[width=0.5\textwidth]{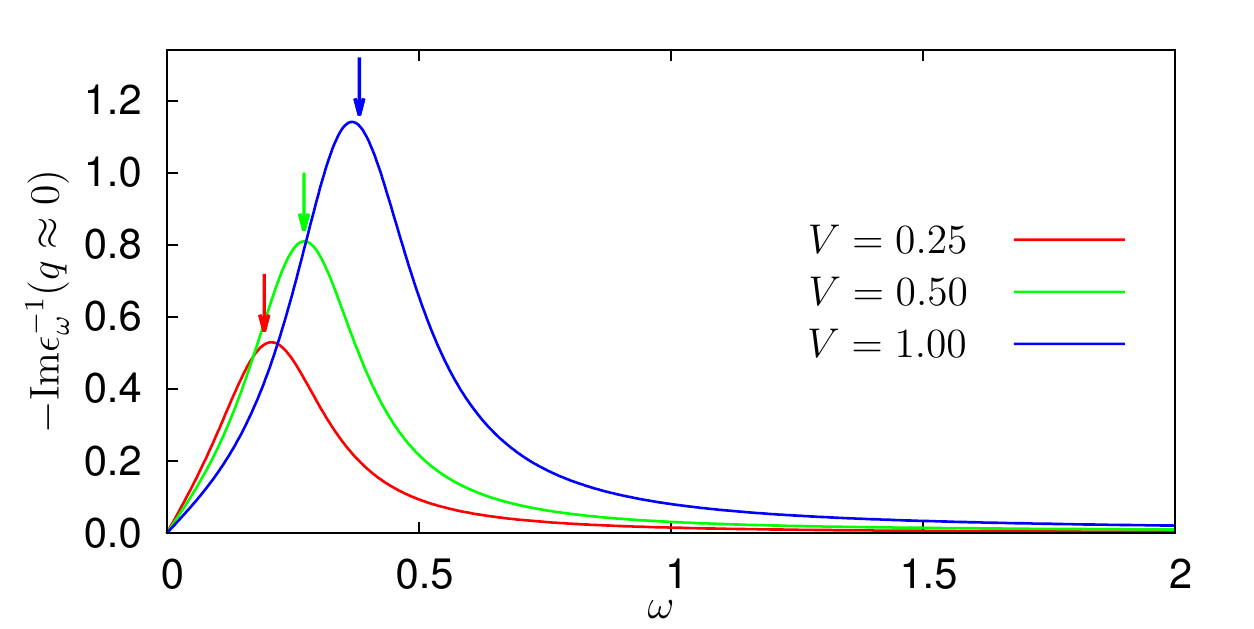}
\caption{\label{fig:epsinv_cut} (Color online) Fixed momentum cuts of the imaginary part of the inverse dielectric function for the smallest momentum $q>0$ and for $U=1$, $T=0.02$ and different values of $V$. The arrows indicate the plasma frequency computed from Eq.~\eqref{wp}.
}
\end{figure}

\begin{figure}[t]
\includegraphics[width=0.5\textwidth]{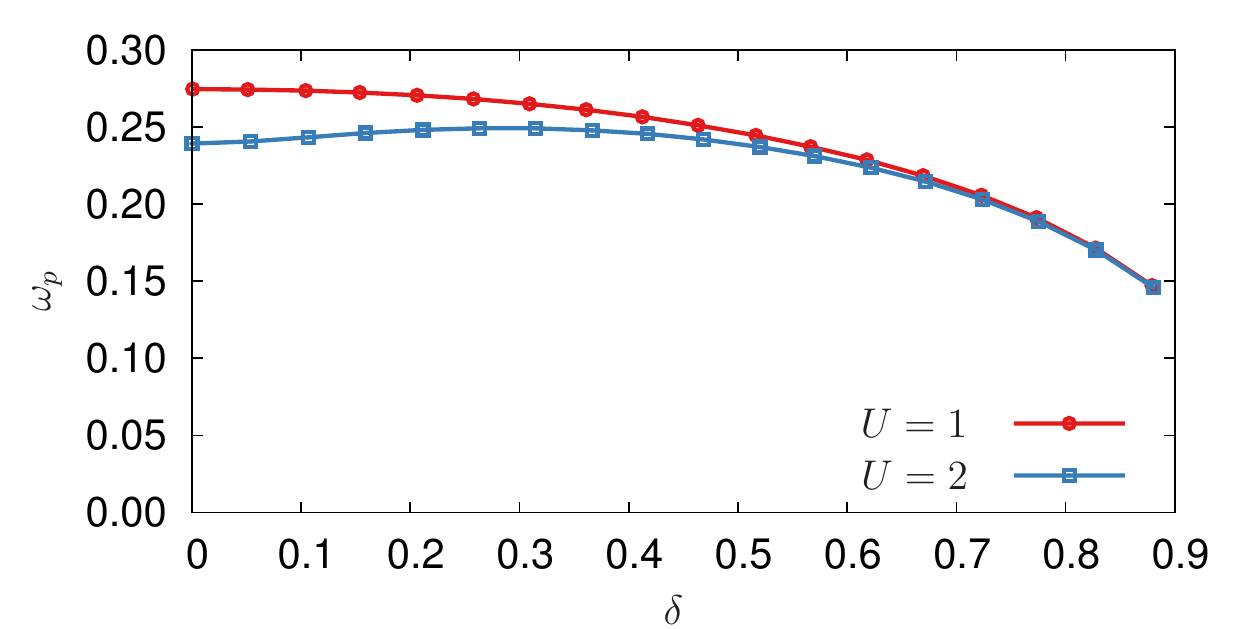}
\caption{\label{fig:wpvsd} (Color online) Doping ($\delta$-)dependence of the plasma frequency fixed interaction $V=0.5$, $T=0.02$, and two values of $U$.}
\end{figure}

In Fig.~\ref{fig:wpvsd}, we finally plot the doping dependence of the plasma frequency. $\omega_{p}$ is seen to decrease with doping but only appreciably so for an almost empty (filled) band. For sufficiently large $U$ there appears to be a shallow maximum in the doping dependence of the plasma frequency.

\section{Conclusions}

In this paper, we have addressed the collective charge excitations of strongly correlated electrons in presence of short- and long- range interactions and discussed the relation between gauge invariance and vertex corrections.
Non-local vertex has been shown to be essential for a qualitatively correct description of the collective modes in both cases.
Both the zero-sound mode in the case of a short-range interaction and the plasmon mode emerge through an RPA-like mechanism and are present up to the Mott transition.
Our results emphasize the importance of including vertex corrections from a fully frequency-dependent irreducible vertex when working with a frequency-dependent self-energy. Respecting gauge invariance is necessary in order to obtain a proper description of the collective modes in correlated media.

On  the technical side, we have proven that the DMFT susceptibility including vertex corrections yields a gauge-invariant charge response in finite dimensions. We have further shown that an alternative expression for the susceptibility that emerges in the dual boson approach is equivalent to the DMFT susceptibility. Such a formulation has the advantage that it circumvents numerical problems due to a divergence of the irreducible vertex close to a metal-insulator transition. It also resolves the ambiguity of calculating the susceptibility in EDMFT.
The approach is straightforwardly generalized to treat spin excitations.

\begin{acknowledgments}
We thank Thomas Ayral, Sergey Brener, Silke Biermann, Junya Otsuki and Alexey Rubtsov for valuable discussions. H.H. would further like to thank Fran\c{c}ois G\'elis for useful discussions on lattice gauge theory. We also thank Lewin Boehnke for providing his maximum entropy code.
E.G.C.P.v.L. and M.I.K. acknowledge support from ERC Advanced Grant No.~338957~FEMTO/NANO, A.I.L. from the DFG (Grant No.~FOR1346) and H.H. and O.P. from the ERC Grant No.~278472--\emph{MottMetals}.
This research used high-performance computing resources of GENCI-CCRT under Grant No.~t2014056112.
The simulations employed a modified version of an open source implementation of the hybridization-expansion continuous-time quantum impurity solver~\cite{Hafermann13}, based on the ALPS libraries~\cite{ALPS2}.
\end{acknowledgments}

\appendix

\section{Analytical continuation}
\label{app:pade}

The analytical continuation requires accurate input data. For solving the quantum impurity model, we utilize the numerically exact hybridization-expansion continuous-time quantum Monte Carlo method~\cite{Werner06}, which can treat a local retarded interaction without approximation~\cite{Werner07,Werner10}. To maximize accuracy, we employ improved measurements for the local susceptibility $\chi$, self-energy, three-leg vertex $\lambda$, and four-leg vertex function $\gamma$ in the simulation~\cite{Hafermann14}.

For the analytical continuation of the imaginary time data itself, we use a straightforward implementation of the Pad\'e algorithm as presented in Ref.~\onlinecite{Vidberg77}.
Because the analytical continuation is an ill-posed mathematical problem, we performed consistency checks. We compared the analytical continuation of the local part of the susceptibility to the local part of the analytically continued result with good agreement for all values of $U$ (the momentum sum and analytical continuation should commute if the latter was exact). As an additional check for the Pad\'e algorithm, we numerically integrated the imaginary part of the analytically continued susceptibility to verify that the Kramers-Kronig relation
\begin{align}
\chi'(\omega) &=-\mathcal{P}\int_{-\infty}^{\infty}\frac{d\omega'}{\pi}\frac{\chi''(\omega')}{\omega-\omega'}
\end{align}
is well fulfilled for $\omega=0$.
In the calculations, the vertex function was determined for up to $128$ fermionic Matsubara frequencies (including positive and negative), which corresponds to a cutoff of four times the bandwidth. We took the same number of positive bosonic frequencies and checked that the results do not change appreciably for a smaller number of frequencies. For the single-particle Green's function and self-energy, we took a larger cutoff of 192 frequencies, which is sufficient at this temperature.
The analytical continuation of the Matsubara data turns out to be robust when varying the number of input frequencies. Despite these checks, we cannot exclude a qualitative deviation from the real spectra, such as a splitting into multiple peaks. The analytical continuation by Pad\'e approximants tends to give a single-peak spectrum when used in conjunction with data afflicted with statistical errors~\cite{Huang14}.
For the long-wavelength excitations we are mainly interested in, the Pad\'e results are verified independently: by computing the dispersion from a fit of the Matsubara data in the case of short-range forces (Sec.~\ref{sec:results}) and by evaluating the plasma frequency from the density in the case of long-range interaction (Sec.~\ref{sec:lrresults}).

\section{Current operator}
\label{app:current}

In order to discuss local charge conservation, we require a gauge theory on the lattice. While it can be formulated more generally, it is sufficient for our purposes to consider the case of weak and slowly varying fields.\footnote{By slowly varying we mean that the vector potential does not change appreciably over an interatomic distance, so that we can write $\int_{\rv}^{\rv+\dv}\Av d\rv \approx \Av_{\rv}\dv$ We only discuss quantities in absence of external fields, i.e. in the limit $\Av\to\vc{0}$.} We introduce the coupling of the Hamiltonian \eqref{hmlt} to a vector potential via the Peierls substitution~\cite{Peierls33}:
\begin{align}
\hat{T}=-\tti\sum_{\rv\dv\sigma} c^{\dagger}_{\rv\sigma} e^{\i e\Av_{\rv}\dv}c_{\rv-\dv\sigma} + c_{\rv-\dv\sigma}^{\dagger} e^{-\i e\Av_{\rv}\dv}c_{\rv\sigma}.
\end{align}
For a discussion on the validity of the Peierls substitution and the consequences of the above assumptions, see Refs.~\cite{Alexandrov91,Graf95}.
The coupling only affects the kinetic energy $\hat{T}$.
Under a gauge transformation, $\Av_{\rv}\dv\to\Av_{\rv}\dv+\Lambda_{\rv-\dv}-\Lambda_{\rv}$, $c_{\rv}^{\dagger}\to c_{\rv}^{\dagger}e^{\i\Lambda_{\rv}}$, $c_{\rv}\to c_{\rv}e^{-\i\Lambda_{\rv}}$, the Hamiltonian remains invariant.
The current is determined in the usual way as the functional derivative $\jv_{\rv}=-\delta H/\delta \Av_{\rv}$. Within linear response, the exponential is expanded up to second order in the vector potential. For the current, we obtain
\begin{align}
\label{current}
\jv_{\rv} =& \i e\tti\sum_{\dv\sigma}\left(c_{\rv\sigma}^{\dagger}c_{\rv-\dv\sigma}-c_{\rv-\dv\sigma}^{\dagger}c_{\rv\sigma}\right)\dv\notag\\
& -e^{2}\tti\sum_{\dv\sigma} \left(c_{\rv\sigma}^{\dagger}(\Av_{\rv}\dv)c_{\rv-\dv\sigma}+c_{\rv-\dv\sigma}^{\dagger}(\Av_{\rv}\dv)c_{\rv\sigma}\right)\dv.
\end{align}
The first term is the paramagnetic current $\jv_{\rv}^{p}$ and the second is the diamagnetic contribution $\jv_{\rv}^{d}$.
The momentum representation of the paramagnetic current is
\begin{align}
\vc{j}_{\vc{q}}^{p} &= \sum_{\rv} \vc{j}^p_{\rv} e^{-\i\qv\rv}\notag\\
&=\frac{\i e\tti}{N} \sum_{\kv\dv\sigma}c_{\kv\sigma}^{\dagger}\left(e^{-\i(\kv+\qv)\dv}- e^{\i\kv\dv}\right)\dv\, c_{\kv+\qv\,\sigma}.
\end{align}
The individual spatial components can be written
\begin{align}
\label{jpalpha}
[\vc{j}_{\vc{q}}^{p}]_{\alpha}=\frac{\i e\tti a}{N}\sum_{\kv\sigma} c_{\kv\sigma}^{\dagger} (e^{-\i (k_{\alpha}+q_{\alpha})a}-e^{\i k_{\alpha}a})c_{\kv+\qv\,\sigma}.
\end{align}
This expression can be cast into in the symmetrical form
\begin{align}
[\vc{j}_{\vc{q}}^{p}]_{\alpha}=\frac{\i e\tti a}{N} \sum_{\kv\sigma} c_{\kv-\qv/2\,\sigma}^{\dagger} (e^{-\i k_{\alpha}a}-e^{\i k_{\alpha}a})e^{-\i q_{\alpha}/2}c_{\kv+\qv/2\,\sigma}.
\end{align}
Symbolically, this can be expressed in terms of a derivative of the dispersion
\begin{align}
\label{epsk}
\epsilon_{\kv} = -\tti\sum_{\alpha}\left(e^{-\i k_{\alpha} a} + e^{\i k_{\alpha} a} \right)
\end{align}
in the form
\begin{align}
[\vc{j}_{\vc{q}}^{p}]_{\alpha} =& \frac{e}{N}\sum_{\kv\sigma} c_{\kv-\qv/2\,\sigma}^{\dagger} \frac{\partial \varepsilon_{\kv}}{\partial k_{\alpha}} e^{-\i q_{\alpha}a/2} c_{\kv+\qv/2\,\sigma}.
\end{align}
In the long-wavelength ($q\to 0$) limit, this reduces to
\begin{align}
\vc{j}^{p} =&  \frac{e}{N}\sum_{\kv\sigma} c_{\kv\sigma}^{\dagger} (\nabla \varepsilon_{\kv}) c_{\kv\sigma}.
\end{align}
For completeness, we provide the result for the diamagnetic contribution to the current. In momentum space, we obtain
\begin{align}
[\vc{j}^{d}_{\qv}]_{\alpha}
&=-\frac{e^{2}a^{2}\tti}{N^{2}}\sum_{\kv\kv'\sigma}c_{\kv\sigma}^{\dagger} A^{\alpha}(\kv-\kv') \big[e^{-\i(k_{\alpha}'+q_{\alpha})a}\notag\\
&\qquad\qquad\qquad + e^{\i k_{\alpha}a}\big] c_{\kv'+\qv\,\sigma}\notag\\
&=-\frac{e^{2}a^{2}\tti}{N^{2}}\sum_{\kv\qv'\sigma}c_{\kv\sigma}^{\dagger}A^{\alpha}(\qv') \big[e^{-\i(k_{\alpha}+q_{\alpha}-q'_{\alpha})a}\notag\\
&\qquad\qquad\qquad+ e^{\i k_{\alpha}a}\big] c_{\kv'+\qv-\qv'\,\sigma}\notag\\
&=-\frac{e^{2}a^{2}\tti}{N^{2}}\sum_{\kv\qv'\sigma}c_{\kv-\qv/2+\qv'/2\,\sigma}^{\dagger} A^{\alpha}(\qv')\big[e^{-\i k_{\alpha}a}+ e^{\i k_{\alpha}a}\big]\notag\\
&\qquad\qquad\qquad\times e^{-\i(q_{\alpha}/2-q_{\alpha}'/2)a} c_{\kv+\qv/2-\qv'/2\,\sigma}.
\end{align}
Similarly as for the paramagnetic current, we can rewrite this as
\begin{align}
[j^{d}_{\qv}]_{\alpha}&=-\frac{e^{2}}{N^{2}}\sum_{\kv\qv'\sigma}c_{\kv-\qv/2+\qv'/2\,\sigma}^{\dagger} A^{\alpha}(\qv')\Big[\sum_{\beta}\frac{\partial^{2} \varepsilon_{\kv}}{\partial k_{\alpha}\partial k_{\beta}}\Big]\notag\\
&\qquad\qquad\qquad\times e^{-\i(q_{\alpha}/2-q_{\alpha}'/2)a} c_{\kv+\qv/2-\qv'/2\,\sigma}.
\end{align}

In the following, we are interested in the continuity equation in absence of an electromagnetic field and will henceforth focus on the paramagnetic contribution to the current.
We introduce the four-vector notation $j^{\mu}_{\qv}=(j^{0}_{\qv},\jv^{\alpha}_{\qv})$, with $j^{0}_{\qv}=en_{\qv}$  and define the bare current vertex through
\begin{align}
j^{\mu}_{\qv} =  \frac{e}{N}\sum_{\kv\sigma}c_{\kv\sigma}^{\dagger} \gamma^{\mu}(\kv,\qv) c_{\kv+\qv\,\sigma}.
\end{align}
Using \eqref{jpalpha}, we obtain
\begin{align}
\label{gammamubare}
\gamma^{\mu} = \left\{\begin{array}{ccl}
\i\tti a\left(e^{-\i(k_{\alpha}+q_{\alpha})a}-e^{\i k_{\alpha}a}\right), & \mu=&\alpha=x,y,z\\
1, & \mu=&0
\end{array} \right. .
\end{align}

\section{Continuity equation}

The Hamiltonian including the coupling to the electromagnetic field fulfills the continuity equation
\begin{align}
\label{eom}
e\frac{\partial n_{\rv}}{\partial t} =& -\i e[n_{\rv},H]\notag\\
=&\i e \tti \sum_{\dv\sigma} \left(c_{\rv\sigma}^{\dagger}c_{\rv+\dv\sigma}\!+\!c_{\rv\sigma}^{\dagger}c_{\rv-\dv\sigma}\!-\!c_{\rv+\dv\sigma}^{\dagger}c_{\rv\sigma}\!-\! c_{\rv-\dv\sigma}^{\dagger}c_{\rv\sigma} \right)\notag\\
&-e^{2}\tti\sum_{\dv\sigma}\Big(c_{\rv\sigma}^{\dagger}(\Av_{\rv}\dv)
c_{\rv+\dv\sigma}\!+\!c_{\rv\sigma}^{\dagger}(\Av_{\rv}\dv)c_{\rv-\dv\sigma}\notag\\
&\quad\qquad\qquad-c_{\rv+\dv\sigma}^{\dagger}(\Av_{\rv}\dv)c_{\rv\sigma}\!-\! c_{\rv-\dv\sigma}^{\dagger}(\Av_{\rv}\dv)c_{\rv\sigma} \Big).
\end{align}
We can write the right hand side as a divergence of the current operator \eqref{current}. To this end, we define the \emph{forward derivative} of the current
\begin{align}
\nabla^{F}\cdot\jv_{\rv} \Let \frac{\jv_{\rv+\dv}-\jv_{\rv}}{a},
\end{align}
which should be understood such that the finite difference of the $x$-direction of the current is formed by displacement $\delta$ in the $x$-direction, etc.
Using Eqs.~\eqref{eom} and \eqref{current}, it is easy to see that the continuity equation can then be written in the form
\begin{align}
e\frac{\partial n_{\rv}}{\partial t} + \nabla^{F}\cdot\jv_{\rv} = 0.
\end{align}
In order to recast this equation into momentum space, one defines $\qv^{F}$ as the eigenvalue of the operator $\nabla^{F}=\sum_{\alpha}\partial^{F}_{\alpha}\ev_{\alpha}$ acting on a plane wave $\phi_{\qv}(\rv)\sim e^{\i\qv\rv}$. Denoting $q_{\alpha}=\qv_{\alpha}\ev_{\alpha}$ as the $\alpha$-component of $\qv$, which takes the discrete values $q_{\alpha}^{(n)} = 2\pi\i n/(N a)$ and correspondingly $r_{\alpha}=\rv\ev_{\alpha}$, $r_{l_{\alpha}}= a l_{\alpha}$, we have 
\begin{align}
\nabla^{F}e^{\i\qv\rv} & = \frac{1}{a} \sum_{\alpha} \left(e^{\i q_{\alpha}(l_{\alpha}+1)a}\!\!-e^{\i q_{\alpha}l_{\alpha}a}\right)\!\ev_{\alpha}\notag\\
&=\frac{1}{a} \sum_{\alpha} \left(e^{\i q_{\alpha}a}-1\right)e^{\i q_{\alpha}l_{\alpha}a} \equiv \i\qv^{F}e^{\i\qv\rv}
\end{align}
so that
\begin{align}
\label{qf}
\qv^{F} = \sum_{\alpha} q^{F}_{k_{\alpha}}\ev_{\alpha} = -\frac{\i}{a}\sum_{\alpha}\left(e^{\i q_{\alpha}a}-1\right)\ev_{\alpha}.
\end{align}
In the long-wavelength limit, this simplifies to
\begin{align}
\qv^{F} \approx -\frac{\i}{a}\sum_{\alpha}\left(\i q_{\alpha}a\right)\ev_{\alpha} \equiv\qv.
\end{align}
We note that by defining the \emph{backward derivative} as $\nabla^{B}\cdot\jv_{\rv}\Let(\jv_{\rv-\dv}-\jv_{\rv})/a$, we obtain $\qv^{B}=(\qv^{F})^{*}$. The dispersion can be expressed in terms of the product $\qv^{F}\cdot\qv^{B}$.

\section{Ward identity}
\label{app:sec:ward}

The derivation of the Ward identity can be found in textbooks. See, e.g., Ref. \onlinecite{Schrieffer99}. Here, we sketch the derivation as required for the subsequent discussion. It is convenient to use four-vector notation throughout.

With $k=(\i\omega,\kv)$, $z=(\tau,\rv)$, $\partial^{F}_{\mu}=(\partial_{\tau},\nabla^{F})$,  $q_{\mu}^{F}=(\i\omega,\qv^{F})$, and the metric $(-1,1,1,1)$, we have $k_{\mu}z_{\mu} = \kv\rv - i\omega\tau$ and the continuity equation becomes $\partial^{F}_{\mu}j_{\mu}(z) = 0$. The Ward identity is obtained by applying the four-divergence to the three-leg correlation function
\begin{align}
\Lambda_{\mu}(x,y,z) = \av{\Ttau c(x)c^{\dagger}(y)j_{\mu}(z)}.
\end{align}
We have
\begin{align}
\label{fourdivlambda}
\partial_{\mu}^{F}\Lambda_{\mu} = &\av{\Ttau c(x)c^{\dagger}(y)\left[ \partial^{F}_{z_{\alpha}} j_{\alpha} + \partial_{z_{0}} j_{0} \right]}\notag\\
&+\av{\Ttau c(x)[j_{0}(z),c^{\dagger}(y)]\delta(y_0-z_0)}\notag\\
&+\av{\Ttau c^{\dagger}(y)[c(x),j_{0}(z)]\delta(x_0-z_0)},
\end{align}
where the term in angular brackets in the first line vanishes by virtue of the continuity equation and the two other lines emerge due to time differentiation while accounting for time-ordering.
With $j_{0}\equiv en$ the commutators are
\begin{align}
[j_{0}(z),c^{\dagger}(y)]\delta(y_0-z_0) &= e c^{\dagger}(y)\delta(y-z),\notag\\
[c(x),j_{0}(z)]\delta(x_0-z_0) &= e c(x)\delta(x-z).
\end{align}
Inserting this back into \eqref{fourdivlambda} and using the definition of Green's function, $G(x-y)\Let-\av{\Ttau c(x)c^{\dagger}(y)}$, yields
\begin{align}
\label{wardrealsp}
\partial_{\mu}^{F}\Lambda_{\mu} = e[\delta(x-z)-\delta(y-z)]G(x-y).
\end{align}
Noting that $\Lambda_{\mu}$ can be expressed in terms of the generalized susceptibility (see following), this equation is recognized as the lattice formulation of Eq.~(13) in Ref. \onlinecite{Baym62}.
Defining the current vertex $\Gamma_{\mu}$ through
\begin{align}
\Lambda_{\mu}(x,y,z)=e\int dx'\int dy' G(x-x')\Gamma_{\mu}(x',y',z)G(y'-y)
\end{align}
and introducing the lattice Fourier transform $G(k) =\int dx e^{-\i kx}G(x)$, where $\int dx\Let \sum_{\vc{x}}\int_{0}^{\beta}d\tau$, one straightforwardly obtains the momentum space representation of \eqref{wardrealsp},
\begin{align}
G(k)q^{F}_{\mu}\Gamma_{\mu}(k,q)G(k+q) = G(k+q) - G(k).
\end{align}
It is commonly written in the form
\begin{align}
\label{app:ward}
q^{F}_{\mu}\Gamma_{\mu}(k,q) = G^{-1}(k) - G^{-1}(k+q),
\end{align}
which is the Ward identity.

\section{Current vertex}
\label{app:Gmu}

\subsection{Noninteracting case}

In the noninteracting case, the bare current vertex \eqref{gammamubare} has to fulfill the Ward identity
\begin{align}
q^{F}_{\mu}\gamma_{\mu}(k,q) = G_{0}^{-1}(k) - G_{0}^{-1}(k+q).
\end{align}
In this case, we have $G_{0}^{-1}(k)=k_{0}-\varepsilon_{\kv}$ so the right hand side of \eqref{ward} becomes $\varepsilon_{\kv+\qv}-\varepsilon_{\kv}-q_{0}$, while the left-hand side reads $q^{F}_{\alpha}\gamma_{\alpha}-q_{0}$.
Using the \eqref{epsk} and \eqref{qf}, we see that indeed
\begin{align}
q_{\alpha}^{F}\gamma_{\alpha}&=\tilde{t}\sum_{\alpha}\left(e^{\i q_{\alpha}a}-1\right) \left(e^{-\i(k_{\alpha}+q_{\alpha})a}-e^{\i k_{\alpha}a}\right)\notag\\
&=\tilde{t}\sum_{\alpha}\left(e^{-\i k_{\alpha}a} - e^{\i(k_{\alpha}+q_{\alpha})a}-e^{-\i(k_{\alpha}+q_{\alpha})a} + e^{-\i k_{\alpha}a}\right)\notag\\
&=\varepsilon_{\kv+\qv}-\varepsilon_{\kv}.
\end{align}

\subsection{Interacting case}

\begin{figure}[t]
\includegraphics[scale=0.4,angle=0]{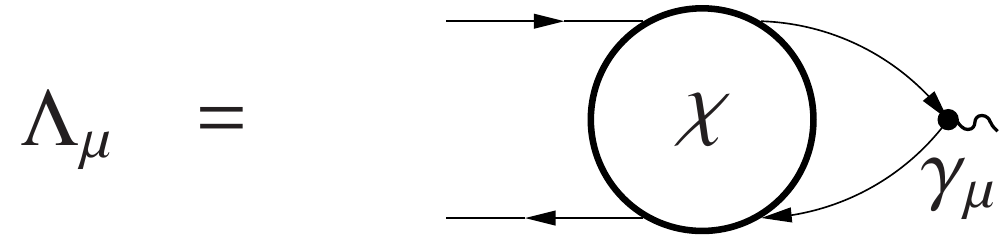} 
\caption{\label{fig:lambdamu} Diagrammatic representation of the three-leg correlation function $\Lambda_{\mu}$ in terms of the generalized susceptibility.
}
\end{figure}

In the interacting case, we have
\begin{align}
\Lambda_{\mu}^{\sigma}(k,q)&=\av{c_{k\sigma}c^{\dagger}_{k+q,\sigma}j^{\mu}_{q}}\notag\\
&=-\frac{e}{N}\sum_{k'\sigma'}\av{c_{k\sigma}c^{\dagger}_{k+q,\sigma}c_{k'\sigma'} c^{\dagger}_{k'+q,\sigma'}}\gamma^{\mu}(k',q),
\end{align}
where the correlation function in the second line is the generalized susceptibility (see Fig.~\ref{fig:lambdamu}):
\begin{align}
\label{lambdathroughl}
\Lambda_{\mu}^{\sigma}(k,q)=-\frac{e}{N}\sum_{k'\sigma'}\chi_{kk'q}^{\sigma\sigma'}\gamma^{\mu}(k',q).
\end{align}
The latter can be expressed in terms of the vertex function as follows:
\begin{align}
\label{lthroughgamma}
\chi_{kk'q}^{\sigma\sigma'} =-&\frac{1}{T} G_{k\sigma}G_{k+q\sigma}\delta_{kk'}\delta_{\sigma\sigma'}\notag\\
&+G_{k\sigma}G_{k+q\sigma}\Gamma_{kk'q}^{\sigma\sigma'}G_{k'\sigma'}G_{k'+q\sigma'}.
\end{align}
Inserting \eqref{lthroughgamma} into \eqref{lambdathroughl} and using the definition of the current vertex, $\Lambda_{\mu}(k,q) = eG_{k}\Gamma_{\mu}(k,q)G_{k+q}$, one obtains
\begin{align}
\label{gammamuthroughgamma}
\Gamma_{\mu}^{\sigma}(k,q) = \gamma_{\mu}(k,q) - \frac{T}{N}\sum_{k'\sigma'}\Gamma_{kk'q}^{\sigma\sigma'}G_{k'\sigma'}G_{k'+q,\sigma'}\gamma_{\mu}(k',q).
\end{align}
Inserting the Bethe-Salpeter equation for the vertex
\begin{align}
\Gamma_{kk'q}^{\sigma\sigma'} = \Gamma_{kk'q}^{\text{irr}\sigma\sigma'} - \frac{T}{N}\sum_{k''\sigma''}\Gamma_{kk''q}^{\text{irr}\sigma\sigma''}G_{k''\sigma''}G_{k''+q,\sigma''}\Gamma_{k''k'q}^{\sigma''\sigma'}
\end{align}
into \eqref{gammamuthroughgamma}, we obtain the ladder equation for the current vertex,
\begin{align}
\Gamma_{\mu}^{\sigma}(k,q) = \gamma_{\mu}(k,q) &- \frac{T}{N}\sum_{k'\sigma'}\Gamma_{kk'q}^{\text{irr}\,\sigma\sigma'}G_{k'\sigma'}G_{k'+q,\sigma'}\Gamma_{\mu}^{\sigma'}(k',q).
\end{align}

\section{Electromagnetic response kernel}
\label{app:k}

Within linear response, the electromagnetic response kernel is defined by
\begin{align}
J_{\mu}(q) = K_{\mu\nu}(q)A_{\nu}(q),
\end{align}
where $J_{\mu}(q)$ is the expectation value of the current. Demanding the invariance of the kernel under a gauge transformation $A_{\nu}(q)\to A_{\nu}(q)+\i q^{F}_{\nu}\Lambda(q)$ implies $K_{\mu\nu}q^{F}_{\nu}=0$. On the other hand, the fact that the expectation value $J_{\mu}$ fulfills the continuity equation $q^{F}_{\mu}J_{\mu}(q)=0$ implies the condition $q_{\mu}^{F}K_{\mu\nu}(q)=0$ imposed by charge conservation.

Following the standard derivation, i.e., by expressing the kernel as a functional derivative of the current with respect to the vector potential, we obtain the following result for the kernel on the discrete lattice:
\begin{align}
K_{\mu\nu}(\rv,\tau;\rv',\tau') =& \av{T_{\tau}j_{\mu}(\rv,\tau)j_{\nu}(\rv',\tau')}\notag\\
&-e^{2}a^{2} \tilde{t} \delta_{\mu\nu}(1-\delta_{0\mu})\delta(\tau-\tau')\delta_{\rv',\rv-\dv}\notag\\
&\times\sum_{\dv\sigma}\av{c_{\rv\sigma\tau}^{\dagger}c_{\rv-\dv\sigma\tau'} + c_{\rv-\dv\sigma\tau}^{\dagger}c_{\rv\sigma\tau'}},
\end{align}
where the second term originates from the derivative of the diamagnetic current [see Eq.~\eqref{current}] with respect to the vector potential and expectation values are taken in the absence of an external field.
Note that the field does not couple directly to the local density $n_{\rv}\equiv\sum_{\sigma}\av{c^{\dagger}_{\rv\sigma\tau}c_{\rv\sigma\tau}}$, because the electromagnetic potential is a link variable.
In momentum space, the kernel becomes
\begin{align}
K_{\mu\nu}(\qv,\iom) =&\av{T_{\tau}j_{\mu}(\qv,\iom)j_{\nu}(-\qv,-\iom)} \notag\\
&- e^{2}a^{2} \tti \delta_{\mu\nu}(1-\delta_{0\nu})\mathcal{N},
\end{align}
where for simplicity we have defined
\begin{align}
\mathcal{N}\Let&\frac{1}{N}\sum_{\kv\sigma}\left(e^{-\i k_{\nu}a}+e^{\i k_{\nu}a}\right)\av{n_{\kv\sigma}}.
\end{align}
In order to show that the exact kernel obeys the constraint imposed by charge conservation, we form the expression
\begin{widetext}
\begin{align}
\label{qkmunu}
q^{F}_{\mu}K_{\mu\nu}(\qv,\iom) =&\av{T_{\tau}[q_{\mu}^{F}j_{\mu}(\qv,\iom)]j_{\nu}(-\qv,-\iom)} + \av{[j_{0}(\qv),j_{\nu}(-\qv)]}\notag\\
&-e^{2}a^{2} \tti (1-\delta_{0\nu})q_{\nu}^{F}\frac{1}{N}\sum_{\kv\sigma}\left(e^{-\i k_{\nu}a}+e^{\i k_{\nu}a}\right)\av{n_{\kv\sigma}}\notag\\
=& \av{[j_{0}(\qv),j_{\nu}(-\qv)]}
+\i e^{2} a\tti (1-\delta_{0\nu})\frac{1}{N}\sum_{\kv\sigma}\left(e^{-\i (k_{\nu}-q_{\nu})a}-e^{-\i k_{\nu}a} + e^{\i (k_{\nu}+q_{\nu})a} -e^{\i k_{\nu}a}\right)\av{n_{\kv\sigma}},
\end{align}
where $\av{n_{\kv\sigma}}=\langle c_{\kv\sigma}^{\dagger}c_{\kv\sigma}\rangle$. The first term on the right-hand side vanishes for the exact kernel because of the continuity equation, $q_{\mu}^{F}j_{\mu}=0$. To obtain the last line, the explicit expression \eqref{qf} was substituted for $\qv^F$. The commutator arises because the time derivative does not commute with the time-ordering symbol. It evaluates to
\begin{align}
\av{[j_{0}(\qv),j_{\nu}(-\qv)]} &= e^{2}(1-\delta_{0\nu})\frac{1}{N}\sum_{\kv\sigma}[\gamma_{\nu}(\kv+\qv,-\qv)-\gamma_{\nu}(\kv,-\qv)]\av{n_{\kv\sigma}}\notag\\
&=\i e^{2}a\tti (1-\delta_{0\nu})\frac{1}{N}\sum_{\kv\sigma} \left(e^{-\i k_{\nu}a}-e^{\i (k_{\nu}+q_{\nu})a} - e^{-\i(k_{\nu}-q_{\nu})a} + e^{\i k_{\nu}}\right)\av{n_{\kv\sigma}},
\end{align}
\end{widetext}
so that the terms on the right-hand side of \eqref{qkmunu} cancel, leading to the required result $q_{\mu}^{F}K_{\mu\nu}(q)=0$. In the same way, one shows that the kernel is gauge invariant, i.e. $K_{\mu\nu}(q)q_{\nu}^{F}=0$.

\section{Plasma frequency}
\label{app:plasmafreq}

The plasma frequency is determined by the uniform response, i.e., in the limit $q_{\alpha}\to 0$. For small momenta, we replace $q_{\alpha}^{F}\to q_{\alpha}$. From the gauge invariance condition $K_{\mu\nu}q_{\nu}^{F}=0$ we have $K_{\mu0} =K_{\mu\alpha}q_{\alpha}/q_{0}$. We can take the direction of the field parallel to the $z$ axis. The transverse response vanishes by symmetry. Hence, we have $K_{00}=K_{0z}q/q_{0}$, where $q\equiv q_{z}$. Similarly, we obtain $K_{0z}=K_{zz}q/q_{0}$, which yields the continuity equation
\begin{align}
K_{00} = \frac{q^{2}}{q_{0}^{2}}K_{zz}.
\end{align}
Now, we use that the same relation holds when the density-density and current-current correlation functions in the kernel, which are reducible in the interaction, are replaced by the corresponding quantities irreducible in the interaction $V(\qv)$ (see, e.g., Ref.~\onlinecite{Nozieres64}). This is possible because the former are related to the latter through simple geometric series. We denote the resulting quantities $\tilde{K}_{00}$ and $\tilde{K}_{zz}$. Hence, we can identify $\tilde{K}_{00}$ with $e^{2}\Pi$ where $\Pi$ contains all polarization diagrams irreducible with respect to $V(\qv)$. 
In the long-wavelength limit, only the diamagnetic term in $\tilde{K}_{zz}$ contributes.  With the Coulomb interaction $V(\qv)=e^{2}V/q^{2}$ for $q\neq 0$, we have in the limit $q_{\alpha}\to 0$ and for $q_{0}\to\omega+i0^{+}$,
\begin{align}
\epsilon_{\omega}=\lim_{q\to 0}1+V(\qv)\Pi_{\omega}(\qv) = 1-\omega_{p}^{2}/\omega^{2}
\end{align}
with the plasma frequency
\begin{align}
\omega_{p} = ea\sqrt{\tti V\mathcal{N}}.
\end{align}
Alternatively, the same result may be obtained by relating the response kernel to the conductivity and using the $f$-sum rule.

\section{Equivalence of Eqs.~\eqref{chi} and \eqref{chialt}}
\label{app:proof}

\begin{figure}[h]
\includegraphics[scale=0.4,angle=0]{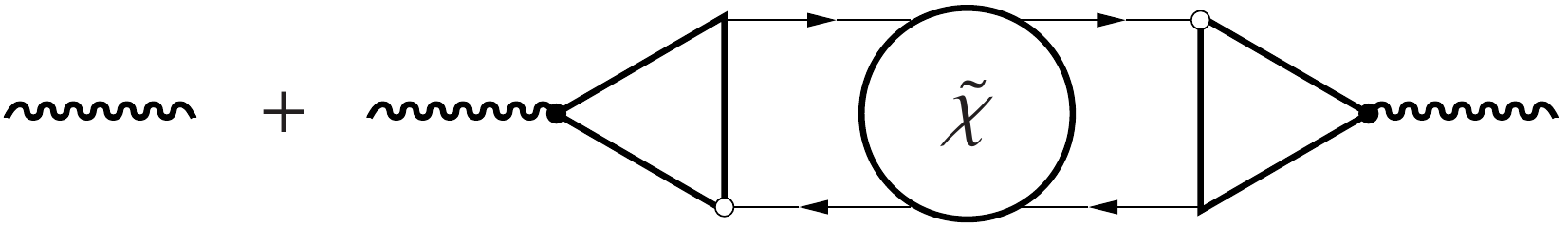} 
\caption{\label{fig:chialt} Diagrammatic representation of Eq.~\eqref{app:chialt}.
}
\end{figure}

We would like to establish the equivalence between the usual expression for the susceptibility
\begin{align}
\label{app:chi}
\chi_{\omega}(\qv) = 2T\sum_{\nu}\chi^{0}_{\nu\omega}(\qv) - 2T^{2}\sum_{\nu\nu'}\chi^{0}_{\nu\omega}(\qv) \Gamma_{\nu\nu'\omega}(\qv)\chi^{0}_{\nu\omega}(\qv)
\end{align}
and the alternative form given by the equations (see Fig.~\ref{fig:chialt})
\begin{align}
\label{app:chialt}
\chi_{\omega}(\qv)&=\chi_{\omega} + \chi_{-\omega}2T^{2}\sum_{\nu\nu'}\lambda_{\nu+\omega,-\omega} \tilde{\chi}_{\nu\nu'\omega}(\qv)\lambda_{\nu'\omega}\chi_{\omega} ,\\
\label{app:tchigen}
\tilde{\chi}_{\nu\nu'\omega}(\qv)&= \frac{1}{T}\tilde{\chi}^{0}_{\nu\omega}(\qv)\delta_{\nu\nu'} -\tilde{\chi}^{0}_{\nu\omega}(\qv)\Gamma_{\nu\nu'\omega}(\qv)\tilde{\chi}^{0}_{\nu'\omega}(\qv).
\end{align}
For simplicity, we consider the paramagnetic case and the charge susceptibility only. Correspondingly, spin labels are omitted and the vertex functions are taken in the charge channel (e.g., $\Gamma^{\text{ch}}=\Gamma^{\uparrow\uparrow}+\Gamma^{\uparrow\downarrow}$). The label ``ch'' is also suppressed in the following.
We further use the following definitions:
\begin{align}
\label{app:chi0lat}
\chi^{0}_{\nu\omega}(\qv) &= \frac{1}{N}\sum_{\kv}G_{\nu}(\kv)G_{\nu+\omega}(\kv+\qv),\\
\tilde{\chi}^{0}_{\nu\omega}(\qv) &= \frac{1}{N}\sum_{\kv}\tilde{G}_{\nu}(\kv)\tilde{G}_{\nu+\omega}(\kv+\qv),\\
\chi_{\nu\omega}^{0} &= g_{\nu}g_{\nu+\omega},\\
\label{app:gtilde}
\tilde{G}_{\nu}(\kv) &= G_{\nu}(\kv)-g_{\nu},
\end{align}
where $G$ is the lattice Green's function and $g$ is the impurity Green's function.
The impurity Green's function, charge susceptibility, as well as of the three-leg charge vertex and the four-leg vertex of the impurity are defined in terms of impurity correlation functions as follows:
\begin{align}
g_{\nu\sigma} &\Let -\av{c_{\nu\sigma} c^{*}_{\nu\sigma}},\\
\label{chidef}
\chi_{\omega} &\Let -\Big(\av{n_{\omega}n_{-\omega}}-\av{n}\av{n}\delta_{\omega}\Big),\\
\label{lambdadef}
\lambda_{\nu\omega}^{\sigma} &\Let \frac{g_{\nu\omega}^{\sigma(3)} - g_{\nu\sigma}\av{n}\delta_{\omega}/T}{g_{\nu\sigma}g_{\nu+\omega,\sigma}\chi_{\omega}},\\
\label{gammadef}
\gamma_{\nu\nu'\omega}^{\sigma\sigma'} & \Let \frac{g^{(4)\sigma\sigma'}_{\nu\nu'\omega}- (g_{\nu\sigma}g_{\nu'\sigma'}\delta_{\omega} - g_{\nu+\omega,\sigma}g_{\nu\sigma}\delta_{\nu\nu'}\delta_{\sigma\sigma'})/T
}{g_{\nu\sigma}g_{\nu+\omega,\sigma}g_{\nu'+\omega\sigma'}g_{\nu'\sigma'}}.
\end{align}
The three- and four-point functions $g^{(3)}$ and $g^{(4)}$ in turn are given by the averages
\begin{align}
g_{\nu\omega}^{(3)\sigma}&\Let -\av{c_{\nu\sigma}c^{*}_{\nu+\omega,\sigma}n_{\omega}},\\
g_{\nu\nu'\omega}^{(4)\sigma\sigma'} & \Let +\av{c_{\nu\sigma}c^{*}_{\nu+\omega,\sigma}c_{\nu'+\omega,\sigma'}c^{*}_{\nu'\sigma'}}.
\end{align}
In the paramagnetic state, $g$, $g^{(3)}$ and $\lambda$ are independent of spin. Here, $\gamma$ is the reducible impurity charge vertex.

\begin{figure}[t]
\includegraphics[scale=0.5,angle=0]{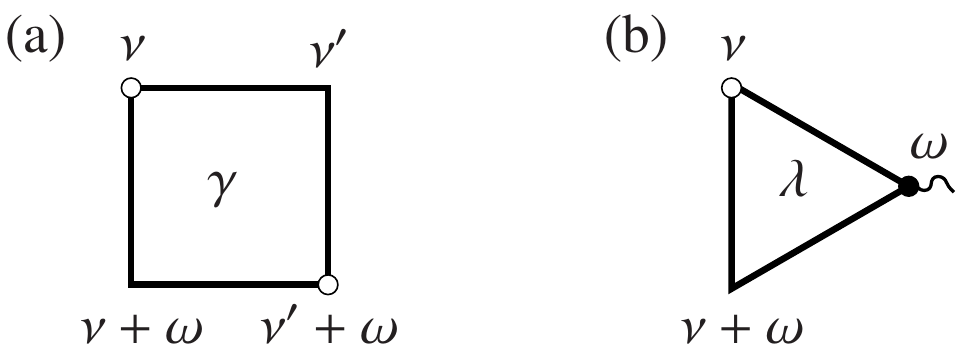} 
\caption{\label{fig:vertices} Definition of the impurity vertex functions.
}
\end{figure}

The vertices $\lambda$ and $\gamma$ (see Fig.~\ref{fig:vertices}) are closely related~\cite{Rubtsov12}. Using $n_{\omega}=T\sum_{\nu'\sigma'} c^{*}_{\nu'\sigma'}c_{\nu'+\omega,\sigma'}$, the fact that the Grassmann numbers anticommute, together with the above definitions of the correlation functions, one finds that
\begin{align}
g_{\nu\sigma}g_{\nu+\omega,\sigma}\lambda_{\nu\omega}^{\sigma}\chi_{\omega} =& g_{\nu\omega}^{(3)\,\sigma}-\frac{1}{T}g_{\nu\sigma}\av{n}\delta_{\omega}\notag\\
=& T\sum_{\nu'\sigma'} \av{c_{\nu\sigma}c^{*}_{\nu+\omega,\sigma}c_{\nu'+\omega,\sigma'}c^{*}_{\nu'\sigma'}}\notag\\
& + g_{\nu\sigma}\sum_{\nu'\sigma'}\av{c_{\nu'\sigma'}c^{*}_{\nu'+\omega,\sigma'}}\delta_{\omega}\notag\\
=&T\sum_{\nu'\sigma'} g_{\nu\nu'\omega}^{(4)\,\sigma\sigma'} -g_{\nu\sigma}\sum_{\nu'\sigma'}g_{\nu'\sigma'}\delta_{\omega}.
\end{align}
Similarly, one obtains
\begin{align}
&g_{\nu\sigma}g_{\nu+\omega,\sigma}T\sum_{\nu'\sigma'}\gamma_{\nu\nu'\omega}^{\sigma\sigma'}g_{\nu'+\omega\sigma'}g_{\nu'\sigma'}\notag\\
=& T\sum_{\nu'\sigma'}g_{\nu\nu'\omega}^{(4)\,\sigma\sigma'} - g_{\nu\sigma}\sum_{\nu'\sigma'}g_{\nu'\sigma'} +g_{\nu\sigma}g_{\nu+\omega\sigma}.
\end{align}
Taken together, one finds the relation
\begin{align}
&g_{\nu\sigma}g_{\nu+\omega,\sigma}\left(T\sum_{\nu'\sigma'}\gamma_{\nu\nu'\omega}^{\sigma\sigma'}g_{\nu'+\omega\sigma'}g_{\nu'\sigma'}-1\right)\notag\\
=& g_{\nu\sigma}g_{\nu+\omega,\sigma}\lambda_{\nu\omega}^{\sigma}\chi_{\omega}.
\end{align}
Further using that
\begin{align}
n_{-\omega} &=T\sum_{\nu'\sigma'}c^{*}_{\nu'\sigma'}c_{\nu'-\omega,\sigma'} = T\sum_{\nu'\sigma'}c^{*}_{\nu'+\omega,\sigma'}c_{\nu',\sigma'}
\end{align}
and
\begin{align}
\av{c_{\nu+\omega,\sigma}c^{*}_{\nu\sigma}n_{-\omega}}&=\av{n_{-\omega}c_{\nu+\omega,\sigma}c^{*}_{\nu\sigma}},
\end{align}
we have
\begin{align}
\chi_{-\omega}\lambda_{\nu+\omega,-\omega}^{\sigma}g_{\nu+\omega,\sigma}g_{\nu\sigma} =& g_{\nu+\omega,-\omega}^{(3)\,\sigma}-\frac{1}{T}g_{\nu+\omega\sigma}\av{n}\delta_{\omega}\notag\\
=& T\sum_{\nu'\sigma'} \av{c_{\nu'+\omega,\sigma'}c^{*}_{\nu'\sigma'}c_{\nu\sigma}c^{*}_{\nu+\omega,\sigma}}\notag\\
& + g_{\nu+\omega,\sigma}\sum_{\nu'\sigma'}\av{c_{\nu'\sigma'}c^{*}_{\nu'+\omega,\sigma'}}\delta_{\omega}\notag\\
=&T\sum_{\nu'\sigma'} g_{\nu'\nu\omega}^{(4)\,\sigma'\sigma} -g_{\nu+\omega\sigma}\sum_{\nu'\sigma'}g_{\nu'\sigma'}\delta_{\omega}.
\end{align}
This result can be expressed in terms of the four-leg vertex similarly to the above.
We therefore find the following relations between the three-leg and four-leg vertices:
\begin{align}
\lambda_{\nu\omega}^{\sigma}\chi_{\omega}&=T\sum_{\nu'\sigma'}\gamma_{\nu\nu'\omega}^{\sigma\sigma'}\chi^{0}_{\nu'\omega}-1,\\
\chi_{-\omega}\lambda_{\nu+\omega,-\omega}^{\sigma}&=T\sum_{\nu'\sigma'}\chi^{0}_{\nu'\omega}\gamma_{\nu'\nu\omega}^{\sigma'\sigma}-1.
\end{align}
From Eqs.~\eqref{app:chi0lat}-\eqref{app:gtilde} it is further easy to see that
\begin{align}
\label{app:chi0tilde}
\tilde{\chi}^{0}_{\nu\omega}(\qv) =  \chi^{0}_{\nu\omega}(\qv) - \chi_{\nu\omega}^{0}.
\end{align}
Now recall that the lattice vertex $\Gamma$ is calculated from the Bethe-Salpeter equation (BSE)
\begin{align}
[\Gamma_{\omega}^{-1}(\qv)]_{\nu\nu'} = [\gamma^{\text{irr}\,-1}_{\omega}]_{\nu\nu'} + T\chi^{0}_{\nu\omega}(\qv)\delta_{\nu\nu'},
\end{align}
where in turn the local irreducible vertex is calculated from the BSE of the impurity,
\begin{align}
[\gamma_{\omega}^{-1}]_{\nu\nu'} = [\gamma^{\text{irr}\,-1}_{\omega}]_{\nu\nu'} + T\chi^{0}_{\nu\omega}\delta_{\nu\nu'}.
\end{align}
Combining the BSEs using \eqref{app:chi0tilde}, one can write
\begin{align}
[\Gamma_{\omega}(\qv)^{-1}]_{\nu\nu'} = [\gamma_{\omega}^{-1}]_{\nu\nu'} + T\tilde{\chi}^{0}_{\nu\omega}(\qv)\delta_{\nu\nu'},
\end{align}
from which it follows that
\begin{align}
\label{app:Gamma-gamma}
\gamma_{\nu\nu'\omega} - \Gamma_{\nu\nu'\omega}(\qv) &=
T\sum_{\nu''}\gamma_{\nu\nu''\omega} \tilde{\chi}^{0}_{\nu''\omega}(\qv)\Gamma_{\nu''\nu'\omega}(\qv)\notag\\
&= T\sum_{\nu''} \Gamma_{\nu''\nu'\omega}(\qv)\tilde{\chi}^{0}_{\nu''\omega}(\qv)\gamma_{\nu\nu''\omega}.
\end{align}
The charge susceptibility can be expressed in terms of the local vertex and Green's functions as
\begin{align}
\chi_{\omega} &= 2T\sum_{\nu}\chi^{0}_{\nu\omega} - 2T^{2}\sum_{\nu\nu'}\chi^{0}_{\nu\omega}\gamma_{\nu\nu'\omega}\chi^{0}_{\nu'\omega}.
\end{align}
Using these relations and inserting \eqref{app:tchigen} into \eqref{app:chialt}, one obtains
\begin{widetext}
\begin{align}
\chi_{\omega}(\qv)=&2T\sum_{\nu}(\tilde{\chi}^{0}_{\nu\omega}(\qv) + \chi^{0}_{\nu\omega}) - 2T^{2}\sum_{\nu\nu'}\chi^{0}_{\nu\omega} \gamma_{\nu\nu'\omega}\chi^{0}_{\nu'\omega}
+2T^{3}\sum_{\nu\nu'\nu''}\chi^{0}_{\nu\omega}\gamma_{\nu\nu'}\tilde{\chi}^{0}_{\nu'\omega}(\qv)\gamma_{\nu'\nu''}\chi^{0}_{\nu''\omega} \notag\\
&- 2T^{2}\sum_{\nu\nu'}\chi^{0}_{\nu\omega}\gamma_{\nu\nu'}\tilde{\chi}^{0}_{\nu'\omega}(\qv)
 -2T^{2}\sum_{\nu\nu'}\tilde{\chi}^{0}_{\nu\omega}(\qv)\gamma_{\nu\nu'}\chi^{0}_{\nu'\omega} + 2T^{3} \sum_{\nu\nu'\nu''}\chi^{0}_{\nu\omega}\gamma_{\nu\nu'}\tilde{\chi}^{0}_{\nu'\omega}(\qv)\Gamma_{\nu'\nu''\omega}(\qv)\tilde{\chi}^{0}_{\nu''\omega}(\qv)\notag\\
& + 2T^{3}\sum_{\nu\nu'\nu''}\tilde{\chi}^{0}_{\nu\omega}(\qv)\Gamma_{\nu\nu'\omega}(\qv)\tilde{\chi}^{0}_{\nu'\omega}(\qv)\gamma_{\nu'\nu''}\chi^{0}_{\nu''\omega}
- 2T^{4}\sum_{\nu\nu'\nu''\nu'''}\chi^{0}_{\nu\omega}\gamma_{\nu\nu'}\tilde{\chi}^{0}_{\nu'\omega}(\qv)\Gamma_{\nu'\nu''\omega}(\qv)\tilde{\chi}^{0}_{\nu''\omega}(\qv)\gamma_{\nu''\nu'''}\chi^{0}_{\nu'''\omega}\notag\\
&
- 2T^{2}\sum_{\nu\nu'}\tilde{\chi}^{0}_{\nu\omega}(\qv)\Gamma_{\nu\nu'\omega}(\qv)\tilde{\chi}^{0}_{\nu'\omega}(\qv).
\end{align}
Substituting \eqref{app:Gamma-gamma}, all terms involving the impurity vertex cancel and one is left with
\begin{align}
\chi_{\omega}(\qv)=&2T\sum_{\nu}(\tilde{\chi}^{0}_{\nu\omega}(\qv) + \chi^{0}_{\nu\omega}) - 2T^{2}\sum_{\nu\nu'}\chi^{0}_{\nu\omega}\Gamma_{\nu\nu'\omega}(\qv)\tilde{\chi}^{0}_{\nu'\omega}(\qv)
- 2T^{2}\sum_{\nu\nu'}\tilde{\chi}^{0}_{\nu\omega}(\qv)\Gamma_{\nu\nu'\omega}(\qv)\chi^{0}_{\nu'\omega}\notag\\
& - 2T^{2}\sum_{\nu\nu'}\chi^{0}_{\nu\omega}\Gamma_{\nu\nu'\omega}(\qv)\chi^{0}_{\nu'\omega} - 2T^{2}\sum_{\nu\nu'}\tilde{\chi}^{0}_{\nu\omega}(\qv)\Gamma_{\nu\nu'\omega}(\qv)\tilde{\chi}^{0}_{\nu'\omega}(\qv)\\
=&  2T\sum_{\nu}(\tilde{\chi}^{0}_{\nu\omega}(\qv) + \chi^{0}_{\nu\omega}) - 2T^{2}\sum_{\nu\nu'}(\tilde{\chi}^{0}_{\nu\omega}(\qv) + \chi^{0}_{\nu\omega})\Gamma_{\nu\nu'\omega}(\qv)(\tilde{\chi}^{0}_{\nu\omega}(\qv) + \chi^{0}_{\nu'\omega}).
\end{align}
Using \eqref{app:chi0tilde} this is seen to be equal to \eqref{app:chi}.
\end{widetext}

\bibliography{main}

\begin{thebibliography}{55}%
\makeatletter
\providecommand \@ifxundefined [1]{%
 \@ifx{#1\undefined}
}%
\providecommand \@ifnum [1]{%
 \ifnum #1\expandafter \@firstoftwo
 \else \expandafter \@secondoftwo
 \fi
}%
\providecommand \@ifx [1]{%
 \ifx #1\expandafter \@firstoftwo
 \else \expandafter \@secondoftwo
 \fi
}%
\providecommand \natexlab [1]{#1}%
\providecommand \enquote  [1]{``#1''}%
\providecommand \bibnamefont  [1]{#1}%
\providecommand \bibfnamefont [1]{#1}%
\providecommand \citenamefont [1]{#1}%
\providecommand \href@noop [0]{\@secondoftwo}%
\providecommand \href [0]{\begingroup \@sanitize@url \@href}%
\providecommand \@href[1]{\@@startlink{#1}\@@href}%
\providecommand \@@href[1]{\endgroup#1\@@endlink}%
\providecommand \@sanitize@url [0]{\catcode `\\12\catcode `\$12\catcode
  `\&12\catcode `\#12\catcode `\^12\catcode `\_12\catcode `\%12\relax}%
\providecommand \@@startlink[1]{}%
\providecommand \@@endlink[0]{}%
\providecommand \url  [0]{\begingroup\@sanitize@url \@url }%
\providecommand \@url [1]{\endgroup\@href {#1}{\urlprefix }}%
\providecommand \urlprefix  [0]{URL }%
\providecommand \Eprint [0]{\href }%
\providecommand \doibase [0]{http://dx.doi.org/}%
\providecommand \selectlanguage [0]{\@gobble}%
\providecommand \bibinfo  [0]{\@secondoftwo}%
\providecommand \bibfield  [0]{\@secondoftwo}%
\providecommand \translation [1]{[#1]}%
\providecommand \BibitemOpen [0]{}%
\providecommand \bibitemStop [0]{}%
\providecommand \bibitemNoStop [0]{.\EOS\space}%
\providecommand \EOS [0]{\spacefactor3000\relax}%
\providecommand \BibitemShut  [1]{\csname bibitem#1\endcsname}%
\let\auto@bib@innerbib\@empty
\bibitem [{\citenamefont {Georges}\ \emph {et~al.}(1996)\citenamefont
  {Georges}, \citenamefont {Kotliar}, \citenamefont {Krauth},\ and\
  \citenamefont {Rozenberg}}]{Georges96}%
  \BibitemOpen
  \bibfield  {author} {\bibinfo {author} {\bibfnamefont {A.}~\bibnamefont
  {Georges}}, \bibinfo {author} {\bibfnamefont {G.}~\bibnamefont {Kotliar}},
  \bibinfo {author} {\bibfnamefont {W.}~\bibnamefont {Krauth}}, \ and\ \bibinfo
  {author} {\bibfnamefont {M.~J.}\ \bibnamefont {Rozenberg}},\ }\href {\doibase
  10.1103/RevModPhys.68.13} {\bibfield  {journal} {\bibinfo  {journal} {Rev.
  Mod. Phys.}\ }\textbf {\bibinfo {volume} {68}},\ \bibinfo {pages} {13}
  (\bibinfo {year} {1996})}\BibitemShut {NoStop}%
\bibitem [{\citenamefont {Maier}\ \emph
  {et~al.}(2005{\natexlab{a}})\citenamefont {Maier}, \citenamefont {Jarrell},
  \citenamefont {Pruschke},\ and\ \citenamefont {Hettler}}]{Maier05}%
  \BibitemOpen
  \bibfield  {author} {\bibinfo {author} {\bibfnamefont {T.}~\bibnamefont
  {Maier}}, \bibinfo {author} {\bibfnamefont {M.}~\bibnamefont {Jarrell}},
  \bibinfo {author} {\bibfnamefont {T.}~\bibnamefont {Pruschke}}, \ and\
  \bibinfo {author} {\bibfnamefont {M.~H.}\ \bibnamefont {Hettler}},\ }\href
  {\doibase 10.1103/RevModPhys.77.1027} {\bibfield  {journal} {\bibinfo
  {journal} {Rev. Mod. Phys.}\ }\textbf {\bibinfo {volume} {77}},\ \bibinfo
  {pages} {1027} (\bibinfo {year} {2005}{\natexlab{a}})}\BibitemShut {NoStop}%
\bibitem [{\citenamefont {Toschi}\ \emph {et~al.}(2007)\citenamefont {Toschi},
  \citenamefont {Katanin},\ and\ \citenamefont {Held}}]{Toschi07}%
  \BibitemOpen
  \bibfield  {author} {\bibinfo {author} {\bibfnamefont {A.}~\bibnamefont
  {Toschi}}, \bibinfo {author} {\bibfnamefont {A.~A.}\ \bibnamefont {Katanin}},
  \ and\ \bibinfo {author} {\bibfnamefont {K.}~\bibnamefont {Held}},\ }\href
  {\doibase 10.1103/PhysRevB.75.045118} {\bibfield  {journal} {\bibinfo
  {journal} {Physical Review B (Condensed Matter and Materials Physics)}\
  }\textbf {\bibinfo {volume} {75}},\ \bibinfo {eid} {045118} (\bibinfo {year}
  {2007})}\BibitemShut {NoStop}%
\bibitem [{\citenamefont {Rubtsov}\ \emph {et~al.}(2008)\citenamefont
  {Rubtsov}, \citenamefont {Katsnelson},\ and\ \citenamefont
  {Lichtenstein}}]{Rubtsov08}%
  \BibitemOpen
  \bibfield  {author} {\bibinfo {author} {\bibfnamefont {A.~N.}\ \bibnamefont
  {Rubtsov}}, \bibinfo {author} {\bibfnamefont {M.~I.}\ \bibnamefont
  {Katsnelson}}, \ and\ \bibinfo {author} {\bibfnamefont {A.~I.}\ \bibnamefont
  {Lichtenstein}},\ }\href {\doibase 10.1103/PhysRevB.77.033101} {\bibfield
  {journal} {\bibinfo  {journal} {Phys. Rev. B}\ }\textbf {\bibinfo {volume}
  {77}},\ \bibinfo {pages} {033101} (\bibinfo {year} {2008})}\BibitemShut
  {NoStop}%
\bibitem [{\citenamefont {Rubtsov}\ \emph {et~al.}(2009)\citenamefont
  {Rubtsov}, \citenamefont {Katsnelson}, \citenamefont {Lichtenstein},\ and\
  \citenamefont {Georges}}]{Rubtsov09}%
  \BibitemOpen
  \bibfield  {author} {\bibinfo {author} {\bibfnamefont {A.~N.}\ \bibnamefont
  {Rubtsov}}, \bibinfo {author} {\bibfnamefont {M.~I.}\ \bibnamefont
  {Katsnelson}}, \bibinfo {author} {\bibfnamefont {A.~I.}\ \bibnamefont
  {Lichtenstein}}, \ and\ \bibinfo {author} {\bibfnamefont {A.}~\bibnamefont
  {Georges}},\ }\href {\doibase 10.1103/PhysRevB.79.045133} {\bibfield
  {journal} {\bibinfo  {journal} {Phys. Rev. B}\ }\textbf {\bibinfo {volume}
  {79}},\ \bibinfo {pages} {045133} (\bibinfo {year} {2009})}\BibitemShut
  {NoStop}%
\bibitem [{\citenamefont {Rubtsov}\ \emph {et~al.}(2012)\citenamefont
  {Rubtsov}, \citenamefont {Katsnelson},\ and\ \citenamefont
  {Lichtenstein}}]{Rubtsov12}%
  \BibitemOpen
  \bibfield  {author} {\bibinfo {author} {\bibfnamefont {A.}~\bibnamefont
  {Rubtsov}}, \bibinfo {author} {\bibfnamefont {M.}~\bibnamefont {Katsnelson}},
  \ and\ \bibinfo {author} {\bibfnamefont {A.}~\bibnamefont {Lichtenstein}},\
  }\href {\doibase 10.1016/j.aop.2012.01.002} {\bibfield  {journal} {\bibinfo
  {journal} {Annals of Physics}\ }\textbf {\bibinfo {volume} {327}},\ \bibinfo
  {pages} {1320} (\bibinfo {year} {2012})}\BibitemShut {NoStop}%
\bibitem [{\citenamefont {Rohringer}\ \emph {et~al.}(2013)\citenamefont
  {Rohringer}, \citenamefont {Toschi}, \citenamefont {Hafermann}, \citenamefont
  {Held}, \citenamefont {Anisimov},\ and\ \citenamefont
  {Katanin}}]{Rohringer13}%
  \BibitemOpen
  \bibfield  {author} {\bibinfo {author} {\bibfnamefont {G.}~\bibnamefont
  {Rohringer}}, \bibinfo {author} {\bibfnamefont {A.}~\bibnamefont {Toschi}},
  \bibinfo {author} {\bibfnamefont {H.}~\bibnamefont {Hafermann}}, \bibinfo
  {author} {\bibfnamefont {K.}~\bibnamefont {Held}}, \bibinfo {author}
  {\bibfnamefont {V.~I.}\ \bibnamefont {Anisimov}}, \ and\ \bibinfo {author}
  {\bibfnamefont {A.~A.}\ \bibnamefont {Katanin}},\ }\href {\doibase
  10.1103/PhysRevB.88.115112} {\bibfield  {journal} {\bibinfo  {journal} {Phys.
  Rev. B}\ }\textbf {\bibinfo {volume} {88}},\ \bibinfo {pages} {115112}
  (\bibinfo {year} {2013})}\BibitemShut {NoStop}%
\bibitem [{\citenamefont {Gull}\ \emph {et~al.}(2013)\citenamefont {Gull},
  \citenamefont {Parcollet},\ and\ \citenamefont {Millis}}]{Gull13}%
  \BibitemOpen
  \bibfield  {author} {\bibinfo {author} {\bibfnamefont {E.}~\bibnamefont
  {Gull}}, \bibinfo {author} {\bibfnamefont {O.}~\bibnamefont {Parcollet}}, \
  and\ \bibinfo {author} {\bibfnamefont {A.~J.}\ \bibnamefont {Millis}},\
  }\href {\doibase 10.1103/PhysRevLett.110.216405} {\bibfield  {journal}
  {\bibinfo  {journal} {Phys. Rev. Lett.}\ }\textbf {\bibinfo {volume} {110}},\
  \bibinfo {pages} {216405} (\bibinfo {year} {2013})}\BibitemShut {NoStop}%
\bibitem [{\citenamefont {Kotliar}\ \emph {et~al.}(2006)\citenamefont
  {Kotliar}, \citenamefont {Savrasov}, \citenamefont {Haule}, \citenamefont
  {Oudovenko}, \citenamefont {Parcollet},\ and\ \citenamefont
  {Marianetti}}]{Kotliar06}%
  \BibitemOpen
  \bibfield  {author} {\bibinfo {author} {\bibfnamefont {G.}~\bibnamefont
  {Kotliar}}, \bibinfo {author} {\bibfnamefont {S.~Y.}\ \bibnamefont
  {Savrasov}}, \bibinfo {author} {\bibfnamefont {K.}~\bibnamefont {Haule}},
  \bibinfo {author} {\bibfnamefont {V.~S.}\ \bibnamefont {Oudovenko}}, \bibinfo
  {author} {\bibfnamefont {O.}~\bibnamefont {Parcollet}}, \ and\ \bibinfo
  {author} {\bibfnamefont {C.~A.}\ \bibnamefont {Marianetti}},\ }\href
  {\doibase 10.1103/RevModPhys.78.865} {\bibfield  {journal} {\bibinfo
  {journal} {Rev. Mod. Phys.}\ }\textbf {\bibinfo {volume} {78}},\ \bibinfo
  {pages} {865} (\bibinfo {year} {2006})}\BibitemShut {NoStop}%
\bibitem [{\citenamefont {Si}\ and\ \citenamefont {Smith}(1996)}]{Si96}%
  \BibitemOpen
  \bibfield  {author} {\bibinfo {author} {\bibfnamefont {Q.}~\bibnamefont
  {Si}}\ and\ \bibinfo {author} {\bibfnamefont {J.~L.}\ \bibnamefont {Smith}},\
  }\href {\doibase 10.1103/PhysRevLett.77.3391} {\bibfield  {journal} {\bibinfo
   {journal} {Phys. Rev. Lett.}\ }\textbf {\bibinfo {volume} {77}},\ \bibinfo
  {pages} {3391} (\bibinfo {year} {1996})}\BibitemShut {NoStop}%
\bibitem [{\citenamefont {Kajueter}(1996)}]{Kajueter96}%
  \BibitemOpen
  \bibfield  {author} {\bibinfo {author} {\bibfnamefont {H.}~\bibnamefont
  {Kajueter}},\ }\href@noop {} {Ph.D. thesis},\ \bibinfo  {school} {Rutgers
  University} (\bibinfo {year} {1996})\BibitemShut {NoStop}%
\bibitem [{\citenamefont {Smith}\ and\ \citenamefont {Si}(2000)}]{Smith00}%
  \BibitemOpen
  \bibfield  {author} {\bibinfo {author} {\bibfnamefont {J.~L.}\ \bibnamefont
  {Smith}}\ and\ \bibinfo {author} {\bibfnamefont {Q.}~\bibnamefont {Si}},\
  }\href {\doibase 10.1103/PhysRevB.61.5184} {\bibfield  {journal} {\bibinfo
  {journal} {Phys. Rev. B}\ }\textbf {\bibinfo {volume} {61}},\ \bibinfo
  {pages} {5184} (\bibinfo {year} {2000})}\BibitemShut {NoStop}%
\bibitem [{\citenamefont {Chitra}\ and\ \citenamefont
  {Kotliar}(2001)}]{Chitra01}%
  \BibitemOpen
  \bibfield  {author} {\bibinfo {author} {\bibfnamefont {R.}~\bibnamefont
  {Chitra}}\ and\ \bibinfo {author} {\bibfnamefont {G.}~\bibnamefont
  {Kotliar}},\ }\href {\doibase 10.1103/PhysRevB.63.115110} {\bibfield
  {journal} {\bibinfo  {journal} {Phys. Rev. B}\ }\textbf {\bibinfo {volume}
  {63}},\ \bibinfo {pages} {115110} (\bibinfo {year} {2001})}\BibitemShut
  {NoStop}%
\bibitem [{\citenamefont {Sun}\ and\ \citenamefont {Kotliar}(2002)}]{Sun02}%
  \BibitemOpen
  \bibfield  {author} {\bibinfo {author} {\bibfnamefont {P.}~\bibnamefont
  {Sun}}\ and\ \bibinfo {author} {\bibfnamefont {G.}~\bibnamefont {Kotliar}},\
  }\href {\doibase 10.1103/PhysRevB.66.085120} {\bibfield  {journal} {\bibinfo
  {journal} {Phys. Rev. B}\ }\textbf {\bibinfo {volume} {66}},\ \bibinfo
  {pages} {085120} (\bibinfo {year} {2002})}\BibitemShut {NoStop}%
\bibitem [{\citenamefont {Biermann}\ \emph {et~al.}(2003)\citenamefont
  {Biermann}, \citenamefont {Aryasetiawan},\ and\ \citenamefont
  {Georges}}]{Biermann03}%
  \BibitemOpen
  \bibfield  {author} {\bibinfo {author} {\bibfnamefont {S.}~\bibnamefont
  {Biermann}}, \bibinfo {author} {\bibfnamefont {F.}~\bibnamefont
  {Aryasetiawan}}, \ and\ \bibinfo {author} {\bibfnamefont {A.}~\bibnamefont
  {Georges}},\ }\href {\doibase 10.1103/PhysRevLett.90.086402} {\bibfield
  {journal} {\bibinfo  {journal} {Phys. Rev. Lett.}\ }\textbf {\bibinfo
  {volume} {90}},\ \bibinfo {pages} {086402} (\bibinfo {year}
  {2003})}\BibitemShut {NoStop}%
\bibitem [{\citenamefont {Ayral}\ \emph {et~al.}(2012)\citenamefont {Ayral},
  \citenamefont {Werner},\ and\ \citenamefont {Biermann}}]{Ayral12}%
  \BibitemOpen
  \bibfield  {author} {\bibinfo {author} {\bibfnamefont {T.}~\bibnamefont
  {Ayral}}, \bibinfo {author} {\bibfnamefont {P.}~\bibnamefont {Werner}}, \
  and\ \bibinfo {author} {\bibfnamefont {S.}~\bibnamefont {Biermann}},\ }\href
  {\doibase 10.1103/PhysRevLett.109.226401} {\bibfield  {journal} {\bibinfo
  {journal} {Phys. Rev. Lett.}\ }\textbf {\bibinfo {volume} {109}},\ \bibinfo
  {pages} {226401} (\bibinfo {year} {2012})}\BibitemShut {NoStop}%
\bibitem [{\citenamefont {Ayral}\ \emph {et~al.}(2013)\citenamefont {Ayral},
  \citenamefont {Biermann},\ and\ \citenamefont {Werner}}]{Ayral13}%
  \BibitemOpen
  \bibfield  {author} {\bibinfo {author} {\bibfnamefont {T.}~\bibnamefont
  {Ayral}}, \bibinfo {author} {\bibfnamefont {S.}~\bibnamefont {Biermann}}, \
  and\ \bibinfo {author} {\bibfnamefont {P.}~\bibnamefont {Werner}},\ }\href
  {\doibase 10.1103/PhysRevB.87.125149} {\bibfield  {journal} {\bibinfo
  {journal} {Phys. Rev. B}\ }\textbf {\bibinfo {volume} {87}},\ \bibinfo
  {pages} {125149} (\bibinfo {year} {2013})}\BibitemShut {NoStop}%
\bibitem [{\citenamefont {Jarrell}\ and\ \citenamefont
  {Pruschke}(1993)}]{Jarrell93}%
  \BibitemOpen
  \bibfield  {author} {\bibinfo {author} {\bibfnamefont {M.}~\bibnamefont
  {Jarrell}}\ and\ \bibinfo {author} {\bibfnamefont {T.}~\bibnamefont
  {Pruschke}},\ }\href {\doibase 10.1007/BF02198153} {\bibfield  {journal}
  {\bibinfo  {journal} {Zeitschrift für Physik B Condensed Matter}\ }\textbf
  {\bibinfo {volume} {90}},\ \bibinfo {pages} {187} (\bibinfo {year}
  {1993})}\BibitemShut {NoStop}%
\bibitem [{\citenamefont {Kent}\ \emph {et~al.}(2005)\citenamefont {Kent},
  \citenamefont {Jarrell}, \citenamefont {Maier},\ and\ \citenamefont
  {Pruschke}}]{Kent05}%
  \BibitemOpen
  \bibfield  {author} {\bibinfo {author} {\bibfnamefont {P.~R.~C.}\
  \bibnamefont {Kent}}, \bibinfo {author} {\bibfnamefont {M.}~\bibnamefont
  {Jarrell}}, \bibinfo {author} {\bibfnamefont {T.~A.}\ \bibnamefont {Maier}},
  \ and\ \bibinfo {author} {\bibfnamefont {T.}~\bibnamefont {Pruschke}},\
  }\href {\doibase 10.1103/PhysRevB.72.060411} {\bibfield  {journal} {\bibinfo
  {journal} {Phys. Rev. B}\ }\textbf {\bibinfo {volume} {72}},\ \bibinfo
  {pages} {060411} (\bibinfo {year} {2005})}\BibitemShut {NoStop}%
\bibitem [{\citenamefont {Maier}\ \emph
  {et~al.}(2005{\natexlab{b}})\citenamefont {Maier}, \citenamefont {Jarrell},
  \citenamefont {Schulthess}, \citenamefont {Kent},\ and\ \citenamefont
  {White}}]{Maier05-2}%
  \BibitemOpen
  \bibfield  {author} {\bibinfo {author} {\bibfnamefont {T.~A.}\ \bibnamefont
  {Maier}}, \bibinfo {author} {\bibfnamefont {M.}~\bibnamefont {Jarrell}},
  \bibinfo {author} {\bibfnamefont {T.~C.}\ \bibnamefont {Schulthess}},
  \bibinfo {author} {\bibfnamefont {P.~R.~C.}\ \bibnamefont {Kent}}, \ and\
  \bibinfo {author} {\bibfnamefont {J.~B.}\ \bibnamefont {White}},\ }\href
  {\doibase 10.1103/PhysRevLett.95.237001} {\bibfield  {journal} {\bibinfo
  {journal} {Phys. Rev. Lett.}\ }\textbf {\bibinfo {volume} {95}},\ \bibinfo
  {pages} {237001} (\bibinfo {year} {2005}{\natexlab{b}})}\BibitemShut
  {NoStop}%
\bibitem [{\citenamefont {Rohringer}\ \emph {et~al.}(2012)\citenamefont
  {Rohringer}, \citenamefont {Valli},\ and\ \citenamefont
  {Toschi}}]{Rohringer12}%
  \BibitemOpen
  \bibfield  {author} {\bibinfo {author} {\bibfnamefont {G.}~\bibnamefont
  {Rohringer}}, \bibinfo {author} {\bibfnamefont {A.}~\bibnamefont {Valli}}, \
  and\ \bibinfo {author} {\bibfnamefont {A.}~\bibnamefont {Toschi}},\ }\href
  {\doibase 10.1103/PhysRevB.86.125114} {\bibfield  {journal} {\bibinfo
  {journal} {Phys. Rev. B}\ }\textbf {\bibinfo {volume} {86}},\ \bibinfo
  {pages} {125114} (\bibinfo {year} {2012})}\BibitemShut {NoStop}%
\bibitem [{\citenamefont {Kinza}\ and\ \citenamefont
  {Honerkamp}(2013)}]{Kinza13}%
  \BibitemOpen
  \bibfield  {author} {\bibinfo {author} {\bibfnamefont {M.}~\bibnamefont
  {Kinza}}\ and\ \bibinfo {author} {\bibfnamefont {C.}~\bibnamefont
  {Honerkamp}},\ }\href {\doibase 10.1103/PhysRevB.88.195136} {\bibfield
  {journal} {\bibinfo  {journal} {Phys. Rev. B}\ }\textbf {\bibinfo {volume}
  {88}},\ \bibinfo {pages} {195136} (\bibinfo {year} {2013})}\BibitemShut
  {NoStop}%
\bibitem [{\citenamefont {Huang}\ and\ \citenamefont {Wang}(2013)}]{Huang13}%
  \BibitemOpen
  \bibfield  {author} {\bibinfo {author} {\bibfnamefont {L.}~\bibnamefont
  {Huang}}\ and\ \bibinfo {author} {\bibfnamefont {Y.}~\bibnamefont {Wang}},\
  }\href {http://arxiv.org/abs/1303.2818} {\  (\bibinfo {year} {2013})},\
  \Eprint {http://arxiv.org/abs/1303.2818} {arXiv:1303.2818 [cond-mat]}
  \BibitemShut {NoStop}%
\bibitem [{\citenamefont {Rubtsov}\ \emph {et~al.}(2005)\citenamefont
  {Rubtsov}, \citenamefont {Savkin},\ and\ \citenamefont
  {Lichtenstein}}]{Rubtsov05}%
  \BibitemOpen
  \bibfield  {author} {\bibinfo {author} {\bibfnamefont {A.~N.}\ \bibnamefont
  {Rubtsov}}, \bibinfo {author} {\bibfnamefont {V.~V.}\ \bibnamefont {Savkin}},
  \ and\ \bibinfo {author} {\bibfnamefont {A.~I.}\ \bibnamefont
  {Lichtenstein}},\ }\href {\doibase 10.1103/PhysRevB.72.035122} {\bibfield
  {journal} {\bibinfo  {journal} {Phys. Rev. B}\ }\textbf {\bibinfo {volume}
  {72}},\ \bibinfo {pages} {035122} (\bibinfo {year} {2005})}\BibitemShut
  {NoStop}%
\bibitem [{\citenamefont {Werner}\ \emph {et~al.}(2006)\citenamefont {Werner},
  \citenamefont {Comanac}, \citenamefont {de' Medici}, \citenamefont {Troyer},\
  and\ \citenamefont {Millis}}]{Werner06}%
  \BibitemOpen
  \bibfield  {author} {\bibinfo {author} {\bibfnamefont {P.}~\bibnamefont
  {Werner}}, \bibinfo {author} {\bibfnamefont {A.}~\bibnamefont {Comanac}},
  \bibinfo {author} {\bibfnamefont {L.}~\bibnamefont {de' Medici}}, \bibinfo
  {author} {\bibfnamefont {M.}~\bibnamefont {Troyer}}, \ and\ \bibinfo {author}
  {\bibfnamefont {A.~J.}\ \bibnamefont {Millis}},\ }\href {\doibase
  10.1103/PhysRevLett.97.076405} {\bibfield  {journal} {\bibinfo  {journal}
  {Phys. Rev. Lett.}\ }\textbf {\bibinfo {volume} {97}},\ \bibinfo {pages}
  {076405} (\bibinfo {year} {2006})}\BibitemShut {NoStop}%
\bibitem [{\citenamefont {Gull}\ \emph {et~al.}(2008)\citenamefont {Gull},
  \citenamefont {Werner}, \citenamefont {Parcollet},\ and\ \citenamefont
  {Troyer}}]{Gull08}%
  \BibitemOpen
  \bibfield  {author} {\bibinfo {author} {\bibfnamefont {E.}~\bibnamefont
  {Gull}}, \bibinfo {author} {\bibfnamefont {P.}~\bibnamefont {Werner}},
  \bibinfo {author} {\bibfnamefont {O.}~\bibnamefont {Parcollet}}, \ and\
  \bibinfo {author} {\bibfnamefont {M.}~\bibnamefont {Troyer}},\ }\href
  {http://stacks.iop.org/0295-5075/82/i=5/a=57003} {\bibfield  {journal}
  {\bibinfo  {journal} {EPL (Europhysics Letters)}\ }\textbf {\bibinfo {volume}
  {82}},\ \bibinfo {pages} {57003} (\bibinfo {year} {2008})}\BibitemShut
  {NoStop}%
\bibitem [{\citenamefont {Gull}\ \emph {et~al.}(2011)\citenamefont {Gull},
  \citenamefont {Millis}, \citenamefont {Lichtenstein}, \citenamefont
  {Rubtsov}, \citenamefont {Troyer},\ and\ \citenamefont {Werner}}]{Gull11}%
  \BibitemOpen
  \bibfield  {author} {\bibinfo {author} {\bibfnamefont {E.}~\bibnamefont
  {Gull}}, \bibinfo {author} {\bibfnamefont {A.~J.}\ \bibnamefont {Millis}},
  \bibinfo {author} {\bibfnamefont {A.~I.}\ \bibnamefont {Lichtenstein}},
  \bibinfo {author} {\bibfnamefont {A.~N.}\ \bibnamefont {Rubtsov}}, \bibinfo
  {author} {\bibfnamefont {M.}~\bibnamefont {Troyer}}, \ and\ \bibinfo {author}
  {\bibfnamefont {P.}~\bibnamefont {Werner}},\ }\href {\doibase
  10.1103/RevModPhys.83.349} {\bibfield  {journal} {\bibinfo  {journal} {Rev.
  Mod. Phys.}\ }\textbf {\bibinfo {volume} {83}},\ \bibinfo {pages} {349}
  (\bibinfo {year} {2011})}\BibitemShut {NoStop}%
\bibitem [{\citenamefont {Nambu}(1960)}]{Nambu60}%
  \BibitemOpen
  \bibfield  {author} {\bibinfo {author} {\bibfnamefont {Y.}~\bibnamefont
  {Nambu}},\ }\href {\doibase 10.1103/PhysRev.117.648} {\bibfield  {journal}
  {\bibinfo  {journal} {Phys. Rev.}\ }\textbf {\bibinfo {volume} {117}},\
  \bibinfo {pages} {648} (\bibinfo {year} {1960})}\BibitemShut {NoStop}%
\bibitem [{\citenamefont {Baym}\ and\ \citenamefont {Kadanoff}(1961)}]{Baym61}%
  \BibitemOpen
  \bibfield  {author} {\bibinfo {author} {\bibfnamefont {G.}~\bibnamefont
  {Baym}}\ and\ \bibinfo {author} {\bibfnamefont {L.~P.}\ \bibnamefont
  {Kadanoff}},\ }\href {\doibase 10.1103/PhysRev.124.287} {\bibfield  {journal}
  {\bibinfo  {journal} {Phys. Rev.}\ }\textbf {\bibinfo {volume} {124}},\
  \bibinfo {pages} {287} (\bibinfo {year} {1961})}\BibitemShut {NoStop}%
\bibitem [{\citenamefont {Baym}(1962)}]{Baym62}%
  \BibitemOpen
  \bibfield  {author} {\bibinfo {author} {\bibfnamefont {G.}~\bibnamefont
  {Baym}},\ }\href {\doibase 10.1103/PhysRev.127.1391} {\bibfield  {journal}
  {\bibinfo  {journal} {Phys. Rev.}\ }\textbf {\bibinfo {volume} {127}},\
  \bibinfo {pages} {1391} (\bibinfo {year} {1962})}\BibitemShut {NoStop}%
\bibitem [{\citenamefont {Brener}\ \emph {et~al.}(2008)\citenamefont {Brener},
  \citenamefont {Hafermann}, \citenamefont {Rubtsov}, \citenamefont
  {Katsnelson},\ and\ \citenamefont {Lichtenstein}}]{Brener08}%
  \BibitemOpen
  \bibfield  {author} {\bibinfo {author} {\bibfnamefont {S.}~\bibnamefont
  {Brener}}, \bibinfo {author} {\bibfnamefont {H.}~\bibnamefont {Hafermann}},
  \bibinfo {author} {\bibfnamefont {A.~N.}\ \bibnamefont {Rubtsov}}, \bibinfo
  {author} {\bibfnamefont {M.~I.}\ \bibnamefont {Katsnelson}}, \ and\ \bibinfo
  {author} {\bibfnamefont {A.~I.}\ \bibnamefont {Lichtenstein}},\ }\href
  {\doibase 10.1103/PhysRevB.77.195105} {\bibfield  {journal} {\bibinfo
  {journal} {Phys. Rev. B}\ }\textbf {\bibinfo {volume} {77}},\ \bibinfo {eid}
  {195105} (\bibinfo {year} {2008})}\BibitemShut {NoStop}%
\bibitem [{\citenamefont {Jani\ifmmode~\check{s}\else \v{s}\fi{}}\ and\
  \citenamefont {Pokorn\'y}(2014)}]{Janis14}%
  \BibitemOpen
  \bibfield  {author} {\bibinfo {author} {\bibfnamefont {V.}~\bibnamefont
  {Jani\ifmmode~\check{s}\else \v{s}\fi{}}}\ and\ \bibinfo {author}
  {\bibfnamefont {V.}~\bibnamefont {Pokorn\'y}},\ }\href {\doibase
  10.1103/PhysRevB.90.045143} {\bibfield  {journal} {\bibinfo  {journal} {Phys.
  Rev. B}\ }\textbf {\bibinfo {volume} {90}},\ \bibinfo {pages} {045143}
  (\bibinfo {year} {2014})}\BibitemShut {NoStop}%
\bibitem [{\citenamefont {Sch\"afer}\ \emph {et~al.}(2013)\citenamefont
  {Sch\"afer}, \citenamefont {Rohringer}, \citenamefont {Gunnarsson},
  \citenamefont {Ciuchi}, \citenamefont {Sangiovanni},\ and\ \citenamefont
  {Toschi}}]{Schaefer13}%
  \BibitemOpen
  \bibfield  {author} {\bibinfo {author} {\bibfnamefont {T.}~\bibnamefont
  {Sch\"afer}}, \bibinfo {author} {\bibfnamefont {G.}~\bibnamefont
  {Rohringer}}, \bibinfo {author} {\bibfnamefont {O.}~\bibnamefont
  {Gunnarsson}}, \bibinfo {author} {\bibfnamefont {S.}~\bibnamefont {Ciuchi}},
  \bibinfo {author} {\bibfnamefont {G.}~\bibnamefont {Sangiovanni}}, \ and\
  \bibinfo {author} {\bibfnamefont {A.}~\bibnamefont {Toschi}},\ }\href
  {\doibase 10.1103/PhysRevLett.110.246405} {\bibfield  {journal} {\bibinfo
  {journal} {Phys. Rev. Lett.}\ }\textbf {\bibinfo {volume} {110}},\ \bibinfo
  {pages} {246405} (\bibinfo {year} {2013})}\BibitemShut {NoStop}%
\bibitem [{\citenamefont {Khurana}(1990)}]{Khurana90}%
  \BibitemOpen
  \bibfield  {author} {\bibinfo {author} {\bibfnamefont {A.}~\bibnamefont
  {Khurana}},\ }\href {\doibase 10.1103/PhysRevLett.64.1990} {\bibfield
  {journal} {\bibinfo  {journal} {Phys. Rev. Lett.}\ }\textbf {\bibinfo
  {volume} {64}},\ \bibinfo {pages} {1990} (\bibinfo {year}
  {1990})}\BibitemShut {NoStop}%
\bibitem [{\citenamefont {Wilson}(1974)}]{Wilson74}%
  \BibitemOpen
  \bibfield  {author} {\bibinfo {author} {\bibfnamefont {K.~G.}\ \bibnamefont
  {Wilson}},\ }\href {\doibase 10.1103/PhysRevD.10.2445} {\bibfield  {journal}
  {\bibinfo  {journal} {Phys. Rev. D}\ }\textbf {\bibinfo {volume} {10}},\
  \bibinfo {pages} {2445} (\bibinfo {year} {1974})}\BibitemShut {NoStop}%
\bibitem [{\citenamefont {Graf}\ and\ \citenamefont {Vogl}(1995)}]{Graf95}%
  \BibitemOpen
  \bibfield  {author} {\bibinfo {author} {\bibfnamefont {M.}~\bibnamefont
  {Graf}}\ and\ \bibinfo {author} {\bibfnamefont {P.}~\bibnamefont {Vogl}},\
  }\href {\doibase 10.1103/PhysRevB.51.4940} {\bibfield  {journal} {\bibinfo
  {journal} {Phys. Rev. B}\ }\textbf {\bibinfo {volume} {51}},\ \bibinfo
  {pages} {4940} (\bibinfo {year} {1995})}\BibitemShut {NoStop}%
\bibitem [{\citenamefont {Bl\"umer}(2002)}]{bluemer02}%
  \BibitemOpen
  \bibfield  {author} {\bibinfo {author} {\bibfnamefont {N.}~\bibnamefont
  {Bl\"umer}},\ }\emph {\bibinfo {title} {Mott-Hubbard Metal-Insulator
  Transition and Optical Conductivity in High Dimensions}},\ \href@noop {}
  {Ph.D. thesis},\ \bibinfo  {school} {Universit\"at Augsburg} (\bibinfo {year}
  {2002})\BibitemShut {NoStop}%
\bibitem [{\citenamefont {Mahan}(2000)}]{Mahan00}%
  \BibitemOpen
  \bibfield  {author} {\bibinfo {author} {\bibfnamefont {G.~D.}\ \bibnamefont
  {Mahan}},\ }\href@noop {} {\emph {\bibinfo {title} {Many-Particle
  Physics}}},\ Physics of Solids and Liquids\ (\bibinfo  {publisher} {Plenum},\
  \bibinfo {year} {2000})\ \bibinfo {note} {3rd ed.}\BibitemShut {Stop}%
\bibitem [{\citenamefont {Nozières}(1964)}]{Nozieres64}%
  \BibitemOpen
  \bibfield  {author} {\bibinfo {author} {\bibfnamefont {P.}~\bibnamefont
  {Nozières}},\ }\href@noop {} {\emph {\bibinfo {title} {{Theory of
  interacting Fermi systems}}}},\ Advanced book classics\ (\bibinfo
  {publisher} {Addison-Wesley},\ \bibinfo {address} {Reading, MA},\ \bibinfo
  {year} {1964})\BibitemShut {NoStop}%
\bibitem [{\citenamefont {Platzman}\ and\ \citenamefont
  {Wolff}(1973)}]{Platzman73}%
  \BibitemOpen
  \bibfield  {author} {\bibinfo {author} {\bibfnamefont {P.~M.}\ \bibnamefont
  {Platzman}}\ and\ \bibinfo {author} {\bibfnamefont {P.~A.}\ \bibnamefont
  {Wolff}},\ }\href@noop {} {\emph {\bibinfo {title} {Waves and Interactions in
  Solid State Plasmas}}}\ (\bibinfo  {publisher} {New York: Academic Press},\
  \bibinfo {year} {1973})\BibitemShut {NoStop}%
\bibitem [{\citenamefont {Giuliani}\ and\ \citenamefont
  {Vignale}(2005)}]{Giuliani05}%
  \BibitemOpen
  \bibfield  {author} {\bibinfo {author} {\bibfnamefont {G.}~\bibnamefont
  {Giuliani}}\ and\ \bibinfo {author} {\bibfnamefont {G.}~\bibnamefont
  {Vignale}},\ }\href@noop {} {\emph {\bibinfo {title} {Quantum Theory of the
  Electron Liquid}}}\ (\bibinfo  {publisher} {Cambridge University Press},\
  \bibinfo {year} {2005})\BibitemShut {NoStop}%
\bibitem [{\citenamefont {Chitra}\ and\ \citenamefont
  {Kotliar}(2000)}]{Chitra00}%
  \BibitemOpen
  \bibfield  {author} {\bibinfo {author} {\bibfnamefont {R.}~\bibnamefont
  {Chitra}}\ and\ \bibinfo {author} {\bibfnamefont {G.}~\bibnamefont
  {Kotliar}},\ }\href {\doibase 10.1103/PhysRevLett.84.3678} {\bibfield
  {journal} {\bibinfo  {journal} {Phys. Rev. Lett.}\ }\textbf {\bibinfo
  {volume} {84}},\ \bibinfo {pages} {3678} (\bibinfo {year}
  {2000})}\BibitemShut {NoStop}%
\bibitem [{\citenamefont {van Loon}\ \emph {et~al.}(2014)\citenamefont {van
  Loon}, \citenamefont {Lichtenstein}, \citenamefont {Katsnelson},
  \citenamefont {Parcollet},\ and\ \citenamefont {Hafermann}}]{vanLoon2014}%
  \BibitemOpen
  \bibfield  {author} {\bibinfo {author} {\bibfnamefont {E.~G. C.~P.}\
  \bibnamefont {van Loon}}, \bibinfo {author} {\bibfnamefont {A.~I.}\
  \bibnamefont {Lichtenstein}}, \bibinfo {author} {\bibfnamefont {M.~I.}\
  \bibnamefont {Katsnelson}}, \bibinfo {author} {\bibfnamefont
  {O.}~\bibnamefont {Parcollet}}, \ and\ \bibinfo {author} {\bibfnamefont
  {H.}~\bibnamefont {Hafermann}},\ }\href {http://arxiv.org/abs/1406.2150} {\
  (\bibinfo {year} {2014})},\ \Eprint {http://arxiv.org/abs/1406.2150}
  {arXiv:1406.2150 [cond-mat]} \BibitemShut {NoStop}%
\bibitem [{\citenamefont {Anderson}(1997)}]{anderson97}%
  \BibitemOpen
  \bibfield  {author} {\bibinfo {author} {\bibfnamefont {P.}~\bibnamefont
  {Anderson}},\ }\href {http://books.google.fr/books?id=dDViQgAACAAJ} {\emph
  {\bibinfo {title} {The Theory of Superconductivity in the High-TC Cuprate
  Superconductors}}},\ Princeton Series in Physics\ (\bibinfo  {publisher}
  {Princeton University Press},\ \bibinfo {year} {1997})\BibitemShut {NoStop}%
\bibitem [{\citenamefont {Abrikosov}\ \emph {et~al.}(1965)\citenamefont
  {Abrikosov}, \citenamefont {Gor'kov},\ and\ \citenamefont
  {Dzyaloshinskii}}]{Abrikosov65}%
  \BibitemOpen
  \bibfield  {author} {\bibinfo {author} {\bibfnamefont {A.~A.}\ \bibnamefont
  {Abrikosov}}, \bibinfo {author} {\bibfnamefont {L.~P.}\ \bibnamefont
  {Gor'kov}}, \ and\ \bibinfo {author} {\bibfnamefont {I.~E.}\ \bibnamefont
  {Dzyaloshinskii}},\ }\href@noop {} {\emph {\bibinfo {title} {Methods of
  Quantum Field Theory in Statistical Physics}}}\ (\bibinfo  {publisher}
  {Pergamon Press, New York},\ \bibinfo {year} {1965})\BibitemShut {NoStop}%
\bibitem [{\citenamefont {Hafermann}\ \emph {et~al.}(2013)\citenamefont
  {Hafermann}, \citenamefont {Werner},\ and\ \citenamefont
  {Gull}}]{Hafermann13}%
  \BibitemOpen
  \bibfield  {author} {\bibinfo {author} {\bibfnamefont {H.}~\bibnamefont
  {Hafermann}}, \bibinfo {author} {\bibfnamefont {P.}~\bibnamefont {Werner}}, \
  and\ \bibinfo {author} {\bibfnamefont {E.}~\bibnamefont {Gull}},\ }\href
  {\doibase http://dx.doi.org/10.1016/j.cpc.2012.12.013} {\bibfield  {journal}
  {\bibinfo  {journal} {Computer Physics Communications}\ }\textbf {\bibinfo
  {volume} {184}},\ \bibinfo {pages} {1280 } (\bibinfo {year}
  {2013})}\BibitemShut {NoStop}%
\bibitem [{\citenamefont {Bauer}\ \emph {et~al.}(2011)\citenamefont {Bauer},
  \citenamefont {Carr}, \citenamefont {Evertz}, \citenamefont {Feiguin},
  \citenamefont {Freire}, \citenamefont {Fuchs}, \citenamefont {Gamper},
  \citenamefont {Gukelberger}, \citenamefont {Gull}, \citenamefont {Guertler},
  \citenamefont {Hehn}, \citenamefont {Igarashi}, \citenamefont {Isakov},
  \citenamefont {Koop}, \citenamefont {Ma}, \citenamefont {Mates},
  \citenamefont {Matsuo}, \citenamefont {Parcollet}, \citenamefont
  {Pawłowski}, \citenamefont {Picon}, \citenamefont {Pollet}, \citenamefont
  {Santos}, \citenamefont {Scarola}, \citenamefont {Schollwöck}, \citenamefont
  {Silva}, \citenamefont {Surer}, \citenamefont {Todo}, \citenamefont {Trebst},
  \citenamefont {Troyer}, \citenamefont {Wall}, \citenamefont {Werner},\ and\
  \citenamefont {Wessel}}]{ALPS2}%
  \BibitemOpen
  \bibfield  {author} {\bibinfo {author} {\bibfnamefont {B.}~\bibnamefont
  {Bauer}}, \bibinfo {author} {\bibfnamefont {L.~D.}\ \bibnamefont {Carr}},
  \bibinfo {author} {\bibfnamefont {H.~G.}\ \bibnamefont {Evertz}}, \bibinfo
  {author} {\bibfnamefont {A.}~\bibnamefont {Feiguin}}, \bibinfo {author}
  {\bibfnamefont {J.}~\bibnamefont {Freire}}, \bibinfo {author} {\bibfnamefont
  {S.}~\bibnamefont {Fuchs}}, \bibinfo {author} {\bibfnamefont
  {L.}~\bibnamefont {Gamper}}, \bibinfo {author} {\bibfnamefont
  {J.}~\bibnamefont {Gukelberger}}, \bibinfo {author} {\bibfnamefont
  {E.}~\bibnamefont {Gull}}, \bibinfo {author} {\bibfnamefont {S.}~\bibnamefont
  {Guertler}}, \bibinfo {author} {\bibfnamefont {A.}~\bibnamefont {Hehn}},
  \bibinfo {author} {\bibfnamefont {R.}~\bibnamefont {Igarashi}}, \bibinfo
  {author} {\bibfnamefont {S.~V.}\ \bibnamefont {Isakov}}, \bibinfo {author}
  {\bibfnamefont {D.}~\bibnamefont {Koop}}, \bibinfo {author} {\bibfnamefont
  {P.~N.}\ \bibnamefont {Ma}}, \bibinfo {author} {\bibfnamefont
  {P.}~\bibnamefont {Mates}}, \bibinfo {author} {\bibfnamefont
  {H.}~\bibnamefont {Matsuo}}, \bibinfo {author} {\bibfnamefont
  {O.}~\bibnamefont {Parcollet}}, \bibinfo {author} {\bibfnamefont
  {G.}~\bibnamefont {Pawłowski}}, \bibinfo {author} {\bibfnamefont {J.~D.}\
  \bibnamefont {Picon}}, \bibinfo {author} {\bibfnamefont {L.}~\bibnamefont
  {Pollet}}, \bibinfo {author} {\bibfnamefont {E.}~\bibnamefont {Santos}},
  \bibinfo {author} {\bibfnamefont {V.~W.}\ \bibnamefont {Scarola}}, \bibinfo
  {author} {\bibfnamefont {U.}~\bibnamefont {Schollwöck}}, \bibinfo {author}
  {\bibfnamefont {C.}~\bibnamefont {Silva}}, \bibinfo {author} {\bibfnamefont
  {B.}~\bibnamefont {Surer}}, \bibinfo {author} {\bibfnamefont
  {S.}~\bibnamefont {Todo}}, \bibinfo {author} {\bibfnamefont {S.}~\bibnamefont
  {Trebst}}, \bibinfo {author} {\bibfnamefont {M.}~\bibnamefont {Troyer}},
  \bibinfo {author} {\bibfnamefont {M.~L.}\ \bibnamefont {Wall}}, \bibinfo
  {author} {\bibfnamefont {P.}~\bibnamefont {Werner}}, \ and\ \bibinfo {author}
  {\bibfnamefont {S.}~\bibnamefont {Wessel}},\ }\href@noop {} {\bibfield
  {journal} {\bibinfo  {journal} {Journal of Statistical Mechanics: Theory and
  Experiment}\ }\textbf {\bibinfo {volume} {2011}},\ \bibinfo {pages} {P05001}
  (\bibinfo {year} {2011})}\BibitemShut {NoStop}%
\bibitem [{\citenamefont {Werner}\ and\ \citenamefont
  {Millis}(2007)}]{Werner07}%
  \BibitemOpen
  \bibfield  {author} {\bibinfo {author} {\bibfnamefont {P.}~\bibnamefont
  {Werner}}\ and\ \bibinfo {author} {\bibfnamefont {A.~J.}\ \bibnamefont
  {Millis}},\ }\href {\doibase 10.1103/PhysRevLett.99.146404} {\bibfield
  {journal} {\bibinfo  {journal} {Phys. Rev. Lett.}\ }\textbf {\bibinfo
  {volume} {99}},\ \bibinfo {pages} {146404} (\bibinfo {year}
  {2007})}\BibitemShut {NoStop}%
\bibitem [{\citenamefont {Werner}\ and\ \citenamefont
  {Millis}(2010)}]{Werner10}%
  \BibitemOpen
  \bibfield  {author} {\bibinfo {author} {\bibfnamefont {P.}~\bibnamefont
  {Werner}}\ and\ \bibinfo {author} {\bibfnamefont {A.~J.}\ \bibnamefont
  {Millis}},\ }\href {\doibase 10.1103/PhysRevLett.104.146401} {\bibfield
  {journal} {\bibinfo  {journal} {Phys. Rev. Lett.}\ }\textbf {\bibinfo
  {volume} {104}},\ \bibinfo {pages} {146401} (\bibinfo {year}
  {2010})}\BibitemShut {NoStop}%
\bibitem [{\citenamefont {Hafermann}(2014)}]{Hafermann14}%
  \BibitemOpen
  \bibfield  {author} {\bibinfo {author} {\bibfnamefont {H.}~\bibnamefont
  {Hafermann}},\ }\href {\doibase 10.1103/PhysRevB.89.235128} {\bibfield
  {journal} {\bibinfo  {journal} {Phys. Rev. B}\ }\textbf {\bibinfo {volume}
  {89}},\ \bibinfo {pages} {235128} (\bibinfo {year} {2014})}\BibitemShut
  {NoStop}%
\bibitem [{\citenamefont {Vidberg}\ and\ \citenamefont
  {Serene}(1977)}]{Vidberg77}%
  \BibitemOpen
  \bibfield  {author} {\bibinfo {author} {\bibfnamefont {H.~J.}\ \bibnamefont
  {Vidberg}}\ and\ \bibinfo {author} {\bibfnamefont {J.~W.}\ \bibnamefont
  {Serene}},\ }\href {\doibase 10.1007/BF00655090} {\bibfield  {journal}
  {\bibinfo  {journal} {J. Low Temp. Phys.}\ }\textbf {\bibinfo {volume}
  {29}},\ \bibinfo {pages} {179} (\bibinfo {year} {1977})}\BibitemShut
  {NoStop}%
\bibitem [{\citenamefont {Huang}\ \emph {et~al.}(2014)\citenamefont {Huang},
  \citenamefont {Ayral}, \citenamefont {Biermann},\ and\ \citenamefont
  {Werner}}]{Huang14}%
  \BibitemOpen
  \bibfield  {author} {\bibinfo {author} {\bibfnamefont {L.}~\bibnamefont
  {Huang}}, \bibinfo {author} {\bibfnamefont {T.}~\bibnamefont {Ayral}},
  \bibinfo {author} {\bibfnamefont {S.}~\bibnamefont {Biermann}}, \ and\
  \bibinfo {author} {\bibfnamefont {P.}~\bibnamefont {Werner}},\ }\href
  {\doibase 10.1103/PhysRevB.90.195114} {\bibfield  {journal} {\bibinfo
  {journal} {Phys. Rev. B}\ }\textbf {\bibinfo {volume} {90}},\ \bibinfo
  {pages} {195114} (\bibinfo {year} {2014})}\BibitemShut {NoStop}%
\bibitem [{\citenamefont {Peierls}(1933)}]{Peierls33}%
  \BibitemOpen
  \bibfield  {author} {\bibinfo {author} {\bibfnamefont {R.~E.}\ \bibnamefont
  {Peierls}},\ }\href@noop {} {\bibfield  {journal} {\bibinfo  {journal} {Z.
  Phys.}\ }\textbf {\bibinfo {volume} {80}},\ \bibinfo {pages} {763–791}
  (\bibinfo {year} {1933})}\BibitemShut {NoStop}%
\bibitem [{\citenamefont {Alexandrov}\ and\ \citenamefont
  {Capellmann}(1991)}]{Alexandrov91}%
  \BibitemOpen
  \bibfield  {author} {\bibinfo {author} {\bibfnamefont {A.~S.}\ \bibnamefont
  {Alexandrov}}\ and\ \bibinfo {author} {\bibfnamefont {H.}~\bibnamefont
  {Capellmann}},\ }\href {\doibase 10.1103/PhysRevLett.66.365} {\bibfield
  {journal} {\bibinfo  {journal} {Phys. Rev. Lett.}\ }\textbf {\bibinfo
  {volume} {66}},\ \bibinfo {pages} {365} (\bibinfo {year} {1991})}\BibitemShut
  {NoStop}%
\bibitem [{\citenamefont {Schrieffer}(1999)}]{Schrieffer99}%
  \BibitemOpen
  \bibfield  {author} {\bibinfo {author} {\bibfnamefont {J.}~\bibnamefont
  {Schrieffer}},\ }\href@noop {} {\emph {\bibinfo {title} {Theory of
  Superconductivity}}},\ Advanced Book Program Series\ (\bibinfo  {publisher}
  {Advanced Book Program, Perseus Books},\ \bibinfo {year} {1999})\BibitemShut
  {NoStop}%
\end{thebibliography}%

\end{document}